\documentclass[11pt,a4paper]{article}

\usepackage{float}
\usepackage{subcaption}
\usepackage[a4paper, total={6in, 9in}]{geometry}

\usepackage{amsmath,latexsym,amssymb,slashed}
\usepackage{graphicx}		
\usepackage{amssymb}			
\usepackage{tabularx}	
\usepackage{hhline}
\usepackage[latin1]{inputenc}
\usepackage[T1]{fontenc}
\usepackage{slashed}
\usepackage{cite}
\usepackage{comment}
\usepackage{lipsum}

\usepackage[pdftex,hyperref,x11names]{xcolor}
\usepackage[pdftex,colorlinks=true,
pdfstartview=FitV,
pdfnewwindow=true,
linktoc = page,
linkcolor= RoyalBlue3,
citecolor= Red3,
urlcolor= RoyalBlue3,
hyperindex=true,
hyperfigures=false]{hyperref}

\usepackage{diagbox}
\usepackage{subcaption}

\newcommand{\eq}[1]{\begin{equation}#1\end{equation}}
\newcommand{\spl}[1]{\begin{split}#1\end{split}}

\def\bea{\begin{eqnarray}}
\def\eea{\end{eqnarray}}

\def\nn{\nonumber}

\def\d{\text{d}}

\def\a{\alpha}
\def\b{\beta}
\def\c{\gamma}

\newcommand{\om}{\omega}

\newcommand{\Sd}{\dot{S}}
\newcommand{\Sdd}{\ddot{S}}

\def \ctu{\tilde{c}_1}
\def \ctd{\tilde{c}_2}

\begin{document}

\begin{titlepage}
	\vspace{5cm}
	
	\begin{center}
    {\huge \bf
		Universal accelerating  cosmologies \\
		from 10d supergravity}
		\vspace{1.2cm}
		
		{\sc \large Paul Marconnet} {and} {\sc \large Dimitrios Tsimpis}
		
		\vspace{0.3cm}
		
		{\it   Institut de Physique des Deux Infinis de Lyon   }\\
		{\it Universit\'e de Lyon, UCBL, UMR 5822, CNRS/IN2P3 }\\
		{\it 4 rue Enrico Fermi, 69622 Villeurbanne Cedex, France  }
		
		\vspace{0.3 cm}
		
		\texttt{ \href{mailto:marconnet@ipnl.in2p3.fr}{marconnet@ipnl.in2p3.fr}, \; \href{mailto:tsimpis@ipnl.in2p3.fr}{tsimpis@ipnl.in2p3.fr}}
			
		\vspace{1.2 cm}
		
%
\end{center}
\begin{abstract}
\normalsize\noindent 
We study  4d Friedmann-Lema\^{i}tre-Robertson-Walker cosmologies obtained from 
 time-dependent compactifications of Type IIA 10d supergravity on various classes of 6d manifolds (Calabi-Yau, Einstein, Einstein-K\"{a}hler).~The cosmologies we present are {\it universal} in that they do not depend on the detailed features of the compactification manifold, but only on the properties which are common to all the manifolds belonging to that class.~Once the equations of motion are rewritten as an appropriate dynamical system, the existence of solutions featuring a phase of accelerated expansion is made manifest.~The fixed points of this dynamical system, as well as the trajectories on the boundary of the phase space, correspond to analytic solutions which we determine explicitly.~Furthermore, some of the resulting cosmologies exhibit eternal or semi-eternal acceleration, whereas others allow for a parametric control on the number of e-foldings.~At future infinity, one can achieve both large volume and weak string coupling.~Moreover, we find several smooth accelerating cosmologies without Big Bang singularities: the universe is contracting in the cosmological past ($T<0$), expanding in the future ($T>0$), while in the vicinity of $T=0$ 
 it becomes de Sitter in hyperbolic slicing.~We also obtain several cosmologies featuring an infinite number of cycles of alternating  periods of accelerated and decelerated expansions. 
\end{abstract}

\end{titlepage}
\newgeometry{total={6in, 9in}}
\tableofcontents

\newpage

\section{Introduction and summary}
\label{sec:intro}

\setcounter{footnote}{0}

Obtaining realistic 4d cosmologies from the ten-dimensional 
supergravities that capture the  low-energy limit of superstring theory  
has proven notoriously difficult. At the turn of the century, it was thought that accelerating cosmologies 
were as difficult to achieve as de Sitter space itself, being subject to 
a famous no-go theorem  first discovered by Gibbons \cite{Gibbons1,Gibbons:2003gb} and later rediscovered by 
Maldacena-Nu\~{n}ez \cite{Maldacena:2000mw} in a string-theory context. 
More precisely,   a matter whose stress-energy tensor  in the higher-dimensional theory implies  ${R}_{00}\geq0$,   as a consequence of the Einstein equations, 
  is said to satisfy the strong energy condition (SEC).  The SEC  implies  that time-independent compactifications of the higher-dimensional theory can never lead to 4d cosmologies with 
accelerated expansion (which includes de Sitter space as a special case).\footnote{The no-go was also extended in  \cite{Maldacena:2000mw} to the case of massive IIA supergravity, a theory which does {\it not} obey the SEC. }

However, as  was first pointed out in \cite{Townsend:2003fx}, time-dependent compactifications evade the no-go and can 
lead to 4d Einstein-frame accelerated expansion  for some 
period of time \cite{Ohta:2003pu,Ohta:2003ie,Ohta:2004wk,Roy:2003nd,Gutperle:2003kc,Emparan:2003gg}.  
Such transient acceleration is in fact generic in flux compactifications, see \cite{Townsend:2003qv} for a review, 
although de Sitter space is still ruled out by the SEC, if the time-independence of the 4d Newton's constant is obeyed  in a conventional way \cite{Russo:2018akp}. 
If instead the time-independence of the 4d Newton's constant is  obeyed in an averaged way, 
even de Sitter space is not ruled out by the SEC, although the so-called dominant energy condition does rule out non-singular de Sitter compactifications \cite{Russo:2019fnk} (see also \cite{Bernardo:2022ony, Bernardo:2021zxo} for a recent discussion on energy conditions). 

In \cite{Russo:2018akp}, Russo and Townsend greatly refined the no-go of \cite{Gibbons1,Gibbons:2003gb, Maldacena:2000mw}: one of their conclusions 
is that, even if one imposes the SEC and the time-independence of the 4d Newton's constant in a conventional way, 
late-time accelerating cosmologies are {\it not} ruled out. 
However, no late-time accelerating cosmologies from compactification of the  ten- or eleven-dimensional supergravities arising as  low-energy effective actions of string theory 
have ever been constructed.~Indeed, the  eternal accelerating cosmologies of \cite{Chen:2003dca,Andersson:2006du} are such that the acceleration of the scale factor tends to zero at 
future asymptotic infinity, so that there is no cosmological horizon.

Ref.~\cite{Emparan:2003gg} reinterpreted the accelerating solutions of \cite{Townsend:2003fx} from the point of view of a 4d theory with a 
scalar potential.~It was found that there is always a big-bang singularity near which the scale factor behaves as a power law: $S(T)\sim T^{\frac13}$, 
which does not lead to enough e-foldings for inflation \cite{Chen:2003dca,Wohlfarth:2003kw}.  
 The transient acceleration of the solutions could, however, 
 be used  to describe the current cosmological epoch \cite{Gutperle:2003kc}. The general characteristics of one-field inflation with an exponential potential 
were studied in \cite{Halliwell:1986ja}, while  cosmologies from an effective theory with multiple 
 scalar fields were studied in \cite{Collinucci:2004iw,Bergshoeff:2008be,Bergshoeff:2008zza}.  In particular, the analysis of \cite{Townsend:2004zp}  could be  relevant for the 
 swampland conjecture \cite{Obied:2018sgi}.~Cosmological solutions of gauged supergravities and F-theory have been studied in 
 \cite{Caviezel:2008tf,Flauger:2008ad,Danielsson:2011au,Fre:2002pd,Catino:2013syn,deRoo:2002jf,deRoo:2006ms,Kleinschmidt:2005gz, Blaback:2013sda, Heckman:2018mxl, Heckman:2019dsj, Hebecker:2019csg, Calderon-Infante:2022nxb}. \\
 
In the present paper we re-examine some of the previous statements in the context of {\it universal} cosmologies, obtained by compactification to 4d of 
10d Type IIA supergravity (with or  without Romans mass) on 6d Einstein, Einstein-K\"{a}hler, or Calabi-Yau (CY) manifolds. 
In this context, the term ``universal'' means that the ans\"{a}tze that we consider do not depend on the detailed features of the manifold on which we compactify, but only on the  properties which are common to all the manifolds belonging to that class.~For example, our ansatz for the CY compactification exploits the existence of a holomorphic three-form and a K\"{a}hler form, but does not assume the existence of 
any additional harmonic forms on the manifold. 

Our  universal cosmologies   are obtained by solving the {\it ten-dimensional} supergravity equations:~we do {\it not} start from a 4d effective action.~Nevertheless, it turns out that all of the resulting 10d equations of motion are obtainable from a 1d action where all fields only depend on a time coordinate.~In other words, there is a 1d consistent truncation of the theory, $S_{1\d}$.~The fields 
in question are the dilaton $\phi$ and two warp factors $A$, $B$ (one for the internal and one for the external space), while, by virtue of our ansatz,  all fluxes manifest themselves 
as constant coefficients in the potential of $S_{1\d}$, cf.~Table~\ref{tab:constants}.

Moreover we show that 
a two-scalar 4d {\it cosmological} consistent truncation, $S_{4\d}$, of the 10d theory to $\phi$, $A$ is possible in certain special cases.~In other words, every solution of cosmological Friedmann-Lema\^{i}tre-Robertson-Walker (FLRW) type of $S_{4\d}$ uplifts to a solution of the ten-dimensional equations of motion.~Again, the different fluxes 
show up as constant coefficients in the potential of $S_{4\d}$. At least for a certain range of the parameters in the potential, we expect our  $S_{4\d}$ 
to  coincide with the universal sector of the effective 4d theory of the compactification, see \cite{Andriot:2018tmb} for a recent discussion on consistent truncations vs effective actions. 
Much of the literature on string-theory cosmological models uses a 4d effective action and a 4d potential. We make contact with 
the potential description in Section \ref{sec:4dpotential}. \\

Whenever a single ``species'' of flux is turned on,\footnote{Here we use an extended notion of ``flux'' that also includes non-trivial curvature, but excludes the fields $A$, $B$, $\phi$ (the warp factors and the dilaton).} 
we are able to provide analytic cosmological solutions to the equations of motion.~These are described in detail in Section \ref{sec:analytical}. Whenever multiple fluxes are simultaneously turned on, 
one cannot  give an analytic solution in general, although this can still be possible for certain special values of the flux parameters. 
Whenever exactly two different species of flux are turned on, although an analytic solution is not possible in general, 
a powerful tool becomes available:~as we show in Section \ref{sec:dsa}, the equations of motion can be cast in the form of an autonomous dynamical system of three first-order equations and one constraint. 
The use of dynamical-system techniques in general relativity is of course not new: refs.~\cite{Gregory:1999gv, Chodos:2000tf, Townsend:2003fx, Jarv:2004uk, Sonner:2006yn, Sonner:2007cp, Russo:2018akp, Russo:2022pgo} in particular are closely related to the strategy employed here. 
One of the novelties of the present paper is to give an exhaustive analysis of universal cosmological solutions from compactification on the 
previously mentioned classes of manifolds.

The resulting dynamical system description is rather intuitive and captures several generic features of cosmological solutions coming from models of flux compactification. 
Each solution is represented by a trajectory in a three-dimensional phase space.~The constraint forces the trajectories to lie in the interior of a sphere, with expanding  
cosmologies corresponding to trajectories in the northern hemisphere.~Whenever the system of equations admits fixed points, these correspond to analytic solutions.

The boundary of the allowed phase space corresponding to expanding cosmologies, 
i.e.~the equatorial disc and the surface of the northern hemisphere, are both invariant surfaces. This implies that trajectories 
that do not lie entirely on these surfaces can only approach them asymptotically.~Trajectories that do lie entirely on either of these surfaces also correspond to analytic solutions 
(namely analytic solutions with a single species of flux turned on: indeed restricting the trajectory to either the surface of the sphere or the equatorial disc, corresponds to 
turning off one of the two species of flux). 
Moreover, the intersection of these two invariant surfaces -- the equator -- constitutes a circle of fixed points, and corresponds to a family of analytic solutions with all flux turned off, described in Section \ref{sec:special}. The dynamical system 
generally possesses a third invariant surface, an invariant plane $\mathcal{P}$, whose disposition depends on the fluxes of the solution. 

The analytic solutions corresponding to fixed points are all cosmologies with a power-law scale factor, $S(T)\sim T^{a}$, with $\tfrac13\leq a\leq1$.~The fixed points at the equator correspond to power-law 
(scaling) cosmologies with $a=\tfrac13$, while $a=1$ corresponds to either a fixed point at the origin of the sphere (whenever the dynamical system 
admits such a fixed point), or a fixed point in the bulk of the sphere and on the boundary of the acceleration region. In the former case the corresponding cosmology is that of a regular Milne universe, whereas in the latter case it is a Milne universe with angular defect.~In addition, we find fixed points corresponding to  cosmologies with $a=\tfrac34$, $\tfrac{19}{25}$, or $\tfrac{9}{11}$. The list of 
analytic solutions is given in Table  \ref{tab:analyticsol}.

Trajectories interpolating between two different fixed points asymptote the respective scaling solutions in the past and future infinity.~Note in particular that we find no fixed 
points with   $a>1$ (which would correspond to eternally accelerating scaling cosmologies). 

The question of acceleration becomes particularly transparent in the dynamical system description: the acceleration period of the solution corresponds to the 
portion of the trajectory that lies in a certain region in phase space. This ``acceleration region'' is entirely fixed by the type of flux that is turned on, with the different 
cases summarized in Table \ref{tab:regions}. 
\begin{table}[H]
    \centering
\begin{tabular}{|c||c|c|c|}
\hline
\diagbox{$\beta_1$}{$\beta_2$} & 0 & 4 & 6\\
\hline
\hline
0 & $\varnothing$ & $\varnothing$ & \begin{minipage}[c][2cm][c]{.25\textwidth}
\centering
      \includegraphics[width=20mm]{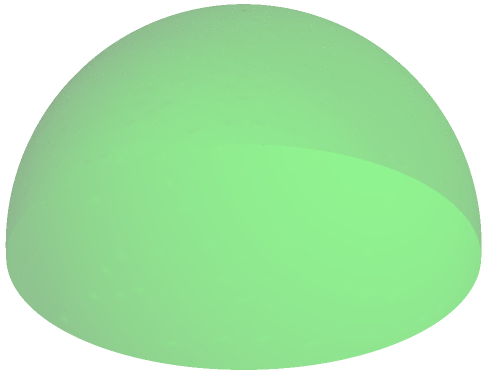}
    \end{minipage} \\
\hline
4 & $\varnothing$ & $\varnothing$ & \begin{minipage}[c][3cm][c]{.25\textwidth}
\centering
      \includegraphics[width=20mm]{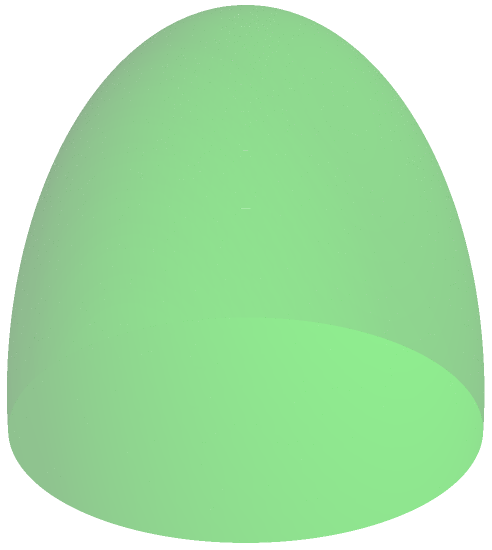}
    \end{minipage} \\
\hline
6 & \begin{minipage}{.25\textwidth}
\centering
      \includegraphics[width=35mm]{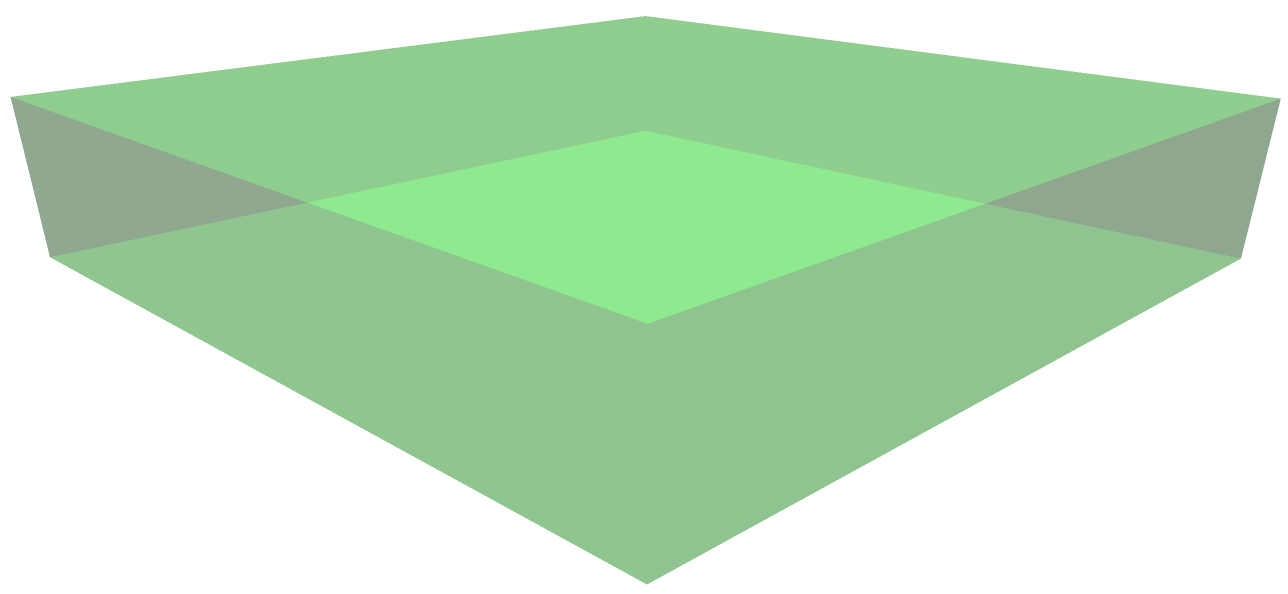}
    \end{minipage} & \begin{minipage}[c][2cm][c]{.25\textwidth}
\centering
      \includegraphics[width=30mm]{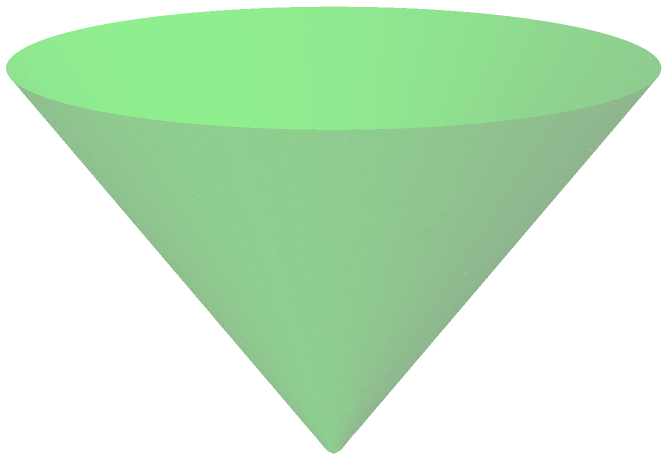}
    \end{minipage} & \begin{minipage}[c][3cm][c]{.25\textwidth}
\centering
      \includegraphics[width=20mm]{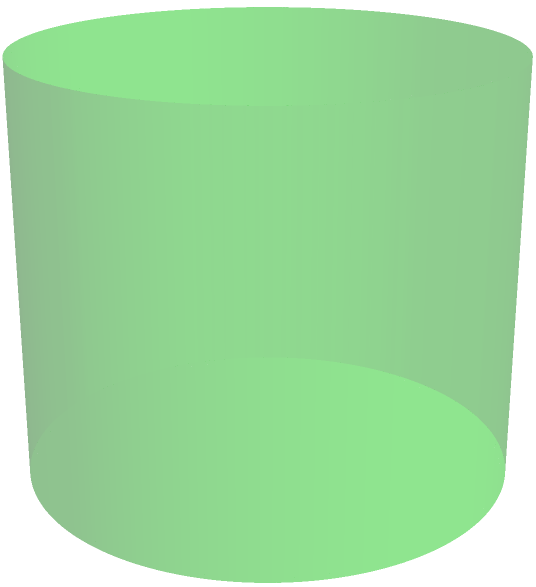}
    \end{minipage} \\
\hline
\end{tabular}
    \caption{
    Different types of universal two-flux compactifications and their corresponding acceleration regions in the phase space. The values of $\beta_{1,2}$ -- corresponding to the two species of flux that are turned on in the solution -- depend on the type of flux and its number of legs along the 
external-space,  internal-space and time  directions, cf.~\eqref{29azerty}. The absence of an acceleration region is denoted by the empty set $\varnothing$. 
}
    \label{tab:regions}
\end{table}
  
This clarifies why (transient) accelerated expansion is essentially generic in flux compactifications:~there is always a trajectory that passes by any given point in phase space (corresponding to some particular choice of initial conditions).~By making sure that  a trajectory passes by some point in the acceleration region  of the northern hemisphere,  
we can thus obtain a cosmological solution that 
will necessarily feature a period of accelerated expansion. Moreover,  the only non-trivial requirement, namely the existence of an acceleration region in phase space,  is not difficult to satisfy, as can be seen from Table \ref{tab:regions}.

\begin{figure}[H]
\begin{subfigure}{.5\textwidth}
  \centering
  \includegraphics[width=0.9\linewidth]{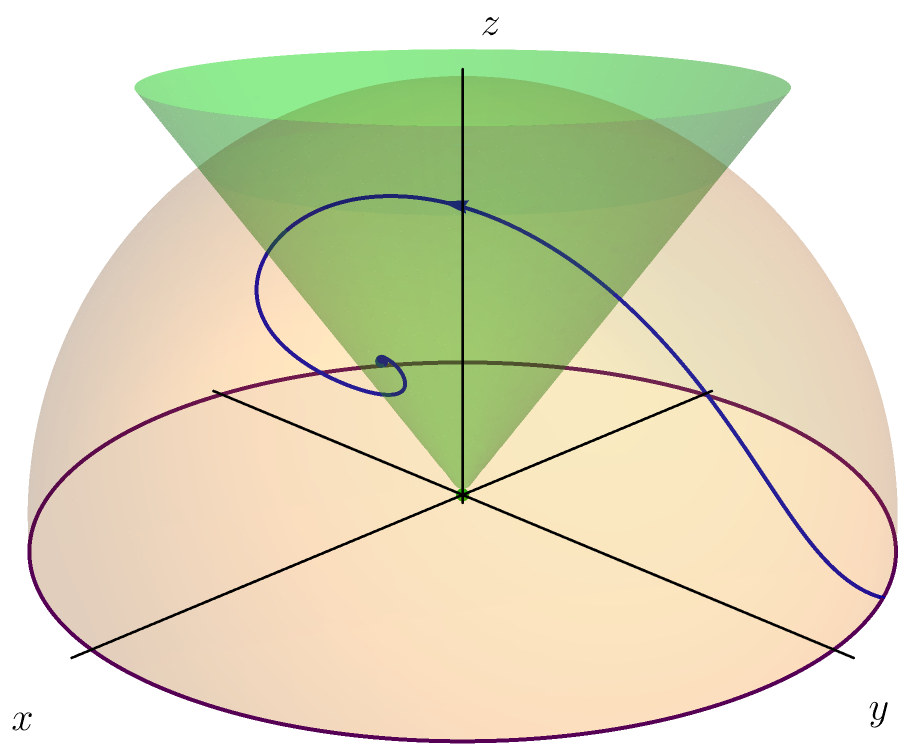}
  \caption{$(m,k)$}
  \label{fig:test1}
\end{subfigure}%
\begin{subfigure}{.5\textwidth}
  \centering
  \includegraphics[width=0.9\linewidth]{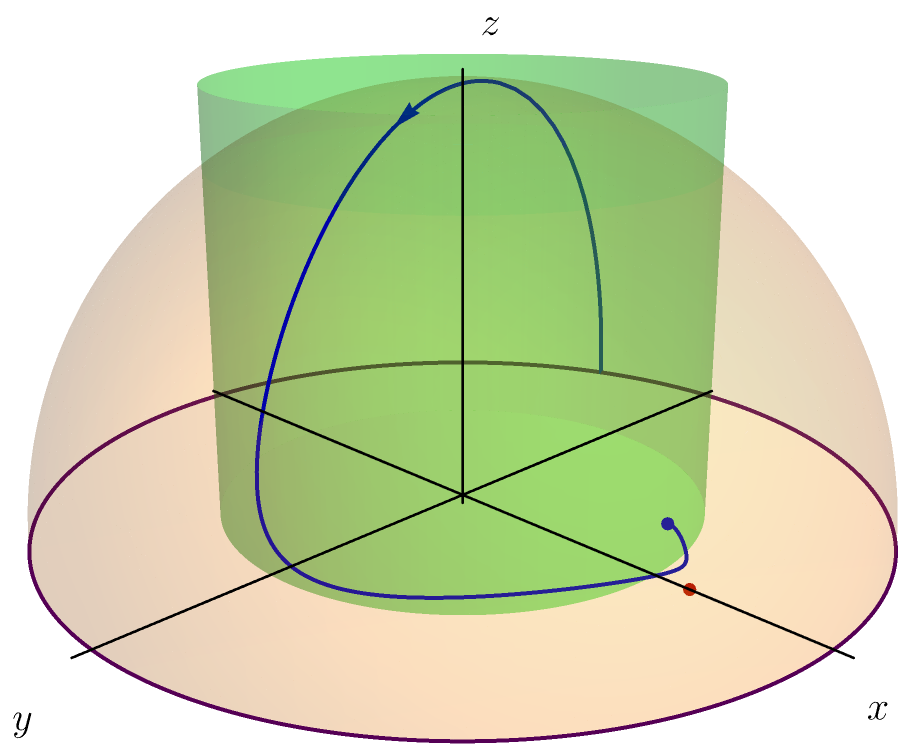}
  \caption{$(m,\lambda)$}
  \label{fig:test2}
\end{subfigure}
\begin{subfigure}{.5\textwidth}
  \centering
  \includegraphics[width=0.9\linewidth]{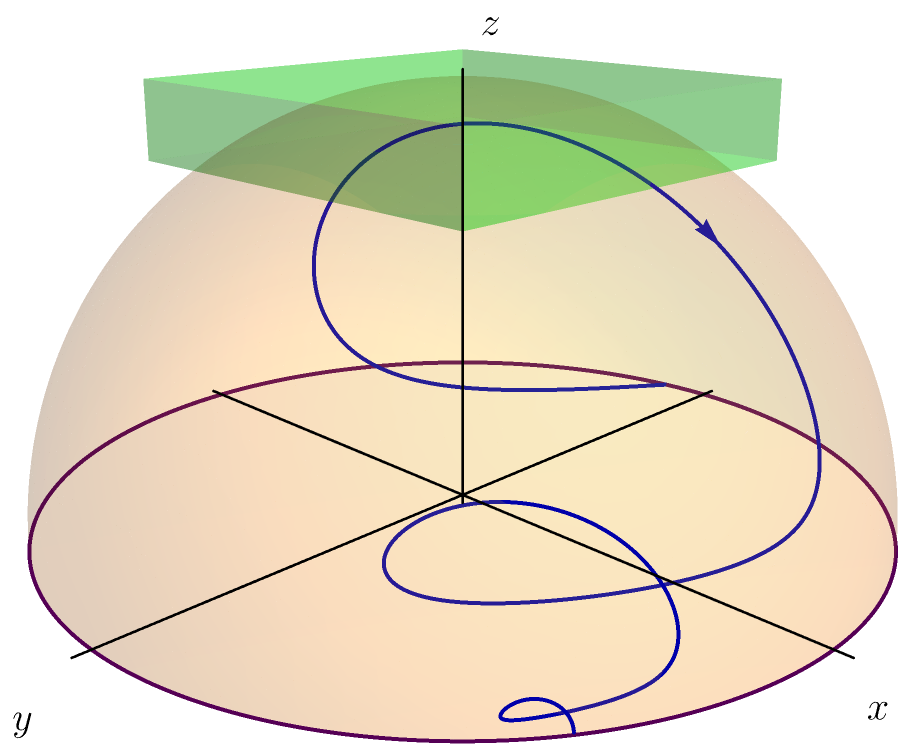}
  \caption{$(\varphi,\chi)$}
  \label{fig:test3}
\end{subfigure}
\begin{subfigure}{.5\textwidth}
  \centering
  \includegraphics[width=0.9\linewidth]{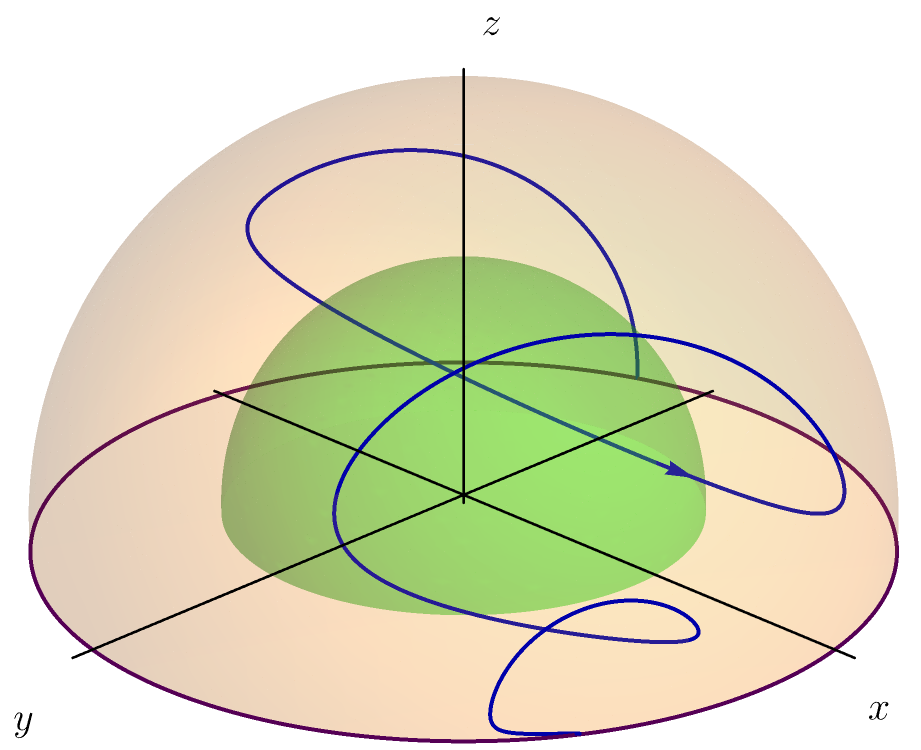}
  \caption{$(\chi,\varphi)$}
  \label{fig:test4}
\end{subfigure}
\caption{Trajectories in phase space corresponding to two-flux cosmological solutions with accelerated expansion. The corresponding dynamical systems are given in Sections \ref{case2}, \ref{sec:34d}, \ref{case4}. 
The pair of fluxes that are turned on in each case is indicated below each subfigure.   
The trajectories (blue lines) interpolate between a fixed point  on the equator in the past infinity, corresponding to a cosmology with a power-law scale factor  $S\sim T^{\frac13}$, 
and another fixed point at future infinity.
For cases \ref{fig:test3} and \ref{fig:test4} the second fixed point is also on the equator, and thus the solution asymptotes  the same scaling cosmology in the past and future infinities. Cases \ref{fig:test1} and \ref{fig:test2}  asymptote at future infinity a fixed point in the interior of the sphere which corresponds to a cosmology with a power-law scale factor $S \sim T$ and  $S\sim T^{\frac{19}{25}}$  respectively. Cases \ref{fig:test3} and \ref{fig:test4}   describe the same system, but employ different parametrizations  which interchange the role of the two fluxes that are turned on.  The transient accelerated expansion corresponds to the portion of the trajectory within the acceleration region depicted in green.
}
\label{fig:fig}
\end{figure}

In the dynamical system description, 
the calculation of the number of e-foldings, $N$, also becomes particularly transparent, since the flow parameter is simply the logarithm of the scale factor.~Most 
instances of transient acceleration are such that the number of e-foldings  is of order one, $N\sim\mathcal{O}(1)$, as was mentioned earlier.  
The  exceptions that we find all correspond to dynamical systems with fixed points on the boundary of the accelerated region, and they all give rise 
to hyperbolic FLRW cosmologies ($k<0$). 
In addition, there is a 
flux turned on corresponding to one of the following being non-zero:  $\lambda$, $m$, $c_f$, $c_0$, or $c_\varphi$; see~Table~\ref{tab:constants} for the  10d origin of these coefficients.  

The first of these solutions is examined in Section \ref{case1} and is obtained from compactification on a 6d Einstein manifold with negative curvature ($\lambda<0$). 
The second solution occurs in IIA supergravity with non-vanishing Romans mass ($m\neq0$) compactified on a Ricci-flat space, and is described in Section \ref{case2}. Despite  having vanishing Romans mass, 
the three remaining 
solutions ($k<0$ and $c_f$, $c_0$, or $c_\varphi\neq0$)
are all qualitatively very similar to the one of Section \ref{case2}. \\

\begin{figure}[H]
\begin{center}
\includegraphics[width=1\textwidth]{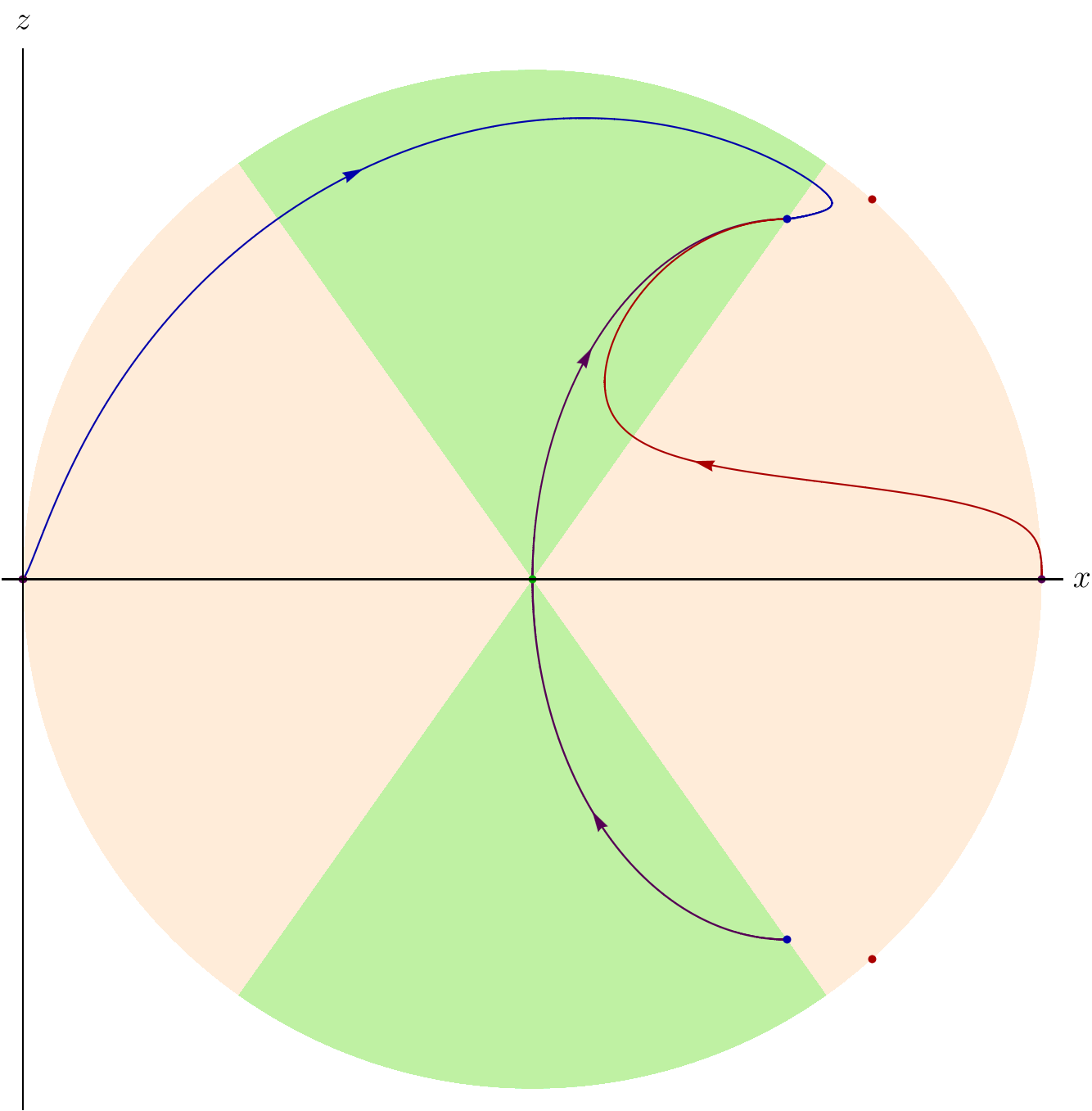}
\caption{Three trajectories lying on the invariant plane $\mathcal{P}$ of the dynamical system obtained from the compactification with $k,\lambda \neq 0$. The point $p_0$ corresponding to a Milne universe is depicted in green. The point $p_1$ (and its mirror in the southern hemisphere) is drawn in blue and corresponds to a Milne universe with angular defect. The equator fixed points coincide with a scaling cosmology with $S\sim T^{\frac13}$ and are illustrated in purple. The blue trajectory features transient acceleration with tunable number of e-foldings, whereas the red one corresponds to semi-eternal acceleration. Depicted in purple is the unique (fine-tuned) trajectory corresponding to eternal acceleration. 
The solution becomes de Sitter in the vicinity of the origin,  which is reached at finite proper time.~The solution can be geodesically completed  in the past beyond the point at the origin, by gluing together 
its mirror trajectory in the southern hemisphere. }\label{fig:trajkl}
\end{center}
\end{figure}

The $k,\lambda<0$ solution has been studied before in \cite{Chen:2003dca} 
(see also \cite{Andersson:2006du}) with somewhat different methods.~It features a conical accelerating region, with a fixed point $p_0$ at the origin 
of the cone (which coincides with the origin of the sphere), and a second fixed point $p_1$ on the surface of the cone.~Both $p_0$, $p_1$ lie on the invariant plane $\mathcal{P}$.~The invariant manifolds of  $p_0$ are the equatorial disc (in which $p_0$ is an attractor) 
and the vertical $z$-axis (in which $p_0$ is an unstable node).  
There are generic trajectories that asymptote some point on the equator at  past infinity, cross into the acceleration region (the interior of the cone) at some point $p'$, then exit the acceleration region again at some point $p''$,  
and asymptote $p_1$ at future infinity, see Figure \ref{fig:trajkl}.

\begin{figure}[H]
\begin{center}
\includegraphics[width=0.7\textwidth]{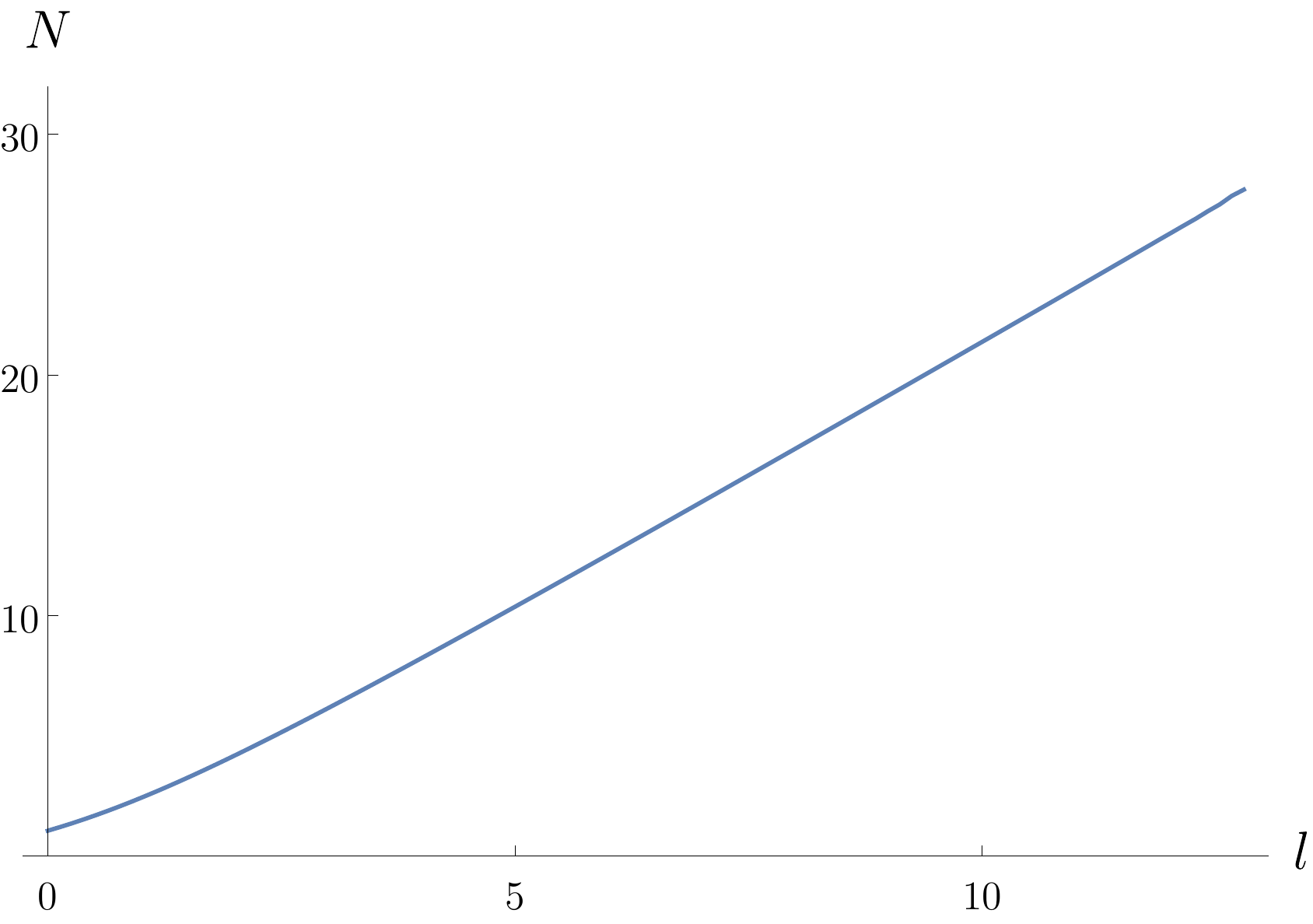}
\caption{Number of e-foldings $N$ as a function of $l$, which parametrizes the distance $\Delta$ between the $x$-coordinates of $p''$ and $p_1$ for the $(\lambda, k)$ model, viz. $l= - \log \Delta + \text{const.}$}\label{fig:extra}
\end{center}
\end{figure}

These trajectories correspond to accelerating cosmologies with $N\rightarrow\infty$ as  $p''\rightarrow p_1$. In other words, we can continuously increase the number of e-foldings  and make it as large as desired, 
by choosing the trajectory such that the exit point from the acceleration region is sufficiently close to $p_1$, see Figure \ref{fig:extra}.

Note also the presence of semi-eternal trajectories that enter the acceleration region at some point $p'$, and reach the fixed point 
$p_1$ at future infinity, without ever exiting the cone.   
The limiting case,  $p'\rightarrow p_0$, 
is the unique trajectory which reaches  $p_0$ in the past  and $p_1$ at future infinity: it corresponds to eternal acceleration,  with acceleration  vanishing  
asymptotically at future infinity. In the vicinity of  $p_0$ the universe becomes de Sitter in hyperbolic slicing.~This is not asymptotic de Sitter, however, as $p_0$ is reached at finite proper time in the past.~The solution can  be geodesically completed beyond $p_0$ in the past,   by gluing together its mirror trajectory in the southern hemisphere.

\begin{figure}[H]
\begin{center}
\includegraphics[width=0.8\textwidth]{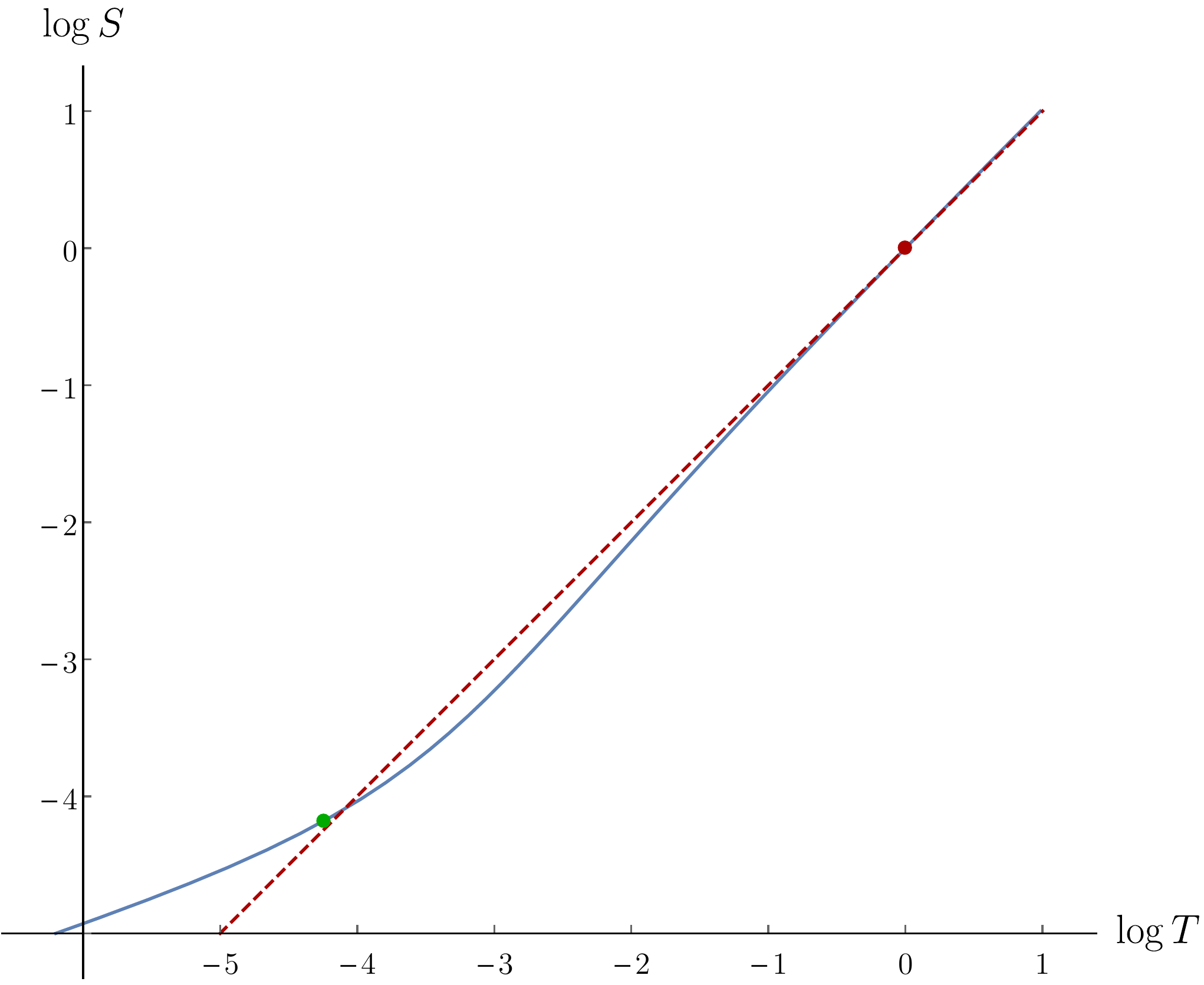}
\caption{Plot log--log of the scale factor $S$ (in blue) as a function of the cosmological time $T$. It corresponds to the $(\lambda,k)$-compactification (with $l=2$). The green dot corresponds to the beginning of inflation; the red one to its end. The red dashed line coincides with the Milne fixed point $S \propto T$ and asymptotes the curve at future infinity.}\label{fig:S}
\end{center}
\end{figure}

From a given trajectory in phase space, it is then straightforward to reconstruct the scale factor $S(T)$ as a function of the cosmological time, as depicted in Figure \ref{fig:S}. This gives access to all the cosmological data (Hubble parameter $H$, slow-roll parameters $\epsilon$ and $\eta$, the spectral tilt $n_s$, or the tensor-to-scalar ratio $r$) needed for phenomenology. \\

\begin{figure}[H]
\begin{center}
\includegraphics[width=0.9\textwidth]{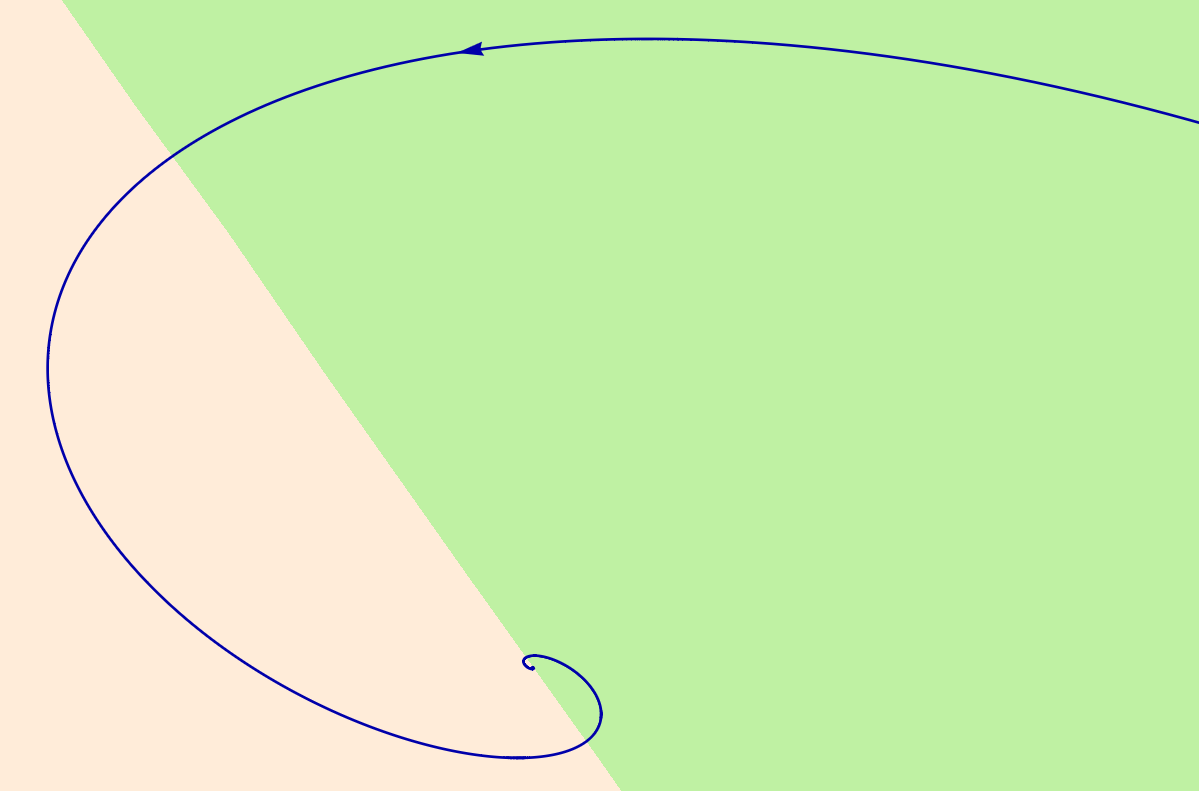}
\caption{Example of a spiraling trajectory around $p_1$ in the $(m,k)$-compactification.}\label{fig:spiral}
\end{center}
\end{figure}

To our knowledge, the $k<0$, $m\neq0$ solution has not appeared before in the literature.~Like the previous solution, it features a conical accelerating region, with a fixed point $p_0$ at the origin,
 and a second fixed point $p_1$ on the surface of the cone. Both $p_0$, $p_1$ lie on the invariant plane $\mathcal{P}$. 
 However, unlike the previous solution, the fixed point $p_1$ is now a stable focus on $\mathcal{P}$, and an attractor in the direction 
 perpendicular to $\mathcal{P}$. This allows for generic trajectories 
that asymptote some point on the equator at  past infinity, cross into the acceleration region (the interior of the cone) at some point $p'$, exit the acceleration region at some point $p''$,  then enter 
again at some point $p'''$, and so on,  
as they spiral into $p_1$ asymptotically at future infinity, see Figure \ref{fig:spiral}.~This asymptotic  spiraling  corresponds to an infinite number of periodic cycles of alternating periods of accelerated and decelerated expansion,  each cycle contributing a finite number of e-foldings, so that $N\rightarrow\infty$.  

As in the previous system, 
there is here a unique (fine-tuned) trajectory which reaches  $p_0$ at finite proper time in the past  and asymptotes $p_1$ at future infinity. In the vicinity of  $p_0$, the universe becomes de Sitter in hyperbolic slicing.~The solution can  be geodesically completed beyond $p_0$ in the past,   by gluing together its mirror trajectory in the southern hemisphere.

\begin{figure}[H]
\begin{center}
\includegraphics[width=1\textwidth]{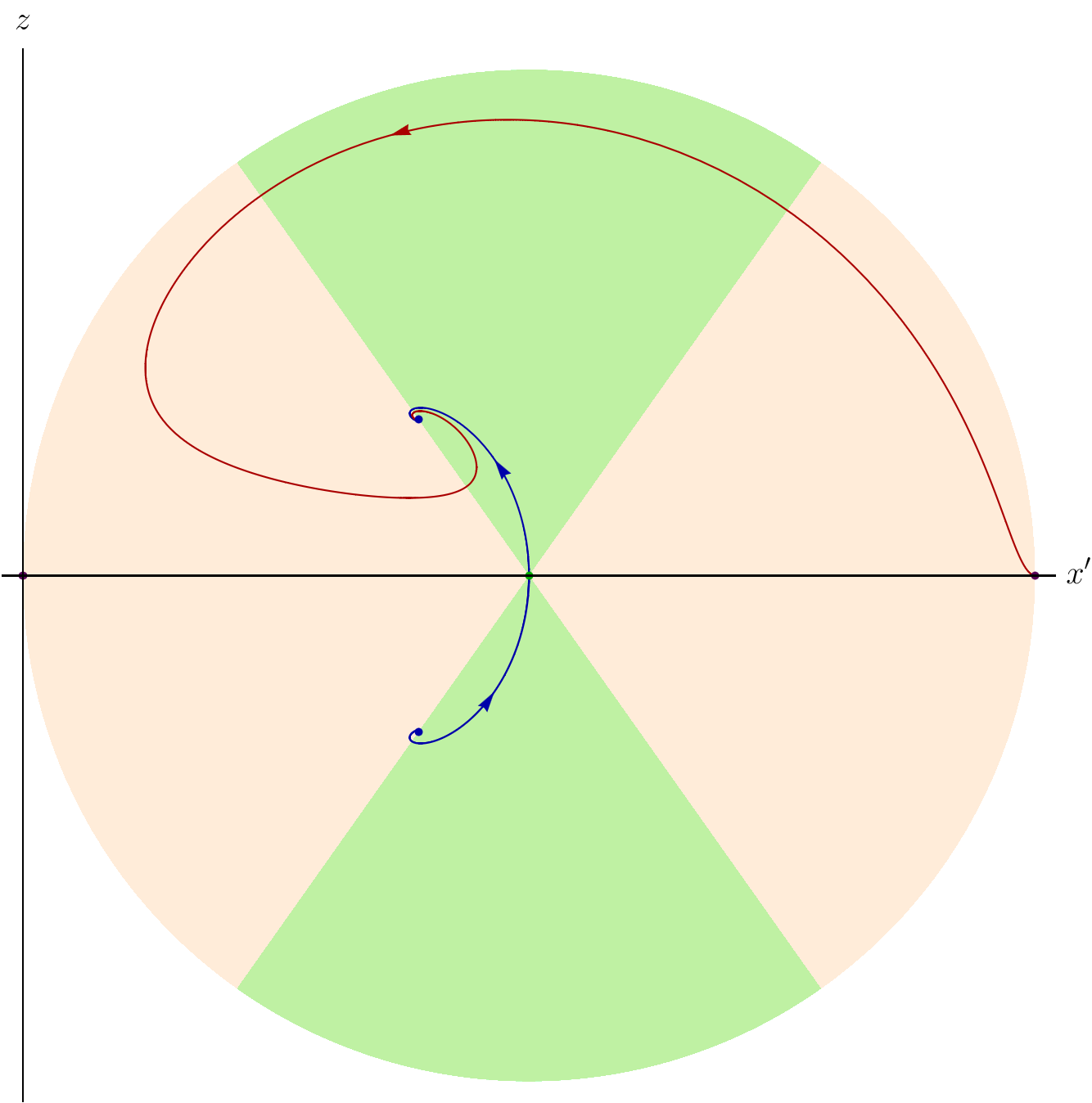}
\caption{
Two trajectories lying on the invariant plane $\mathcal{P}$ of the dynamical system obtained from the compactification with $m,k \neq 0$. The point $p_0$ corresponding to a Milne universe is depicted in green. The point $p_1$ (and its mirror in the southern hemisphere) is drawn in blue and corresponds to a Milne universe with angular defect. The equator fixed points coincide with a scaling cosmology with $S\sim T^{\frac13}$ and are illustrated in purple. The red trajectory enters and exits the acceleration region an infinite number of times, spiraling around $p_1$, see Figure \ref{fig:spiral} for a zoom around this region. Depicted in blue is the unique (fine-tuned) trajectory:  
the solution becomes de Sitter in the vicinity of the origin,  which is reached at finite proper time.~The solution can be geodesically completed  in the past beyond the point at the origin, by gluing together 
its mirror trajectory in the southern hemisphere.
}\label{fig:spiral1}
\end{center}
\end{figure}

\newpage

To our knowledge, none of the remaining systems,  exhibiting a parametric control on the number of e-foldings, 
corresponding to $k<0$ and non-vanishing  $c_\varphi$ \eqref{spk}, $c_0$ \eqref{sczk} or $c_f$ \eqref{skf}, has appeared before in the literature. These are all captured by the 4d consistent truncation \eqref{consttr} with potential given in \eqref{16}.~While they all correspond to solutions with vanishing Romans mass, 
they are all qualitatively very similar to the $(m,k)$ case analyzed above: they all feature a hyperbolic FLRW universe ($k<0$) 
with one type of non-vanishing flux, and they all admit solutions with an infinite number of  cycles of alternating periods of accelerated and decelerated expansion. 
In the ``rollercoaster cosmology'' scenario of \cite{DAmico:2020euu} it was argued that 
an oscillatory behavior of this type could be relevant for inflation. \\

The rest of the paper is organized as follows. In Section 
\ref{sec:setup} we explain the general setup and ansatz  of our
10d  solutions, and make contact with the 4d 
cosmological parameters, such as the FLRW scale factor, etc. In Sections \ref{sec:1d}, \ref{sec:4dpotential} 
 we obtain the 1d, 4d consistent truncations respectively.~Section \ref{sec:analytical} discusses the analytic 
 solutions presented in the paper.~Section \ref{sec:dsa} explains the dynamical system techniques used here, illustrated with some notable cases in subsections \ref{case1} -- \ref{case4}.
 We conclude with some discussion and future directions in Section \ref{sec:concl}. Further  details of all analytical solutions and all 
 two-flux dynamical systems studied here, can be found in the appendices.

\section{The  general setup}
\label{sec:setup}

Our ten-dimensional metric is a warped product of a four-dimensional FLRW cosmological factor and a six-dimensional compact internal space. 
The ansatz for the ten-dimensional metric {\it in 10d Einstein frame} reads
\eq{\label{2}\d s^2_{10} =e^{2A(t)}\left(e^{2B(t)} g_{\mu\nu}\d x^{\mu}\d x^{\nu}+g_{mn}\d y^m\d y^n 
\right)~,
}
where the scalars $A$, $B$ only depend on the conformal time coordinate, while $y^m$ are coordinates of the internal six-dimensional space. 
The unwarped 4d metric  is assumed to be of the form
\eq{\label{metricansatz}
g_{\mu\nu}\d x^\mu\d x^\nu=
-\d t^2+\d\Omega_k^2
~,}
where $t$ is the conformal time, and the spatial 3d part of the metric 
is locally isometric to a maximally-symmetric three-dimensional space of 
scalar curvature $6k$. Explicitly,  
\eq{\label{spatialmetric}
\d\Omega_k^2=
 \gamma_{ij}(\vec{x})\d x^i\d x^j~;~~~
R^{(3)}_{ij}=2k \gamma_{ij}
~,}
with $i,j=1,2,3$, and $R^{(3)}_{ij}$ is the Ricci tensor of the metric $\gamma_{ij}$. 
The 3d metric is thus   locally  a sphere $(k>0)$ or a hyperbolic space $(k<0)$; the case $k=0$ 
corresponds to flat space.

\subsection*{Four-dimensional interpretation}

The correct frame in which the predictions of our model should be compared to the cosmological data is the four-dimensional Einstein frame: 
 this is the frame in which the effective four-dimensional Newton constant becomes time-independent. 
The 4d  Einstein-frame metric  reads
\eq{\label{efm8}
\d s_{4E}^2=-S^6\d\tau^2+S^2\d\Omega_k^2~,
}
where the scale factor is given by
\eq{\label{scfc}
S=e^{4A+B}~,}
and we have introduced of a new time variable $\tau$ defined via
\eq{\label{taudef111}
\frac{\d t}{\d\tau}=e^{8A+2B}
~.}
Note that $\tau$ is neither the conformal nor the cosmological time. It is simply the coordinate with respect to which the equations of motion are simplest to solve.

In terms of the cosmological time coordinate $T$, 
\eq{\label{tcrd}
\frac{\d T}{\d\tau}=S^3
~,}
the metric takes the standard FLRW form, 
\eq{
\d s_{4E}^2=- \d T^2+S^2\d\Omega_k^2
~.}

Suppose we have an explicit solution of our set of equations for all fields, so that in particular we can construct the explicit expression of scale factor. 
We may then always engineer a {4d} perfect fluid of density and pressure $\rho$ and $p$ respectively, such that the scale factor in question 
can be seen as resulting from the {4d} Einstein equations sourced by the stress-energy tensor of that fluid, 
\eq{\spl{
\left( \frac{\dot{S}}{S}\right)^2&=\frac{8\pi G}{3}\rho-\frac{k}{S^2}\\
\left( \frac{\ddot{S}}{S}\right)&=-\frac{4\pi G}{3}(\rho+3p)
~,}}
where the dot refers to differentiation with respect to $T$. 
Solving the above system for $\rho$, $p$ we obtain
\eq{\label{wdef}
w:=\frac{p}{\rho}= -\frac13-\frac{2{S}\ddot{S}}{3(k+\dot{S}^2)}
~.}
In particular, it follows that for $k>-\dot{S}^2$ (and in particular for  $k=0$), the condition for acceleration  is equivalent to $w<-\frac13$.
Of course, unless $w$ is constant, the first equality in \eqref{wdef} is simply a definition, rather than an equation of state. Nevertheless, even 
for non-constant $w$, this quantity is useful insofar as it allows us to compare to the  equations of state of the different cosmological eras. \\

The conditions for expansion and acceleration  read
\eq{\spl{\label{acc1}
0<\dot{S}&=-\frac12 \d_\tau S^{-2}
\\
0<\ddot{S}&=-\frac12 S^{-3}\d^2_\tau S^{-2}
~.}}
The number of e-foldings, $N$, is defined as
\eq{\label{efoldeq1intro}
N=\int \d \ln S
~,}
where the limits of the integral should be taken at the beginning and the end of  the acceleration period.

\section{1d consistent truncation}
\label{sec:1d}

In all of the compactifications  presented here, 
the internal 6d components of the Einstein equations and the dilaton equation reduce  to\footnote{The bosonic sector of type IIA supergravity is given in Appendix \ref{sec:iiasugra}.}
\eq{\spl{\label{eintd}
\d_\tau^2 A&=-\tfrac{1}{48} \left(\partial_AU-4\partial_BU   \right)\\
\d_\tau^2 \phi&=- \partial_{\phi}U
~,}}
where $U$ is a function (potential) that depends on the compactification and on the flux that is turned on, given in \eqref{16u} below. 
The external 4d Einstein equations  
reduce to the following two  equations,  
\eq{\spl{\label{eext12}
\d_\tau^2 B&= \tfrac{1}{12} \left(\partial_AU-3\partial_BU   \right)\\
72(\d_\tau& A)^2+6(\d_\tau B)^2
+48\d_\tau A\d_\tau B-\tfrac12 (\d_\tau\phi)^2  =U
~.}}
The second line above is a {constraint}, consistently propagated by the remaining equations of motion (i.e.~the 
$\tau$-derivative of this equation is automatically satisfied by virtue of the remaining equations). The important point to note here is that, as we shall see,  
the remaining (flux) equations of motion can be solved without imposing any additional conditions.~The above equations of motion ``know'' about the 
flux content of the background which enters via $U$ in the form of constant parameters, cf. Table \ref{tab:constants}. 
The bottom line is that we are left with three second-order equations 
for three unknowns (the two warp factors $A$, $B$ and the dilaton $\phi$), together with a constraint, which only needs to be imposed once at some fixed initial time, 
and amounts to imposing an algebraic condition on the constants parameterizing the fluxes. 

Moreover,  equations \eqref{eintd}, \eqref{eext12} can be derived from a one-dimensional action given by
\eq{\label{15}
S_{1\d}=\int\d\tau \left\{   \frac{1}{\mathcal{N}}\Big(
-72(\d_\tau A)^2-6(\d_\tau B)^2-48\d_\tau A\d_\tau B+\tfrac12 (\d_\tau \phi)^2
\Big)-\mathcal{N}U(A,B,\phi)
\right\} 
~,
}
where $\mathcal{N}$ is a non-dynamical Lagrange multiplier which should be set to $\mathcal{N}=1$ at the end of the calculation; it can be thought of as originating from the lapse function. 
Variation of the action \eqref{15} with respect to $\mathcal{N}$ imposes the constraint (second line of \eqref{eext12}), while variation with respect to the fields $A$, $B$, $\phi$ is equivalent to their respective equations 
of motion in \eqref{eintd},  \eqref{eext12}. 
The potential $U$ is in general  a function of all three scalars $A$, $B$, $\phi$. 
Explicitly, it is given by \\ 
\eq{ \label{16u}
\hspace{-1cm}
U=\left\{
\begin{array}{lll}
\tfrac12 c^2_{\varphi} e^{-\phi/2+6A+6B}
+\tfrac12 c^2_{h} e^{-\phi+12A}+
\tfrac32c^2_{\chi} e^{\phi+4A}+c^2_{\xi\xi'} e^{-\phi/2+6A}
-6ke^{16A+4B}
&~\text{CY}&\eqref{e1}, \eqref{e2}\\ \\
72 b_0^2  e^{-\phi+12A+6B}
+\tfrac32 c_0^2 e^{\phi/2+10A+6B}&~\text{CY}&\eqref{et3bbbf}, \eqref{red5bbbf}\\ \\
\tfrac12 c^2_{\varphi} e^{-\phi/2+6A+6B}
 +\frac12m^2e^{5\phi/2+18A+6B}-6ke^{16A+4B}-6\lambda e^{16A+6B}&~\text{E}&\eqref{et3bbem}, \eqref{red5bbem} \\ \\
\tfrac12 c^2_{\varphi} e^{-\phi/2+6A+6B}
+\tfrac12 c^2_{h} e^{-\phi+12A}+
\tfrac32c^2_{\chi} e^{\phi+4A}
-6ke^{16A+4B}
-6\lambda e^{16A+6B}&~\text{EK}&\eqref{intek}, \eqref{extek}
 \\ \\
\tfrac32 c_0^2 e^{\phi/2+10A+6B} +\frac12m^2e^{5\phi/2+18A+6B}
-6ke^{16A+4B}-6\lambda e^{16A+6B}
&~\text{EK}&\eqref{intek2}, \eqref{extek2} \\ \\
\tfrac12 c^2_{\varphi} e^{-\phi/2+6A+6B}+\tfrac32c_f^2e^{3\phi/2+14A+6B}
  -6ke^{16A+4B}-6\lambda e^{16A+6B}
 &~\text{EK}& \eqref{f2an}~. 
 \end{array} 
\right.}
In the above we have indicated the type of compactification in which the potentials appear (E stands for Einstein, EK for Einstein-K\"{a}hler, CY for Calabi-Yau), as well as numbers of the respective equations  of motion.

Note that the potential  encodes all the information  about the flux 
(which is generically non-vanishing from the ten-dimensional point of view), as well as the external and internal curvature contributions. 
Indeed, this information is encoded in $U$ via the different constants appearing in \eqref{16u}, whose 10d origin is summarized in Table \ref{tab:constants}.

\begin{table}[H]
\begin{center}
\begin{tabular}{|c|c|}
  \hline
   $m$ & zero-form (Romans mass) \\
    \hline
   $c_f$ & internal two-form  \\
     \hline
   $c_h$ &external  three-form \\
  \hline  
$b_0$ &internal  three-form \\
  \hline  
   $c_\chi$ &mixed  three-form \\
   \hline
$c_\varphi$ & external four-form \\
  \hline
$c_0$ &internal  four-form \\
  \hline
$c_{\xi\xi'}$ &mixed four-form \\
  \hline
$k$ &external curvature \\
    \hline
$\lambda$ &internal curvature \\
  \hline
\end{tabular}
\end{center}
\caption{List of the constant coefficients entering  the potential \eqref{16u} of the 1d consistent truncation, and their 10d origin. A form is called external (resp. internal), if all its legs are along the external (resp. internal) directions; mixed if one of its legs is along the time direction, while all remaining legs are internal.}
\label{tab:constants}
\end{table}

Note also that, in terms of the potential, the derivatives of the scale factor with respect to cosmological time, cf.~equations \eqref{acc1},   take the form
\eq{\spl{\label{14}
S^4\dot{S}^2&=\tfrac{1}{6} U
+4(\d_\tau A)^2+\tfrac{1}{12}(\d_\tau\phi)^2 
\\
S^5\ddot{S}&=\tfrac{1}{12}\left( \partial_B U-4 U\right)
-8(\d_\tau A)^2-\tfrac16(\d_\tau\phi)^2 
~.}}
Moreover, the quantity
\eq{\label{windir}
\frac{{S}\ddot{S}}{k+\dot{S}^2}= 
\frac{   \partial_B \tilde{U}-4 \tilde{U} 
-96(\d_\tau A)^2-2(\d_\tau\phi)^2 }{2 \tilde{U}
+48(\d_\tau A)^2+ (\d_\tau\phi)^2  }
~,}
that appears in \eqref{wdef}, is written in terms of the $k$-independent part of the potential, 
\eq{
\tilde{U} = U+6k S^4
~.}
In particular we see that
\eq{\label{wcond1}
w=-1 \Longleftrightarrow   \partial_B \tilde{U}= 6 \tilde{U}
+144(\d_\tau A)^2+ 3(\d_\tau\phi)^2 
~.}

\section{Cosmological 4d consistent truncation}
\label{sec:4dpotential}

In a subset of the cases we present here, the equations of motion \eqref{eintd}, \eqref{eext12} can be written in terms of a potential $V(A,\phi)$ which only depends on $A$ and $\phi$,
\eq{\spl{\label{eomjp}
\d_\tau^2 A&=-\tfrac{1}{48}e^{24A+6B}\partial_AV\\
\d_\tau^2 B&=e^{24A+6B}\big(\tfrac12 V+\tfrac{1}{12}\partial_AV\big)-2ke^{16A+4B}\\
\d_\tau^2 \phi&=-e^{24A+6B}\partial_{\phi}V\\
-12ke^{16A+4B}
+ 2Ve^{24A+6B}&=
144(\d_\tau A)^2+12(\d_\tau B)^2
+96\d_\tau A\d_\tau B-(\d_\tau\phi)^2  
~.}}
As can be verified, these equations can then be ``integrated'' into a four-dimensional action,
\eq{\label{consttr}
S_{4\d}=\int\d^4 x\sqrt{g}\Big(
R-24g^{\mu\nu}\partial_\mu A\partial_\nu A
-\tfrac12 g^{\mu\nu}\partial_\mu \phi\partial_\nu\phi-V(A,\phi)
\Big)~.
}
Indeed, in terms of the scale factor \eqref{scfc} and the cosmological time coordinate \eqref{tcrd}, taking 
suitable linear combinations thereof,  
equations \eqref{eomjp} can be written equivalently as
\eq{\spl{\label{eomjp1}
\frac{\ddot{S}}{S}&=\frac16 V-8\dot{A}^2-\frac16\dot{\phi}^2\\
2\left( \frac{\dot{S}}{S} \right)^2+\frac{2k}{S^2}&=\frac13 V+8\dot{A}^2+\frac16\dot{\phi}^2\\
 \ddot{A}+\frac{3}{S}\dot{S}\dot{A}&=-\tfrac{1}{48}\partial_AV\\
 \ddot{\phi}+\frac{3}{S}\dot{S}\dot{\phi}&=-\partial_\phi V~.
}}
On the other hand, 
the 4d equations of motion resulting from \eqref{consttr} read
\eq{\spl{\label{eomj}
R_{\mu\nu}&=\tfrac12  g_{\mu\nu} V+24\partial_\mu A\partial_\nu A
+\tfrac12  \partial_\mu \phi\partial_\nu\phi \\
\nabla^\mu\partial_\mu  A&=\tfrac{1}{48}\partial_AV~;~~~\nabla^\mu\partial_\mu  \phi=\partial_\phi V
~.}}
Inserting the Einstein metric \eqref{efm8} into the above, and assuming  that $A$, $B$, $\phi$ only depend on the time coordinate, results precisely in \eqref{eomjp1}. 
Therefore, $S_{4\d}$ of \eqref{consttr} is a two-scalar {\it consistent truncation of IIA for cosmological solutions}:~all cosmological solutions (i.e.~all solutions with a metric of FLRW type and   scalar fields that only depend on time) 
of $S_{4\d}$ lift to ten-dimensional solutions of IIA supergravity.

For the different compactifications admitting a consistent cosmological truncation of the form \eqref{consttr}, the potential reads, 
\eq{
\hspace{-0.5cm}
V=\left\{\label{16}
\begin{array}{ll}72b_0^2  e^{-\phi-12A}
+\frac32c_0^2 e^{\phi/2-14A}&~  \text{CY~with~internal~three-~and~four-form~fluxes} \\ \\
\frac12 c_{\varphi}^2  e^{-\phi/2-18A} +\frac12m^2e^{5\phi/2-6A}-6\lambda e^{-8A}&~  \text{E~with external~four-form~flux} \\ \\
\frac32c_0^2 e^{\phi/2-14A}+\frac12m^2e^{5\phi/2-6A}-6\lambda e^{-8A}&~  \text{EK~with internal~four-form~flux} \\ \\
\frac12 c_{\varphi}^2  e^{-\phi/2-18A}+\frac32c_f^2e^{3\phi/2-10A}-6\lambda e^{-8A}&~  \text{EK~with internal~two-form, external~four-form}.
  \end{array} 
\right.
}
Of course terms with $k$ (the external 4d spatial curvature) cannot appear at the level of the 4d action, but rather they potentially arise  as part of its solutions.

It is known \cite{Terrisse:2019usq, Tsimpis:2020ysl} that there exists 
a consistent 4d truncation in the case of the universal CY sector, i.e.~a consistent truncation to the the gravity multiplet, one vectormultiplet and one hypermultiplet. Remarkably, the action $S_{4\d}$ of \eqref{consttr} is a sub-truncation thereof to the graviton and two scalars, such that all information about the flux 
(which is generically non-vanishing from the ten-dimensional point of view) is carried by the potential $V$ via the constants $m$, $b_0$, $c_0$, $c_\varphi$. The latter correspond respectively to non-vanishing zero-form flux (Romans mass), internal three-form flux, internal four-form flux, and external four-form flux; $\lambda$ is the scalar curvature of the internal Einstein manifold, cf.~\eqref{150}. 

Note that: \textit{(i)} not all compactifications considered in the present paper admit the sub-truncation \eqref{consttr}, \eqref{16} to gravity plus two scalars;  \textit{(ii)} the Einstein and Einstein-K\"{a}hler cases in \eqref{16} reduce to CY in the $\lambda\rightarrow 0$ limit.

\section{Analytic solutions}
\label{sec:analytical}

In the present paper we find three different types of analytic solutions: 

\begin{itemize}

\item {\it Critical (scaling) solutions}. Their metrics are of power-law form,

\eq{ 
\d s^2=-\d T^2+ T^{2a}\d\Omega_k^2 
~,}

in terms of the  cosmological time $T$, with 
$a=\tfrac13$, $\tfrac34$, $\tfrac{19}{25}$,  $\tfrac{9}{11}$, $1$.~They  admit an interpretation as critical points in an appropriately defined phase space, as described in Section \ref{sec:dsa}. 
 
\item {\it Type I solutions}. They interpolate between two asymptotically power-law metrics,

\eq{ \label{intro1}
\d s^2\rightarrow-\d T_-^2+ T_-^{\frac23}\d\Omega_k^2~;~~~\d s^2\rightarrow-\d T_+^2+ T_+^{\frac23}\d\Omega_k^2
~,}

as $\tau\rightarrow-\infty$, $\tau\rightarrow+\infty$ respectively.
In terms of the dynamical system description, they correspond to trajectories interpolating between two fixed points on the equator.

The corresponding coordinate patches are parameterized by coordinates $T_\pm\propto e^{b_\pm\tau}$, for 
certain parameters $b_\pm$ which depend on the particular solution. If $b_-$ is positive, $\tau\rightarrow-\infty$ corresponds to $T_-\rightarrow 0$. 
It follows that the solution reaches a singularity at finite proper time in the past. 
If $b_-$ is  negative, $\tau\rightarrow-\infty$ corresponds to $T_-\rightarrow \infty$, and is therefore reached at infinite proper time in the past.~An identical analysis holds for the patch 
parameterized by $T_+$, corresponding to $\tau\rightarrow+\infty$. 
All four different possibilities are realized in the analytic solutions presented here: $\tau\rightarrow\pm\infty$ corresponding to $T_\pm\rightarrow 0$, or $T_\pm\rightarrow \infty$.
\item {\it Type II solutions}. They interpolate between an asymptotically power-law metric,
\eq{ 
\d s^2\rightarrow-\d T_\pm^2+ T_\pm^{2a_\pm}\d\Omega_k^2
~,}
at either $\tau\rightarrow-\infty$ or $\tau\rightarrow+\infty$, with $a_\pm=\tfrac13$, and a critical solution at $\tau=0$ (which is reached at infinite proper time) with 
$a=1$ or $a=\tfrac{3}{4}$. 
In terms of the dynamical system description, they correspond to trajectories interpolating between one fixed point on the equator and an interior fixed point.

\end{itemize}
Table \ref{tab:analyticsol} below summarizes the different types of analytic solutions constructed here, the corresponding type of compactification, and the section in which the explicit details of the 
 solution can be found.

\begin{table}[H]
  \begin{center}
      \noindent\makebox[\textwidth]{
    \begin{tabular}{|c||c|c|c|}
    \hline
  & Critical & Type I & Type II \\
\hhline{====}
  &  $k<0$ \eqref{67milne} & $k>0$ \eqref{40a}  & $k<0$ \eqref{CY3}  \\

       & $\varphi \neq 0$ and $k<0$ \eqref{242bb} & $\varphi \neq 0$ \eqref{sdjkdflk} & $\varphi \neq 0$ and $k<0$ \eqref{79phi} \\

        CY    & $\chi \neq 0$ and $k<0$ \eqref{CYchi3}  & $\chi \neq 0$ \eqref{CYchi}, $\chi \neq 0$ and $k>0$ \eqref{CYchi2}  & $\chi \neq 0$ and $k<0$ \eqref{CYchi3} \\
                     & $\xi \neq 0$ and $k<0$ \eqref{CYxi3}  & $\xi \neq 0$ \eqref{CYxi1}, $\xi \neq 0$ and $k>0$ \eqref{CYxi2}  & $\xi \neq 0$ and $k<0$ \eqref{CYxi3} \\ 
     & $\xi, \chi \neq 0$ and $k<0$ \eqref{CYxck2}  & $\xi,\chi \neq 0$ and $k>0$ \eqref{CYxck1}  & $\xi, \chi \neq 0$ and $k<0$ \eqref{CYxck2} \\ 
          & $\xi, \chi, h \neq 0$ and $k<0$ \eqref{CYhxc}   &$b_0 \neq 0 $ and $c_0 \neq 0 $ \eqref{CYbg} & $\xi, \chi, h \neq 0$ and $k<0$ \eqref{CYhxc} \\ 
    \hline
 & $\lambda <0$ \eqref{El3} & $\lambda >0$ \eqref{El1}  & $\lambda <0$ \eqref{El2} \\
  & $\lambda < 0$ and $k<0$ \eqref{Elk3} & $\lambda > 0$ and $k>0$ \eqref{Elk1}  & $\lambda < 0$ and $k<0$ \eqref{Elk2} \\
 & $m \neq 0$ and $\lambda < 0$ \eqref{242b} & $m \neq 0$ \eqref{193}  & $\varphi \neq 0$ and $\lambda < 0$ \eqref{Elp2} \\

  E  & $m \neq 0$ and $ k < 0$ \eqref{Ekm2} &$\varphi \neq 0$ and $\lambda > 0$ \eqref{Elp1}  &  $m \neq 0$ and $ k < 0$ \eqref{Ekm1} \\
    & $m \neq 0$, $ k,\lambda < 0$ \eqref{Eklm} & $\varphi \neq 0$ and $m \neq 0$ \eqref{Epm1}  &  \\
                                & $\varphi,m \neq 0$, $k < 0$ \eqref{Ekpm} &   &  \\
                                                                & $\varphi,m \neq 0$, $k,\lambda < 0$ \eqref{kl3red5bbem} &   &  \\
    \hline
 & $h \neq 0$ and $\lambda<0$ \eqref{EK1} &  &  \\
EK & $c_0 \neq 0$ and $k <0$ \eqref{EK2} &  &  \\
  & $c_f \neq 0$ and $k <0$ \eqref{EK27} &  &  \\
    \hline
    \end{tabular}
    }
     \caption{All analytic solutions presented in this paper, besides the zero-flux, flat 4d space solution of Section \ref{sec:special}.~We list the corresponding compactification manifold, the type of non-vanishing flux, together with a reference to the explicit details of the solution. Critical solutions correspond to fixed points in the dynamical-system description. 
     All CY critical points correspond to regular Milne universes except the one with $\varphi\neq0$,  which is a Milne universe with angular defect.  
     The $\lambda<0$ critical point corresponds to 
     a scaling cosmology with $a=\tfrac34$. The $m\neq0$, $\lambda<0$ critical point corresponds to a scaling cosmology with $a=\tfrac{19}{25}$. 
     The $k$, $\lambda$, $m$, $\varphi\neq0$ solution corresponds to AdS$_4$ in hyperbolic slicing. The $h\neq0$, $\lambda<0$ critical point corresponds to a scaling cosmology with $a=\tfrac{9}{11}$.~All the other critical points correspond to Milne universes with angular defects.
     Type I solutions always correspond to trajectories interpolating between two points of the equator. Type II solutions interpolate between a fixed point on the equator and an  interior fixed point. 
      For all type II CY solutions, the internal space warp factor $e^A$ and the dilaton $\phi$ tend  to a constant at future infinity,   
     except for the  $\varphi\neq 0$ solution for which $e^A, e^\phi\rightarrow\infty$.
  For all type II Einstein solutions, $e^A  \rightarrow\infty$, $\phi\rightarrow\text{const.}$ at asymptotic infinity,   
     except for the  $(m,k)$ solution for which  $e^\phi\rightarrow 0$.
     }\label{tab:analyticsol}
  \end{center}
\end{table}

\subsection{Minimal (zero-flux) solution}\label{sec:special}

The simplest  solution to the form equations is given by vanishing flux. Let us consider the remaining equations of motion. 
The internal Einstein  and  dilaton equations give, cf.~\eqref{eintd},
\eq{\label{et1b}
A=c_A\tau+d_A~;~~~\phi= c_\phi
\tau+d_\phi
~,}
for some constants $c_A$, $d_A$, $c_\phi$, $d_\phi$.  
The external Einstein equations  
give, cf.~\eqref{eext12},  
\eq{\label{54constra}
B= c_B\tau+d_B
~;~~~
\tfrac{1}{12}c_\phi^2 =
12c_A^2+c_B^2+8c_A c_B 
~,}
where $c_B$, $d_B$ are constants. The second equation above is the constraint, which implies in particular that the ratio $r$ defined by
\eq{\label{rdef}
r:=\frac{c_A}{c_B}
~,}
must satisfy $r\leq-\tfrac12$ or $r\geq-\tfrac16$. 
The points where the constraint is saturated correspond to constant dilaton ($c_\phi=0$).

This solution thus implies a power-law form for the 4d part of the Einstein-frame metric,
\eq{
\d s^2_{4E}=-e^{ (24c_A+6c_B)\tau }\d \tau^2
+e^{(8c_A+2c_B)\tau  } \d\vec{x}^2
~,
}
where we have appropriately rescaled the $\tau$, $\vec{x}$ coordinates to absorb $d_A$, $d_B$. 
In terms of a coordinate $T\propto e^{ (12c_A+3c_B)\tau }$ we have a power-law expansion, 
\eq{\label{range55simple}
\d s^2_{4E}=-\d T^2+T^{\frac23}\d\vec{x}^2~.
}
The time coordinate $T$ ranges from $T=0$, where we have a singularity,  to $T=\infty$. Depending on the values of 
$c_A$, $c_B$ and $c_\phi$, we can have solutions with constant dilaton such that the warp factor collapses ($e^A\rightarrow0$), or decompactifies ($e^A\rightarrow\infty$) as $T\rightarrow 0$ or $\infty$  respectively. 
The opposite behavior is also possible, i.e.~$e^A\rightarrow \infty, 0$ as $T\rightarrow 0,\infty$  respectively.  We can also have solutions with constant internal space 
warp factor, such that either $\phi\rightarrow \pm\infty$ or $\phi\rightarrow \mp\infty$, as $T\rightarrow 0$ or $\infty$. 

\subsection{Single-flux solutions}\label{sec:oneflux}

In the case where a single type of flux is turned on, it is always possible to solve the equations analytically. 
Let us suppose a potential of the form
\eq{\label{of3.1}
U=c ~\!e^{\alpha A+\beta B+\gamma\phi}
~,}
for some real constants $c$, $\alpha$, $\beta$, $\gamma$, whose precise values depend on the type of flux turned on.  
More specifically, let $n_t$, $n_s$, $n_i$ be the number of legs the form has along the time, 3d space, and internal directions respectively. 
Then, for an RR form we have
\eq{\label{29azerty}
\alpha=18(1-n_t)-2(-1)^{n_t}(n_s+n_i)~;~~~
\beta=6(1-n_t)-2(-1)^{n_t}n_s~;~~~
\gamma=(-1)^{n_t}\frac{5-(n_t+n_s+n_i)}{2}
~.}
For an NS-NS three-form the constants $\alpha$, $\beta$ are as above, but $\gamma=-(-1)^{n_t}$.

In the following, it will be useful to  set
\eq{
q:= c\left[
\tfrac{1}{48}(\alpha-4\beta)^2-\tfrac{1}{12}\beta^2+\gamma^2
\right]
~.}

{\it Case $q\neq0$:} \\\\
Let us set
\eq{
E(\tau):=\left\{ 
\begin{array}{c}  \ln  \big[  \tfrac{ 2c_E^2}{q} ~\!\text{sech}^2({c_E}\tau+d_E)  \big]~,~q>0\\
\\
\ln  \big[  \tfrac{ 2c_E^2}{|q|}   ~\!\text{csch}^2({c_E}\tau+d_E)  \big]  ~,~ q<0
\end{array}\right. 
~,}
where $c_E$, $d_E$ are arbitrary constants. 
The $A$, $B$, $\phi$-equations of motion are then solved by
\eq{
\spl{
A&=\tfrac{c}{48q}(\alpha-4\beta) E+c_A\tau+d_A\\
B&= \tfrac{c}{12q}(3\beta-\alpha)E+c_B\tau+d_B\\
\phi&=\tfrac{c\gamma}{q}E+c_\phi\tau+d_\phi
~,}}
where $c_A$, $d_A$, $c_B$, $d_B$, $c_\phi$, $d_\phi$ are arbitrary constants subject to the conditions
\eq{
\alpha c_A+\beta c_B +\gamma c_\phi=0~;~~~\alpha d_A+\beta d_B +\gamma d_\phi=0
~.}
Moreover, the constraint \eqref{eext12} reduces to
\eq{\label{cst29}
\frac12 c_\phi^2+\frac{2c}{q}c_E^2=72c_A^2+48c_Ac_B+6c_B^2
~.}
For $\tau\rightarrow\pm\infty$, the scale factor $\ln S$ grows linearly in $\tau$. This results in a scaling solution with $S(T)\sim T^{\frac13}$. 
For $q>0$,  $S$ goes to a constant in the $\tau\rightarrow 0$ limit. 
In the case $q<0$ we have instead,  
\eq{
\ln S\rightarrow \delta~\! \ln|\tau|~;~~~\delta:=\frac{8\beta }{(\alpha-6\beta)(\alpha-2\beta)+48\gamma^2}
~,}
in the $\tau\rightarrow 0$ limit, where we took \eqref{cst29} into account. This results in a scaling solution with $S(T)\sim T^{a}$, where
\eq{
a=\frac{\delta}{3\delta+1}~.}
Taking \eqref{29azerty} into account, we see that there are no scaling solutions with $a>1$ (accelerated expansion). 
However,  there is one case which gives $a=1$ (scaling solution with vanishing acceleration): it involves  negative external curvature 
and is described in Section \ref{case1}. \\

{\it Case $q=0$:} \\ \\
For special values of $n_t$, $n_s$, $n_i$ it is possible to have $q=0$. It can be seen that this occurs only if the flux is anisotropic in the 3d spatial dimensions. 
The $A$, $B$, $\phi$-equations of motion are then solved by
\eq{
\spl{
A&=\tfrac{c}{48c_E^2}(4\beta-\alpha) e^E+c_A\tau+d_A\\
B&= \tfrac{c}{12c_E^2}(\alpha-3\beta)e^E+c_B\tau+d_B\\
\phi&=-\tfrac{c\gamma}{c_E^2}e^E+c_\phi\tau+d_\phi
~,}}
where $c_A$, $d_A$, $c_B$, $d_B$, $c_\phi$, $d_\phi$ are arbitrary constants and
\eq{
c_E=\alpha c_A+\beta c_B +\gamma c_\phi
~;~~~
d_E=
\alpha d_A+\beta d_B +\gamma d_\phi
~.
}
Moreover, the constraint \eqref{eext12} reduces to
\eq{\label{cst292}
 c_\phi^2 =144c_A^2+96c_Ac_B+12c_B^2
~.}
The scale factor is given by
\eq{\label{lns}
\ln S=(4c_A+c_B)\tau+(4d_A+d_B)+\frac{\beta c}{12c_E^2}e^{c_E\tau+d_E}
~.}
The conditions for accelerated expansion read
\eq{\label{40acce}
c_1+c_2 e^t>0~;~~~ 2c_1^2+2c_2^2 e^{2t}+c_2 (4c_1-c_E)e^t<0
~,}
where we have set
\eq{\label{cf9}
c_1:=4c_A+c_B~;~~~c_2:= \frac{\beta c}{12c_E}~;~~~t:= c_E\tau+d_E
~.}
For concreteness, let us assume $c_E>0$. The conditions \eqref{40acce} can then be written as
\eq{
\tilde{c}_2 e^t>-\tilde{c}_1~;~~~-\tilde{c}_1+\tfrac14(1-\sqrt{1-8\tilde{c}_1})<\tilde{c}_2 e^t<-\tilde{c}_1+\tfrac14(1+\sqrt{1-8\tilde{c}_1})
~,}
where $\tilde{c}_i:= c_i/c_E$; $i=1,2$. 
Let us also denote $B_{\pm}:= -\ctu + \frac{1}{4}\left(1 \pm \sqrt{1 - 8 \ctu} \right)$, appearing in \eqref{40acce}.  Since $B_{\pm} \geq 0$ for all $\ctu \leq \frac{1}{8}$, one needs $\ctd>0$ for the second condition to be satisfied, so we restrict to that case in the following. Furthermore, it holds for $t_1 < t< t_2$ with, 
\begin{equation}
\label{times}
    t_1 = \ln \frac{B_-}{\ctd}~, \quad  t_2 = \ln \frac{B_+}{\ctd} ~. 
\end{equation}

If $0<\ctu < \frac{1}{8}$: the first condition in \eqref{40acce} is always satisfied and the number of e-folds is thus given by
\begin{equation}
\label{N1}
    N = \ctu(t_2-t_1)+ \ctd \left(e^{t_2}-e^{t_1}\right) \qquad \qquad (\ctu>0)~. 
\end{equation}
For the value $\ctu=\frac{1}{8}$, $B_-=B_+$ and there is no accelerated expansion. 

\begin{figure}[H]
\begin{center}
\includegraphics[width=0.8\textwidth]{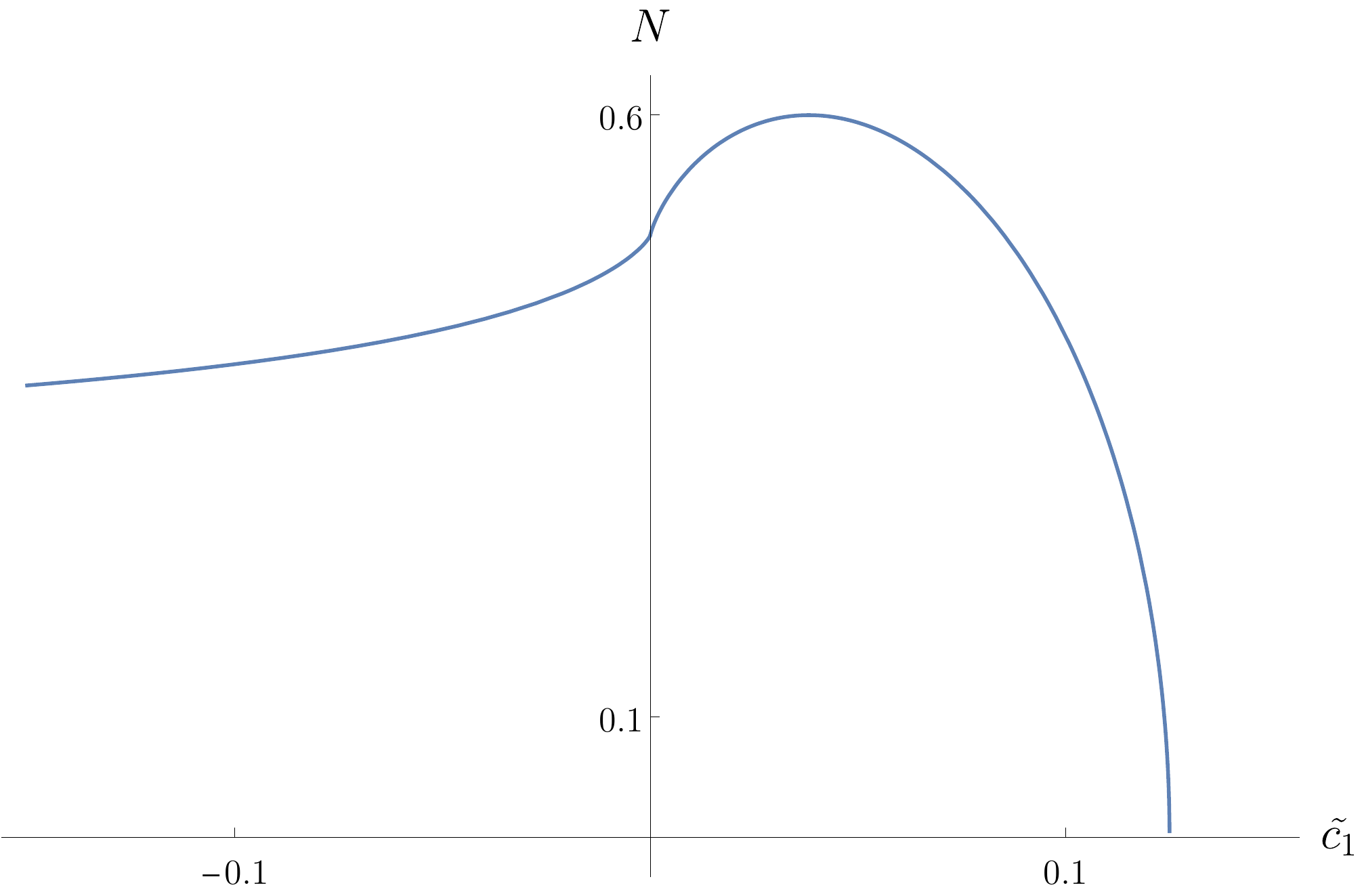}
\caption{Number of e-folds $N$ as a function of the parameter $\ctu$, for one-flux compactifications. $N$ reaches a maximum of $N_\text{max} = 0.59980$, for $\ctu = 0.038148$.}\label{fig:NOneF}
\end{center}
\end{figure}

If $\ctu<0$: the first condition in \eqref{40acce} is satisfied for $t>t_0$ with
\begin{equation}
 t_0 =   \ln - \frac{\ctu}{\ctd} ~. 
\end{equation}
Note that for all $\ctu<0$, $t_1<t_0<t_2$. In this case, the number of e-folds is given by
\begin{equation}
\label{N2}
    N = \ctu(t_2-t_0)+ \ctd \left(e^{t_2}-e^{t_0}\right) \qquad \qquad (\ctu<0)~. 
\end{equation}
For $\ctu \to 0$, $t_1 \to -\infty$ but $N$ is however finite and 
\begin{equation}
\lim_{\ctu \to 0^-} N = \lim_{\ctu \to 0^+} N = \frac{1}{2}~. 
\end{equation}
Plugging \eqref{times} into \eqref{N1} and \eqref{N2}, we observe that $N$ only depends on $\ctu$, and reads explicitly
\begin{equation}
    N(\ctu) = \left\{ 
\begin{array}{ll} \frac{1}{4}(1+\sqrt{1-8\ctu}) + \ctu \ln \left[1- \frac{1}{4 \ctu}\left(1+\sqrt{1-8\ctu}\right) \right]  & \text{if} \; \ctu<0 ~,\\
\\ \frac{1}{2}\sqrt{1-8\ctu} + \ctu \ln \frac{1+\sqrt{1-8\ctu}-4 \ctu}{1-\sqrt{1-8\ctu}-4 \ctu}
  & \text{if} \;  \ctu>0~. 
\end{array}\right. 
\end{equation}
For $\ctu < 0$, $N$ is monotonously increasing from 0 to $\frac{1}{2}$. For $0<\ctu<\frac{1}{8}$, $N$ admits a maximum $N_\text{max} = 0.59980$ at $\ctu = 0.038148$, see Figure \ref{fig:NOneF}. It is the maximal number of e-folds one can obtain in such one-flux solutions.

\section{Two-flux solutions: dynamical system analysis}\label{sec:dsa}

Let us now consider the case where the potential consists of two terms,
\eq{U=\sum_{i=1}^2c_ie^{E_i}~;~~~E_i:=\alpha_iA+\beta_iB+\gamma_i \phi
~,}
where   $c_i$, $\alpha_i$, $\beta_i$, $\gamma_i$ are real constants. 
We shall assume that the potential is not everywhere non-positive, so that 
we can take $c_1>0$, while $c_2$ is unconstrained. 

Let us  define the following phase-space variables,
\eq{\spl{\label{dg206}
v=e^{-E_1/2}\d_\tau A~;~~~u=e^{-E_1/2}\d_\tau B~;~~~w=e^{-E_1/2}\d_\tau \phi
~.}}
In terms of these, the equations of motion become an autonomous dynamical system,
\begin{align}
\d_\sigma v = 
&-\left(\tfrac18 \a_2 - \tfrac12 \b_2\right) u^2 - \left(\a_2 +\tfrac12 \b_1 - 4 \b_2 \right) u v - \left(\tfrac12 \a_1 +\tfrac32  \a_2 - 
    6 \b_2\right) v^2 - \tfrac12 \c_1 v w \nn \\ &- \left(-\tfrac{1}{96}\a_2 +\tfrac{1}{24} \b_2 \right) w^2
+ \tfrac{1}{48}c_1\left[ \alpha_2-\alpha_1+4\left(\beta_1-\beta_2\right)\right] \label{dg201} \\
\d_\sigma u = &-\tfrac12\left(-\a_2 + b_1 + 3 \b_2\right) u^2 - \left(\tfrac12\a_1 - 4 \a_2 - 
    12 \b_2\right) u v - \left(-6 \a_2 + 18 \b_2\right) v^2 - \tfrac12\c_1 u w \nn \\ &- \left(\tfrac{1}{24}\a_2 + \tfrac18\b_2\right) w^2
+\tfrac{1}{12}c_1 \left[\alpha_1-\alpha_2+3\left(\beta_2-\beta_1\right)\right] \nn  \\
\d_\sigma w = &- 
6 \c_2 u^2 - 48 \c_2 u v - 72 \c_2 v^2 - \tfrac12 \b_1 u w - \tfrac12
 \a_1 v w - \tfrac12\left(\c_1 - \c_2\right) w^2 + c_1 \left(\gamma_2-\gamma_1\right) \nn
~,
\end{align}
where $\d\sigma := e^{E_1/2}\d\tau$, 
and we have used the constraint to eliminate the terms with $c_2$ from the equations of motion.  
Moreover, the constraint reads
\eq{\label{dgds1}
72v^2+6u^2
+48vu-\tfrac12 w^2  =
c_1+c_2  e^{E_2-E_1} 
~.}
To better analyse the behavior of the  flow at infinity, it is useful to compactify the phase space as in  \cite{Sonner:2006yn,Russo:2022pgo}. We 
introduce the new variables
\eq{\label{dg212}
x=\frac{2v}{4v+u}~;~~~
y=\frac{w}{2\sqrt{3}(4v+u)}~;~~~
z=\frac{\sqrt{c_1}}{\sqrt{6}(4v+u)}
~.}
Note that in these variables the condition of expansion is equivalent to $z>0$. The system \eqref{dg201} becomes
\eq{\spl{\label{dgeomds2}
x'&=\tfrac14 \Big( [\a_2 + 2 \b_2 (-2 + x)] (-1 + x^2 + y^2+z^2) + [-\a_1  - 
      2 \b_1  (-2 + x)]z^2 \Big)   \\
y'&= \tfrac12 \Big( 
  ( 2 \sqrt{3} \c_2+ \b_2 y)  (-1 + x^2 + y^2 + z^2) -(2 \sqrt{3} \c_1 + \b_1 y) z^2 \Big)      \\
z'&=\tfrac14 z \Big(\a_1 x + 4 \sqrt{3} \c_1 y - 2 \b_1 (-1 + 2 x + z^2) + 
   2 \b_2 (-1 + x^2 + y^2 + z^2) \Big)
~,}}
where $f'=\d_\omega f$ and  $\d\omega := \frac{\sqrt{c_1}}{\sqrt{6}z} \d\sigma$, while the constraint \eqref{dgds1} now reads
\eq{\label{dg214}
{c_1}(1-x^2-y^2-z^2)= {c_2 } ~\!z^2e^{E_2-E_1}
~.}
Let us also note that
\eq{\label{dodtdef}\d\omega=
\frac{\sqrt{c_1}}{\sqrt{6}z}e^{E_1/2}\d\tau~;~~~\d T=\frac{\sqrt{6}z}{\sqrt{c_1}}e^{12A+3B-E_1/2}\d\omega
~,
}
as follows from the previous definitions. The equations of motion \eqref{dgeomds2} imply
\eq{\label{dgsedf}
\begin{aligned}
\tfrac12\left(x^2+y^2+z^2\right)'=
&\tfrac14 \left(-1 + x^2 + y^2 + z^2\right) \\
&\times \left(\a_2 x + 4 \sqrt{3} \c_2 y - 2 \b_1 z^2 + 
   2 \b_2 \left[(-2 + x) x + y^2 + z^2 \right] \right)
~,
\end{aligned}
}
so that the unit  sphere,
\eq{\label{ginvs}
\mathcal{S}=\left\{(x,y,z)\in\mathbb{R}^3~|~x^2+y^2+z^2=1\right\}
~,}
is an invariant surface. This implies  that trajectories which include  some  interior  (resp. exterior) point of the  three-dimensional unit ball  will remain there; such trajectories correspond to 
$c_1$, $c_2$ having the same (resp. opposite) sign. Similarly,  
trajectories that include a point on the sphere $\mathcal{S}$ must lie entirely on $\mathcal{S}$. As can be seen from the constraint \eqref{dg214}, trajectories on $\mathcal{S}$ 
must have $c_2=0$. 

Moreover, it follows immediately from the third line in  \eqref{dgeomds2}  that the plane $z=0$ is  another invariant surface, so trajectories 
which include some   point in the upper (resp. lower) half of the  three-dimensional space $z>0$ (resp. $z<0$) must lie there entirely.~Similarly, 
trajectories which include a point of the $(x,y)$-plane must lie entirely on that plane. The latter  trajectories must have $c_1=0$, as follows  from \eqref{dg214}.

The intersection of the two invariant surfaces above, the unit circle $\mathcal{C}$  in the  $z=0$ plane,  is also 
an invariant surface. 
In fact $\mathcal{C}$ is a circle of fixed points,
\eq{
p_{\mathcal{C}}
\in 
\mathcal{C}:= \left\{(x,y,z)~|~
x^2+y^2=1~\text{and}~z=0\right\}
~.}
Each point $p_{\mathcal{C}}$ corresponds {to} a trajectory (solution) with $c_1, c_2=0$, i.e.~the 
minimal solution of Section \ref{sec:special}.    In particular, the ratio $r$ defined in \eqref{rdef}  is related to the polar angle via, 
\eq{
\tan\theta=\frac{y}{x}=\pm\frac{1}{2r}\sqrt{12r^2+8r-1}
~.}
%


Another consequence of the system \eqref{dgeomds2} is that the plane
\eq{\mathcal{P}:=\left\{(x,y,z)\in\mathbb{R}^3~|~ax+by+c=0\right\}
~,}
is an invariant surface, where  
the constants $a$, $b$, $c$ are obtained as solutions of the system of equations\footnote{Solutions to this system exist, provided $(\a_1\b_2-\a_2\b_1)^2+(\a_1\c_2-\a_2\c_1)^2+(\b_1\c_2-\b_2\c_1)^2\neq0$.}
\eq{\spl{
(\a_2 - 4 \b_2)a  + 4 \sqrt{3}  \c_2 b- 2 \b_2 c&=0\\
 \left[ \a_2-\a_1  - 4 (\b_2-\b_1) \right]a+ 
4\sqrt{3}   (\c_2-\c_1) b - 2 (\b_2-\b_1 ) c &=0
~.}}
Indeed, in this case, \eqref{dgeomds2} implies
\eq{
(ax+by)'=\tfrac12(ax+by+c)\left[\b_2(x^2+y^2-1)+(\b_2-\b_1)z^2\right]
~.}
Allowed trajectories must therefore either lie entirely on $\mathcal{P}$, or be limited on either side of it. 

As follows from \eqref{dgeomds2}, the flow equations are invariant under $(z,\omega) \rightarrow-(z,\omega)$, 
so that each trajectory in the $z>0$ region is paired to a ``mirror'' trajectory in the $z<0$ 
region. As we are ultimately interested in expanding cosmologies, we will restrict our attention to the $z\geq 0$ region. On the other hand, taking \eqref{14}, \eqref{dg214} into account, 
the condition for acceleration is written as\footnote{\label{foot:3}More precisely, the left hand side of the second inequality in \eqref{dgacccylind} is equal to $\frac{3}{c_1}e^{-E_1}z^5S^5\ddot{S}$.}
\eq{\label{dgacccylind}
\ddot{S}(T)>0\Longleftrightarrow (\b_1 - \b_2) z^2 - \b_2 \left(x^2 + y^2\right) + \b_2 - 4>0~,
}
which defines an acceleration region in the phase space. Depending on the values taken by $\beta_1, \beta_2$, it can be a cone, a cylinder, a ball (regular or deformed) or the region above a horizontal plane, see Table \ref{tab:regions} and Figure \ref{fig:vecfield} below. 

\begin{figure}[H]
\begin{center}
\includegraphics[width=1\textwidth]{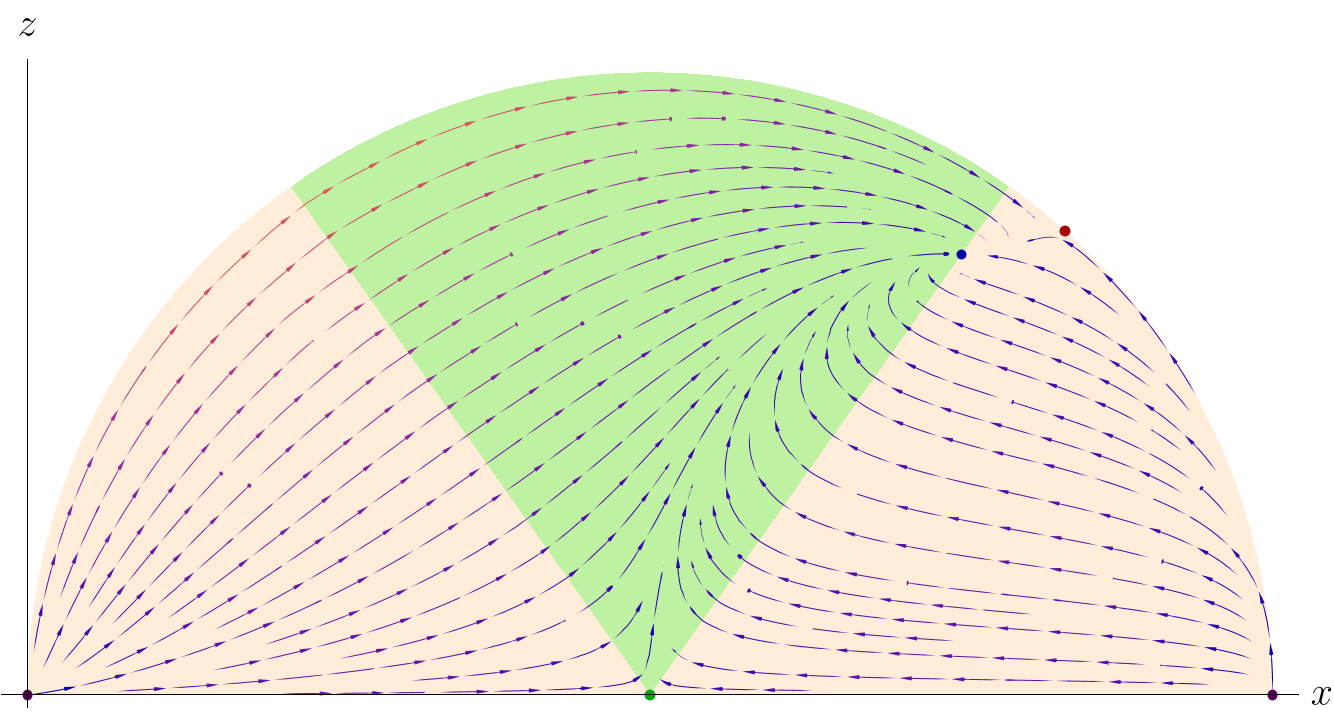}
\caption{Vector field  generated by the dynamical system \eqref{dgeomds2}, plotted in the invariant plane $\mathcal{P}$. Here, the acceleration region is a cone, depicted in green. The generic equator fixed points are illustrated in purple. This example corresponds to the case $\lambda, k \neq 0$, studied extensively in Section \ref{case1}, and coincides with the northern hemisphere of Figure \ref{fig:trajkl}.}\label{fig:vecfield}
\end{center}
\end{figure}
Note that different pairs of $\beta_{1,2}$ lead to acceleration regions of different shape. Although  the solutions are invariant  under $\beta_1\leftrightarrow\beta_2$, this  leads to a reparametrization of the phase space of the corresponding dynamical system. Put differently, although the solutions are invariant under reflections along the diagonal of Table \ref{tab:regions}, the shapes of the acceleration regions are not.

Moreover, using \eqref{wdef}, \eqref{windir} and \eqref{dg214} we have
\begin{equation}
\label{dgw221} 
    w = 1-\tfrac13\b_2 +\tfrac13\b_2\left(x^2+y^2\right)+\tfrac13(\b_2-\b_1)z^2 ~. 
\end{equation}
Note that in the presence of external curvature (with the convention $c_2 = -6k$), the above expression is to be replaced by
\begin{equation}
    w = 1-\tfrac13\b_1~\!\frac{z^2}{x^2+y^2+z^2} ~.
\end{equation}
Eqs.~\eqref{dgacccylind}, \eqref{dgw221} are consistent with the fact that 
 the acceleration condition is equivalent to $w<-\frac13$, if  $k+\dot{S}^2>0$. 

It is also possible to obtain an expression for the number of e-foldings $N$ directly in terms of phase space variables. Indeed, solving \eqref{dg212} for  for $v$, $u$, $w$, and
taking \eqref{dg206} into account, we obtain
\eq{\spl{\label{ABphisl}
\d_\omega A=\tfrac12x~;~~~
\d_\omega B=1-2x~;~~~
\d_\omega \phi=2\sqrt{3}~\! y~.
}}
In particular this   implies  $\d_\omega (4A+B)=1$, which, taking  \eqref{scfc} into account, is equivalent to
\eq{\label{cf70}
\d\ln S=\d\omega
~.}
This means that the flow parameter of the system \eqref{dgeomds2} is simply $\omega=\ln \tfrac{S}{S_0}$, so that $S=S_0$ at $\omega=0$. It follows that
\eq{\label{efoldeq65}
N=\int \d \omega
~,}
where the limits of the integral should be taken as the points of entry and exit of the trajectory into and out of  the acceleration region  \eqref{dgacccylind}, respectively.

\subsection*{The scale factor}

Given a solution $\left(x(\om), y(\om), z(\om) \right)$ of the above dynamical system, it becomes possible to reconstruct the corresponding expression for the scale factor $S(T)$, which is the quantity we are ultimately interested in, in order to construct all the observables related to the cosmological model. 

One can integrate \eqref{dodtdef} to obtain
\begin{equation}
\label{Tomgen}
T(\om)= \sqrt{\frac{6}{c_1}} \int_{-\infty}^{\omega} \d \omega' z(\omega') \exp \left[ \left(12 - \frac{\alpha_1}{2}\right) A(\omega')+ \left(3 - \frac{\beta_1}{2}\right) - \frac{\gamma_1}{2} \phi(\omega') \right]~,
\end{equation}

where $T=0$ corresponds to the lower bound $\omega \to -\infty$ and $S=e^{\om}= 0$. Note that since we restrict to $z>0$, $T$ is ensured to be positive. Here, $A, B, \phi$ are completely determined by the solution (and the data of initial conditions) via \eqref{ABphisl}, and can be computed by numerical integration. 
In practice, we solve the system over a finite range $[\om_{\text{min}},\om_{\text{max}}]$, which gives the bounds to be used in the integrals. One can then numerically invert \eqref{Tomgen} to obtain $\om(T)= \log S$. 

Alternatively, we can compute the parametric curve $(\log T(\om), \om)$ with parameter $\om$ which corresponds to the log--log plot $(\log T, \log S)$, as shown in Figure \ref{fig:S}. In such a plot, the freedom in the parameter $c_1$ can be thought as a freedom to move the curve left and right.\footnote{This may also be thought as a freedom in tuning the value of $H_0$, the Hubble parameter at a given $T_0$.} Furthermore, since the dynamical system is autonomous, meaning it does not depend on $\om$ explicitly, every shift of a solution is also a solution, viz.
\begin{equation}
    \vec{x}(\om) \; \text{is a solution} \; \Longrightarrow \; \vec{x}(\om + \Omega) \; \text{is a solution}~.
\end{equation}
This means that the log--log plot can also be shifted up or down at will. The only freedom left is the choice in the initial conditions $(x_0,y_0,z_0)$. \\

From \eqref{Tomgen}, one can express $\Sd$ and $\Sdd$ as functions of $\om$, 
\begin{equation}
\begin{aligned}
          \Sd&= \frac{\d S}{\d T}= \left(\frac{\d T}{\d S} \right)^{-1} = \left(\frac{\d T}{\d \om} \frac{\d \om}{\d S} \right)^{-1} = \frac{\d S}{\d \om} \left(\frac{\d T}{\d \om} \right)^{-1} = \frac{e^{\om}}{t(\om)}~, \\
              \Sdd&=\frac{\d}{\d T}\Sd = \frac{\d \om}{\d T} \frac{\d}{\d \om} \Sd = \frac{e^{\om}}{t^2}\left(1- \frac{t'}{t} \right)~, \\
              \dddot{S} &= \frac{e^{\om}}{t^5}\left[ t^2 + 3t'^2 - t(3t'+t'')\right]~,
\end{aligned}
\end{equation}
where $t(\om)$ is the integrand in \eqref{Tomgen}, and $t'=\d_{\om}t$. 
From this, one may compute the Hubble parameter $H = \Sd/S$ and its derivatives as functions of $\om$, as well as other quantities such as the tensor--to--scalar ratio $r$ or the scalar spectral index $n_s$. In principle, these could allow to further assess and restrict the viability of the models obtained in this way (beyond the mere number of e-foldings $N$), but this goes beyond the scope of the present paper, although it would be interesting to explore these constraints in future work. \\

In the following four subsections we will study in depth four dynamical systems. 
The first two correspond to  
the two open FLRW cosmologies  described in the Introduction. The one with negative internal curvature is studied in Section \ref{case1}, whereas  the 
one  with non-vanishing Romans mass in Section \ref{case2}. 
Both admit  solutions with infinite or parametrically controlled number of e-foldings.  
The remaining two systems, studied in Sections \ref{sec:34d} and \ref{case4},  were chosen for the richness of their fixed-point structure, allowing for different interpolating solutions. All the other possible two-flux dynamical systems can be found in Appendix \ref{app:sys}, and are summarized in Table \ref{tab:systems} below.

\begin{table}[H]
    \centering
\begin{tabular}{|c||c|c|c|c|c|c|c|c|c|c|}
\hline
 & $m$ & $c_{\varphi}$ & $c_{\chi}$ & $c_{\xi \xi'}$ & $c_f$ &$c_h$ &$b_0$ &$c_0$ &$k$ &$\lambda$ \\
\hline
\hline
$m$ & $\varnothing$ & \eqref{spm} & $\varnothing$ & $\varnothing$ & $\varnothing$ & $\varnothing$ & $\varnothing$ & \eqref{sczm}& \eqref{mksys} & \eqref{eomds2} \\
\hline
$c_{\varphi}$ & \eqref{spm} & $\varnothing$ & \eqref{spc} & \eqref{spx} & \eqref{sfp}  & $\varnothing$ & $\varnothing$ & $\varnothing$ &\eqref{spk} &\eqref{spl} \\
\hline
$c_{\chi}$ & $\varnothing$ & \eqref{spc} & $\varnothing$ & \eqref{scx} & $\varnothing$ &\eqref{sch} & $\varnothing$ & $\varnothing$ &\eqref{sck} &\eqref{scl} \\
\hline
$c_{\xi \xi'}$ & $\varnothing$ & \eqref{spx} & \eqref{scx} & $\varnothing$ & $\varnothing$ &\eqref{sxh} & $\varnothing$ & $\varnothing$ &\eqref{sxk} & $\varnothing$ \\
\hline
$c_f$ & $\varnothing$ & \eqref{sfp} & $\varnothing$ & $\varnothing$ & $\varnothing$ & $\varnothing$ & $\varnothing$ & $\varnothing$ & \eqref{skf} &  \eqref{slf} \\
\hline
$c_h$ & $\varnothing$ & $\varnothing$ & \eqref{sch} & \eqref{sxh} & $\varnothing$ & $\varnothing$ & $\varnothing$ & $\varnothing$ &\eqref{shk} &\eqref{shl} \\
\hline
$b_0$ & $\varnothing$ & $\varnothing$ & $\varnothing$ & $\varnothing$ & $\varnothing$ & $\varnothing$ & $\varnothing$ & \eqref{sczbz} & $\varnothing$ & $\varnothing$ \\
\hline
$c_0$ & \eqref{sczm} & $\varnothing$ & $\varnothing$ & $\varnothing$ & $\varnothing$ & $\varnothing$ & \eqref{sczbz} & $\varnothing$ &\eqref{sczk} &\eqref{sczl} \\
\hline
$k$ & \eqref{mksys} & \eqref{spk} & \eqref{sck} & \eqref{sxk} & \eqref{skf} & \eqref{shk} & $\varnothing$ & \eqref{sczk} & $\varnothing$ & \eqref{kleomds2} \\
\hline
$\lambda$ & \eqref{eomds2} & \eqref{spl} & \eqref{scl} & $\varnothing$ & \eqref{slf} & \eqref{shl} & $\varnothing$ & \eqref{sczl} &  \eqref{kleomds2} & $\varnothing$\\
\hline
\end{tabular}
    \caption{All possible two-flux dynamical systems. A $\varnothing$ indicates that the corresponding pair of fluxes does not relate to a possible subcase of \eqref{16u}.}
    \label{tab:systems}
\end{table}


\subsection{Case study I: $\lambda, k$}
\label{case1}
Let us begin with the case where $\lambda, k\neq0$. 
From \eqref{dgeomds2}, we obtain the dynamical system
\eq{\spl{\label{kleomds2}
x'&=2 x ~\!\left(x^2 + y^2-\tfrac12 z^2-1\right) +2 z^2 \\
y'&=2 y ~\!\left( x^2 + y^2-\tfrac12 z^2-1\right) \\
z'&=z~\!\left[1 + 2 (x-1) x + 2 y^2 - z^2\right]
~,}}
where we have set $c_1=-6\lambda$, $c_2=-6k$, $(\a_1,\b_1,\c_1)=(16,6,0)$, $(\a_2,\b_2,\c_2)=(16,4,0)$. 
Moreover, we assume $\lambda<0$, so that $c_1>0$ as required in Section \ref{sec:dsa}. 
The constraint \eqref{dg214} reduces to
\eq{\label{kldg214}
\lambda\left(1-x^2-y^2-z^2\right)= k  ~\!z^2e^{-2B}
~,}
so that $k<0$ (resp. $k>0$) restricts to trajectories in the interior (resp. exterior) of the invariant surface $\mathcal{S}$, cf.~\eqref{ginvs}. 
Trajectories on $\mathcal{S}$ require $k=0$, while those on the $z=0$ plane require $\lambda=0$. The points on the equator $\mathcal{C}$ of $\mathcal{S}$ require $k$, $\lambda=0$.

Besides  $\mathcal{S}$ and the invariant plane $z = 0$, the plane $y = 0$ is also an invariant surface. Since  eqs.~\eqref{kleomds2} are invariant under $y\rightarrow-y$, we may
restrict our attention to trajectories lying in the $y$, $z\geq 0$ quadrant.

The condition for acceleration  \eqref{dgacccylind} reduces to
\eq{\label{klacccylind}
\ddot{S}(T)>0\Longleftrightarrow z^2>2\left(x^2+y^2\right)
~,
}
so that accelerated expansion occurs in the portion of  the trajectory that lies in the   upper half of the 
cone defined in \eqref{klacccylind}.
Moreover, from  \eqref{dgw221} we have
\eq{\label{klw221} 
w=\frac{x^2+y^2-z^2}{x^2+y^2+z^2}
~,}
so that  $w=-1$ whenever the trajectory passes by the $z$-axis. Equations \eqref{klacccylind}, \eqref{klw221} imply that 
the acceleration condition is equivalent to $w<-\frac13$. 
\subsection*{The critical points} 

For  $z>0$ there are two   critical points of the system \eqref{kleomds2}, given by
\eq{\label{224}
p_{1}=  \tfrac{1}{2} \left(1,0,\sqrt{2} \right)~;~~~p_{2}= \tfrac{1}{3} \left(2,0,\sqrt{5} \right)
~.}
Both $p_{1,2}$ lie on the invariant $y=0$ plane. Moreover,   $p_{1}$  lies on the boundary of the acceleration cone and in the interior of $\mathcal{S}$, while  $p_{2}$ lies outside the cone and  on the boundary of $\mathcal{S}$. 

On the $z=0$ plane, the origin $p_0=(0,0,0)$ is an isolated fixed point.  In addition, we have an  invariant circle of fixed points $p_{\mathcal{C}}$:  the equator of the sphere $\mathcal{S}$. These points require $k$, $\lambda=0$ and correspond to the 
minimal solutions of Section \ref{sec:special}.  

The  linearized system at $p_1$  has   eigenvalues $-1$ (double) and $-2$. The corresponding eigenvectors are along the 
$x$ and $y$ directions, respectively. 
The  linearized system at $p_0$  has   eigenvalues $-2$ (double) and $1$. The corresponding eigenvectors are along the 
$x$, $y$ and $z$ directions, respectively.

The critical points of the dynamical system correspond to solutions that can be given analytically: $p_1$ corresponds to the singular Milne universe given in \eqref{Elk2}; $p_2$ requires $k=0$ (which is consistent with the fact that it lies on $\mathcal{S}$) and corresponds to the critical solution with $\lambda<0$, which is a decelerating power-law expansion. The origin $p_0$ requires $\lambda=0$ (which is consistent with the fact that it lies on the $z=0$ plane) and corresponds to the regular Milne universe of \eqref{CY3}.


\subsubsection*{The invariant surface $\mathcal{S}$} 

Restricting to trajectories on  $\mathcal{S}$, the system \eqref{kleomds2} implies
\eq{
x'=3(x-\tfrac23)(x^2+y^2-1)~;~~~
y'=3y(x^2+y^2-1)
~.}
It follows that the projections of all trajectories on $\mathcal{S}$ to the $z=0$ plane are of the form
\eq{\label{line1lk}
a(x-\tfrac23)+by=0~;~~~a,b\in\mathbb{R}
~,}
i.e. straight lines passing by the point $(x,y)=\left(\tfrac23, 0\right)$, which is the projection of $p_2$ onto the $z=0$ plane. These trajectories correspond precisely to 
the solutions of \eqref{186}, and require $k=0$. More specifically, the slope of the line \eqref{line1lk} is related to the constants in \eqref{186} via
\eq{
\frac{a}{b}= - \frac{ 3\sqrt{3}c_\phi}{20c_A}
~.}

\subsubsection*{The $z=0$ invariant plane}

Restricting to trajectories on the $z=0$ plane, the system \eqref{kleomds2} reduces to
\eq{
x'=2x(x^2+y^2-1)~;~~~
y'=2y(x^2+y^2-1)
~.}
It follows that  all trajectories on  the $z=0$ plane are of the form
\eq{\label{linesd}
ax +by=0~;~~~a,b\in\mathbb{R}
~,}
i.e. straight lines passing by the point $(x,y)=(0, 0)$. These straight lines correspond precisely to the 
solutions of \eqref{40a}, and require $\lambda=0$. More specifically, the slope of the line \eqref{linesd} is related to the constants in \eqref{40a}, \eqref{60bht} via
\eq{
\frac{a}{b}= -\frac{c_\phi}{4\sqrt{3}c_A} 
~.}

\subsubsection*{The  invariant plane $y=0$}

On the plane $y=0$, the system  \eqref{kleomds2} reduces to
\eq{\spl{\label{ip288}
x'&=2 x ~\!\left(x^2 -\tfrac12 z^2-1\right) +2 z^2 \\
z'&=z~\!\left[1 + 2 (x-1) x  - z^2\right]
~.}}
All trajectories of the reduced system are attracted by the stable node $p_1$, except for the two trajectories on $\mathcal{S}$ which start at the two antipodal points 
$(x,y,z)=\pm(1,0,0)\in \mathcal{C}$, 
and end on either side of the unstable node $p_2$. 

Using 
a perturbative analysis, it is possible to obtain an analytic description of the 
trajectory connecting $p_0$ and $p_1$ near the critical endpoints. 
In the neighborhood of   $p_0$ the system \eqref{ip288} admits the solution
\eq{\label{ip288appr}
x=
\tfrac12 c^2 e^{2\omega} - 
 \tfrac34  c^4 e^{4\omega} + 
 \mathcal{O}\left( e^{6\omega}\right)~;~~~
 z=
 c e^{\omega} 
 -  c^3 e^{3\omega} + 
 \mathcal{O}\left( e^{5\omega}\right)
~,}
so that the $(x,z)$ trajectory attains $p_0$ in the $\omega\rightarrow-\infty$ limit, tangentially along  the vertical direction ($z$-axis). 
Moreover the constraint \eqref{kldg214} imposes
\eq{\label{con288}
c=\sqrt{\left|\tfrac{\lambda}{k}\right|}~\!e^{B_0}
~.}
It can  be seen from \eqref{klw221}, \eqref{ip288appr} that $w\rightarrow -1$ as the trajectory tends to  $p_0$. 
However it would be incorrect to conclude that the solution becomes asymptotic de Sitter, since the $p_0$ point is reached at finite cosmological time in the past.~Indeed this can be seen explicitly by reconstructing the 
metric corresponding to the solution \eqref{ip288appr}, \eqref{con288}:  
taking into account the relation between $\d T$ and $\d\omega$, cf.~\eqref{dodtdef}, we obtain
\eq{\label{gfhghhgd}
\d T=S^3\d\tau=\tfrac{1}{\sqrt{|\lambda|}}~\! z~\! e^{4A}\d\omega
~.}
Moreover from \eqref{ABphisl} and \eqref{ip288appr} we obtain the perturbative expression for $A$ and $z$. 
Plugging into \eqref{gfhghhgd} and integrating we obtain\footnote{We have obtained the perturbative solution to a very high order in $e^{2\omega}$, and we have shown that it is consistent with the closed expression, 
\eq{2\sqrt{2|\lambda|}~\!e^{-4A_0}T=c~\!e^\omega+
\text{arcsinh}(c~\!e^\omega)~.}
%
}
\eq{
T= \tfrac{1}{\sqrt{|\lambda|}}~\!c~\! e^{4A_0+\omega} \left[ 1-\tfrac16 c^2e^{2\omega}+\mathcal{O}\left( e^{4\omega}\right)\right]
~,}
where we imposed   $T\rightarrow 0$ as $\omega\rightarrow-\infty$. Taking \eqref{con288} into account gives
\eq{\label{ps298}
\sqrt{|k|} ~\! T=S_0 ~\!  e^\omega \left[ 1-\tfrac16 c^2e^{2\omega}+\mathcal{O}\left( e^{4\omega}\right)\right]
~,}
where we have set $S_0:=e^{4A_0+B_0}$, so that $S=S_0 ~\! e^\omega$, cf.~below \eqref{cf70}. Finally, inverting the perturbative series 
\eqref{ps298} we obtain
\eq{
S=\sqrt{|k|} ~\! T
\left[ 1+\tfrac16 |\lambda|e^{-8A_0}T^{2}+\mathcal{O}\left( T^{4}\right)\right]
~.}
It then follows from \eqref{wdef} that $w=-1+\mathcal{O}(T^2)$, in agreement with our previous result.
Indeed up to and including terms of order $\mathcal{O}(T^4)$ the spacetime metric becomes that of de Sitter space in hyperbolic slicing,
\eq{
\d s^2_{dS}=-\d T^2+|k|\Lambda^2\sinh^2\left(\frac{T}{\Lambda}\right)\d\Omega_k^2~;~~~\Lambda:=\frac{1}{\sqrt{|\lambda|}}e^{4A_0}
~,}
where $\Lambda$ is related to the scalar curvature $R$ of de Sitter via $\Lambda^2=12/R$.~This is not asymptotic de Sitter, however, as $T=0$ is reached at finite proper 
time in the past, where  the space becomes a regular Milne universe.~The solution can thus be geodesically completed in the past to $T<0$, by gluing together its mirror trajectory in the $z<0$ region, cf.~the comment in the paragraph preceding \eqref{dgacccylind}.

A similar analysis can be performed in the neighborhood of $p_1$. In that case we obtain  
$S\rightarrow 2\sqrt{|k|}~\!T$, so that $w\rightarrow -\tfrac13$, as the trajectory 
approaches $p_1$.

\subsubsection*{Comparison with the analytical solutions \eqref{asolu}}

It follows from the system  \eqref{kleomds2} that the ellipse, 
\eq{x=\tfrac12
~;~~~
\tfrac43 y^2+2z^2=1
~,}
is an invariant submanifold. The upper vertex of the ellipse (at $z=\tfrac{1}{\sqrt{2}})$ is the critical point $p_1$, while the left and right vertices (at $y=\pm\tfrac{\sqrt{3}}{2}$) are both in $p_{\mathcal{C}}$.
The two trajectories starting at either the left or the right vertex of the ellipse and ending at the upper vertex, correspond precisely to the analytic solutions around \eqref{asolu}.

\subsubsection*{Potential and kinetic energies}

The acceleration period can in fact be understood as a competition of the kinetic and potential energies of the system, as we would like to illustrate here. 

Let us recall that, by comparing the energy-momentum tensor of a perfect fluid and that of a homogeneous scalar field $\varphi$, one can assign to the latter the following pressure and energy density, 
\begin{equation}
    \begin{aligned}
    p_{\varphi} &= \frac{1}{2} \dot{\varphi}^2 - V(\varphi) \,, \\
        \rho_{\varphi} &= \frac{1}{2} \dot{\varphi}^2 + V(\varphi) \,. 
    \end{aligned}
\end{equation}
The acceleration condition $w=p_{\varphi}/\rho_{\varphi} < -1/3$ then translates to
\begin{equation}
\dot{\varphi}^2 < V(\varphi)~,
\end{equation}
i.e. there is acceleration whenever the potential energy dominates (twice) the kinetic energy. 

\begin{figure}[h]
\begin{center}
\includegraphics[width=0.9\textwidth]{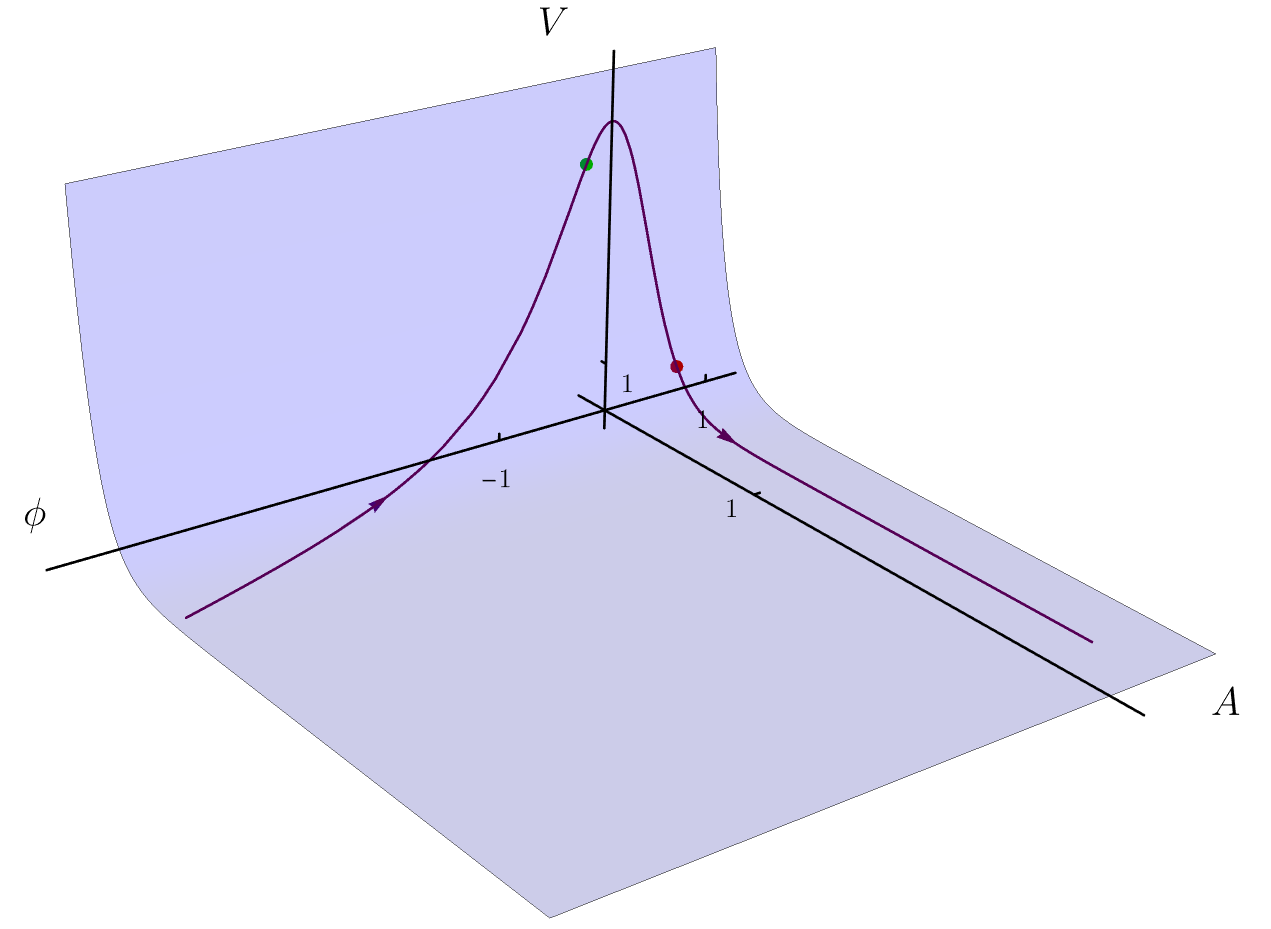}
\caption{The scalar potential $V$ plotted over the field space $(A,\phi)$. On it lies a typical trajectory featuring transient inflation: acceleration starts at the green dot and stops at the red dot. A these two special  positions, the potential energy equals twice the kinetic energy.}\label{fig:potential}
\end{center}
\end{figure}

In our models, the 2-field 4d potential has the shape of an exponential ``wall'', and the system can be thought in field space as thrown against that wall: initially the potential energy $V$ is exponentially small and the kinetic energy $K$ dominates. Eventually the system reaches the wall and starts climbing it, $K$ decreases while $V$ increases; when $2K=V$, acceleration begins, which starts to significantly dissipates energy (the Hubble term in the EOM of $\varphi$ acts a friction term). When the potential energy becomes too important, the trajectory undergo a turnaround and goes back down the wall; the system starts to accelerate again and inflation stops whenever $2K=V$. Such a trajectory is depicted in Figure \ref{fig:potential}.  

More quantitatively, in that case study the 4d potential is given by
\begin{equation}
    V(A)= -6 \lambda e^{-8A} ~.
\end{equation}
To compute the kinetic energy, one first has to canonically normalize the fields,
\begin{equation}
    A \rightarrow \tilde{A} = 4\sqrt{3} A \,, \qquad \phi \rightarrow \phi ~.
\end{equation}
Then, using $g^{\tau \tau} = S(\tau)^{-3} = \exp\left(-13 A - 3B \right)$, the kinetic energy reads
\begin{equation}
    K= \frac{1}{2}S^{-6} \left((\d_{\tau}\tilde{A})^2 + (\d_{\tau}\phi)^2 \right) ~.
\end{equation}
We further use that $\d \om / \d \tau = \sqrt{-\lambda} / z \; e^{8A +3B}$, cf. \eqref{dodtdef}, and 
\begin{equation}
\begin{aligned}
      \d_{\tau} A &= \d_{\om} A \, \frac{\d \om}{\d \tau} =\frac{\sqrt{-\lambda}}{z} \,\frac{1}{2}x \,  e^{8A +3B} ~,   \\
          \d_{\tau} \phi &= \d_{\om} \phi \, \frac{\d \om}{\d \tau} = \frac{\sqrt{-\lambda}}{z} \, 2 \sqrt{3} y \, e^{8A +3B} ~, 
\end{aligned}
\end{equation}

to obtain
\begin{equation}
    K = -6 \lambda e^{-8A} \, \frac{x^2 + y^2}{z^2} = V \times  \frac{x^2 + y^2}{z^2} ~.
\end{equation}
The condition for acceleration $\Sdd > 0 \Leftrightarrow V > 2K$ then precisely recovers the condition \eqref{condacckm}, which defines the boundaries of the acceleration region.


\subsection{Case study II: $k, m$}
\label{case2}
We now turn to the case where $k$, $m\neq 0$. Here, the
system of equations reduces to
\eq{\spl{\label{mksys}
x'&=2 x \left(x^2+y^2-1\right)+\frac{1}{2} (3-2 x) z^2 \\
y'&=2 y \left(x^2+y^2-1\right)-y z^2-\frac{5 \sqrt{3} z^2}{2} \\
z'&=\frac{1}{2} z \left(x (4 x-3)+4 y^2+5 \sqrt{3} y-2 z^2+2\right)
~,}}
while the 
constraint reads
\begin{equation}
12 k z^2 e^{-2 A-2 B-\frac{5 \phi }{2}}= m^2 \left(x^2+y^2+z^2-1\right)~.
\end{equation}
The invariant plane $\mathcal{P}$ 
is given by the equation
\eq{
\label{invkm}
 5 x +  \sqrt{3} y=0
~.}
The acceleration condition reads
\eq{
\label{condacckm}
z^2>2(x^2+y^2)
~,}
with
\eq{
w=\frac{x^2+y^2-z^2}{x^2+y^2+z^2}~.
}
The  critical points are
\eq{
p_\mathcal{C}~;~~~p_0~;~~~
p_1=\left(\frac{1}{14},-\frac{5}{14\sqrt{3}},\sqrt{\frac{2}{21}}\right)~.
}
 Both $p_0$ and $p_1$ lie on the boundary of 
the acceleration region and on the invariant plane. The point $p_0$ requires $m=0$, and  corresponds to the critical solution of \eqref{CY3}; the point $p_1$ corresponds to the critical solution of \eqref{Ekm1}.

The behavior close to the fixed point $p_0$  is similar to that of the $(\lambda,k)$ system analyzed in the previous section.~As the trajectory 
approaches $p_0$, 
we have $w\rightarrow-1$, and 
the solution becomes de Sitter-like.~This however is not an asymptotic de Sitter, as $p_0$ is reached at finite proper time in the past.~At $p_0$ spacetime becomes a regular Milne universe, and 
the solution can be geodesically completed in the past by gluing its mirror trajectory in the 
$z<0$ region.

Close to the fixed point $p_1$ we may linearize and solve \eqref{mksys} analytically.~The solution reads,
up to terms of order $\mathcal{O}(e^{-2\omega} )$, 
\eq{\spl{\label{369}
A&=A_0+\tfrac{1}{28}\omega+\tfrac12 ce^{-\omega} f(\omega)\\
B&=B_0+\tfrac67\omega-2c e^{-\omega}f(\omega)\\
\phi&=\phi_0-\tfrac{5}{7}\omega-10 ce^{-\omega} f(\omega)
~,}}
where $c$ is an integration constant, and we have defined
\eq{f(\omega):= 
\sqrt{17}\cos\left(\sqrt{\tfrac{17}{7}}\omega \right)
+\sqrt{7}\sin \left(\sqrt{\tfrac{17}{7}}\omega \right)
~.}
In addition, the constraint imposes
\eq{
k = -\frac34 m^2e^{2 A_0 + 2 B_0 + 5 \phi_0/2} 
~.}
It can be seen that, up to and including linear terms in $\omega$, the solution \eqref{369} corresponds to the critical solution of \eqref{Ekm1}, with $\tau\sim e^{-2\omega}$.~The acceleration can also be calculated analytically, cf.~\eqref{dgacccylind} and Footnote \ref{foot:3},
\eq{
\ddot{S}=-\frac{12}{5}c~\! e^{-2\omega}\left[
\sqrt{17}\cos \left(\sqrt{\tfrac{17}{7}}\omega \right)
-3\sqrt{7}\sin \left(\sqrt{\tfrac{17}{7}}\omega \right)
\right]
~,}
up to terms of order $\mathcal{O}(e^{-3\omega} )$. There is an infinite number of periodic cycles of  accelerated expansion followed by decelerated expansion, each of which lasts 
a half period, thus contributing a number of e-foldings equal to
\eq{
\Delta N=\sqrt{\tfrac{7}{17}}~\!\pi
~,}
where we have taken 
\eqref{efoldeq65} into account. The scale factor can also be expressed in terms of the cosmological time, 
\eq{
S=\sqrt\frac{7}{6} \sqrt{|k|} T + \frac{7}{85}c S_0 \left(19 \sqrt{17}   \cos \left(\sqrt{\tfrac{17}{7}}\ln T \right)-
      17 \sqrt{7}   \sin \left(\sqrt{\tfrac{17}{7}}\ln T \right)
      \right)
      +\mathcal{O} \left(\frac1T \right)~,
}
where to lowest order, $\ln T\sim \omega+\text{const}$. The fixed point is reached as $T\rightarrow\infty$. \\

\begin{figure}[H]
\begin{center}
\includegraphics[width=0.8\textwidth]{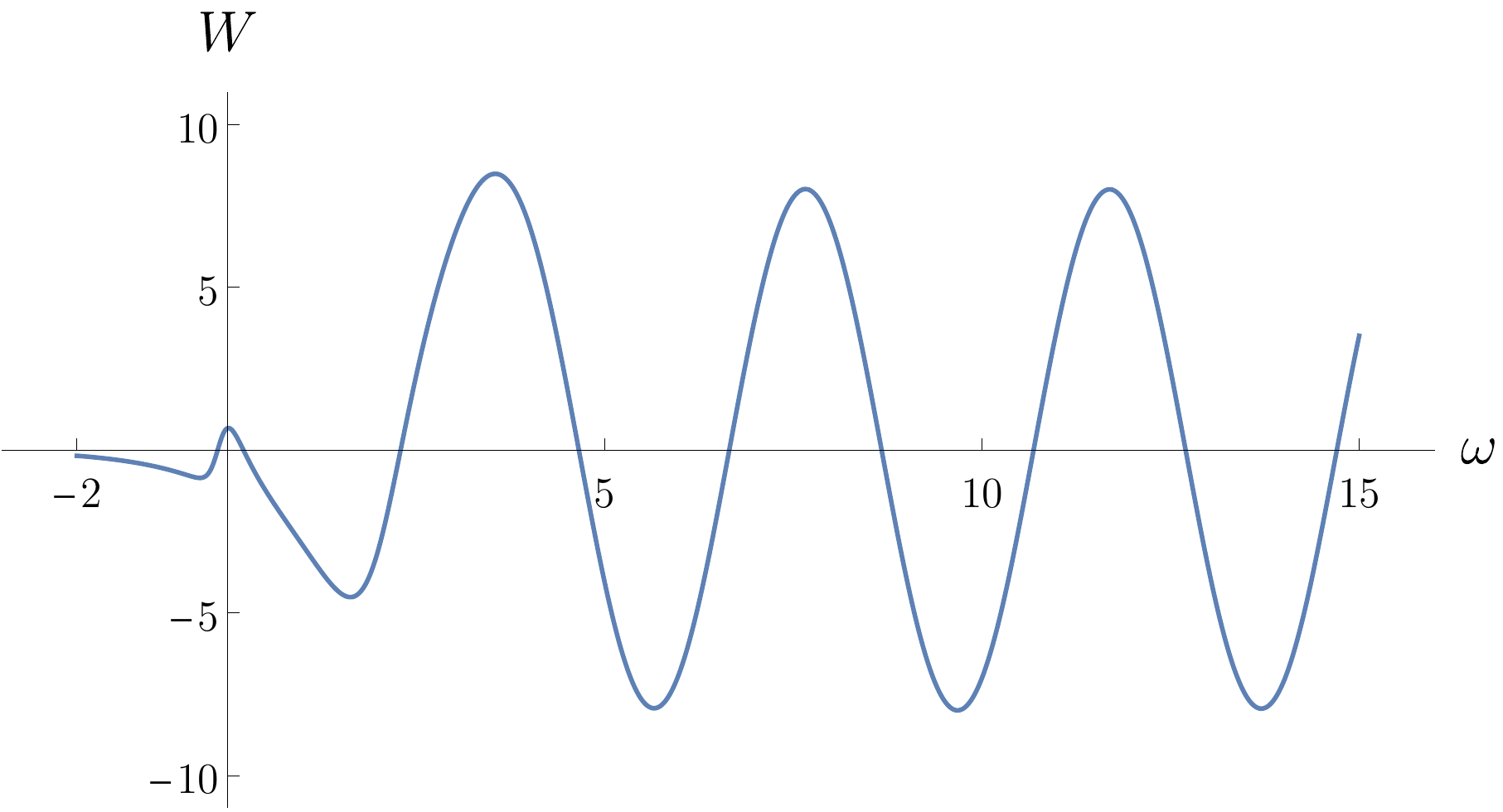}
\caption{Plot of $W$ as a function of $\om$, as defined in \eqref{Weq}. (Half-) periods of acceleration correspond to $W>0$. As $\om$ increases, the duration of these accelerated periods tends to $\sqrt{\frac{7}{17}} \pi$ in the variable $\om$.} \label{fig:W}
\end{center}
\end{figure}

The oscillations of the system can be captured for instance by  the equation of state parameter $w$. Since these are exponentially damped, we rather consider the following quantity,
\begin{equation}
\label{Weq}
    W(\om) := -\left(w(\om) + \frac{1}{3} \right)e^{\om} \,,
\end{equation}
plotted in Figure \ref{fig:W}, which is positive whenever there is acceleration, and negative otherwise.

\subsection{Case study III: $\lambda,m$}
\label{sec:34d}
 We proceed with the case where $\lambda$, $m\neq0$. 
Here, the dynamical system is given by
\eq{\spl{\label{eomds2}
x'&=(6x-4)\left(x^2+y^2-1\right)-z^2\\
y'&=6y~\!(x^2+y^2-1)-5\sqrt{3}z^2\\
z'&=z~\!\left[6x^2-3x+y~\!\left(5\sqrt{3}+6y\right)\right]
~,}}
together with the 
constraint
\eq{\label{214}
1-x^2-y^2-z^2=-12~\!\frac{\lambda z^2}{m^2} e^{-2A-5\phi/2}
~.}
For $\lambda<0$, this forces the trajectories to lie within a unit ball in three-dimensional phase space.  
The invariant  plane is given by
\eq{\mathcal{P}=\left\{(x,y,z)\in\mathbb{R}^3~|~15x-\sqrt{3}y=10\right\}
~.}
The acceleration condition reads
\eq{\label{acccylind}
  x^2+y^2<\frac13~,
}
with,
\eq{\label{w221} 
w=-1+2\left(x^2+y^2\right)~.}
\subsubsection*{The critical points} 

There is a unique  critical point of the system \eqref{eomds2}  away from the $z=0$ plane, given by
\eq{
p_1=\frac{1}{38\sqrt{3}}~\!\left(25\sqrt{3},-5, 8\sqrt{2}\right)
~.}
The critical point $p_1$ lies on the invariant plane $\mathcal{P}$. The linearized system at $p_1$ 
has one real and two complex conjugate eigenvalues: $\tfrac{16}{19}\left(-4, -2\pm i\sqrt{2}\right)$. The eigenvectors corresponding to the real and complex eigenvalues 
are orthogonal and parallel to the invariant plane $\mathcal{P}$, respectively. It follows that for trajectories in $\mathcal{P}$, $p_1$ is a stable focus; for trajectories 
orthogonal to $\mathcal{P}$, $p_1$ is a stable node.

In addition 
we have a circle of fixed points on the equator $\mathcal{C}$ of the sphere $\mathcal{S}$,
\eq{
p_{\mathcal{C}}
\in 
\mathcal{C}:= \left\{(x,y,z)~|~
x^2+y^2=1~;~z=0\right\}
~,}
and an isolated fixed point,
\eq{p_2= \left(\tfrac23,0,0\right)
~,}
which lies on the $x$-axis. 
The  linearized system at $p_2$  has   eigenvalues $\tfrac{2}{3}(-5, -5,1)$. The eigenvectors corresponding to the negative eigenvalue are along the 
$x$ and $y$ directions, whereas the eigenvector corresponding to the positive eigenvalue is along the 
$z$ direction. It follows that for trajectories in the $z=0$ plane, $p_2$ is a stable singular node; for trajectories orthogonal to  the $z=0$ plane, $p_2$ is an unstable node. 

The critical points of the dynamical system correspond to solutions that can be given analytically: the point $p_1$ corresponds to the  solution \eqref{242b}. 
 The  point  $p_2$ corresponds to the critical solution with $\lambda<0$, and requires $m=0$. The critical points $p_{\mathcal{C}}$ correspond to the minimal 
 solution of Section \ref{sec:special}, and require $m,\lambda=0$.

\subsubsection*{The invariant surface $\mathcal{S}$} 

Restricting to trajectories on  $\mathcal{S}$, the system \eqref{eomds2} implies
\eq{
x'=6(x-\tfrac12)\left(x^2+y^2-1\right)~;~~~
y'=6\left(y+\tfrac{5}{2\sqrt{3}}\right)\left(x^2+y^2-1\right)
~.}
It follows that the projections of all trajectories on $\mathcal{S}$ to the $z=0$ plane are of the form
\eq{\label{line1}
a\left(x-\tfrac12\right)+b\left(y+\tfrac{5}{2\sqrt{3}}\right)=0~;~~~a,b\in\mathbb{R}
~,}
i.e. straight lines passing by the point $(x,y)=\left(\tfrac12, -\tfrac{5}{2\sqrt{3}}\right)$. These trajectories correspond precisely to 
the solutions of \eqref{193}, and require $\lambda=0$. More specifically, the slope of the line \eqref{line1} is related to the constants in \eqref{193} via
\eq{
\frac{a}{b}=  \frac{  64c_A+13c_B}{5\sqrt{3}c_B}
~.}

\subsubsection*{The $z=0$ invariant plane}

Restricting to trajectories on the $z=0$ plane, the system \eqref{eomds2} reduces to
\eq{
x'=6\left(x-\tfrac23\right)\left(x^2+y^2-1\right)~;~~~
y'=6y\left(x^2+y^2-1\right)
~.}
It follows that  all trajectories on  the $z=0$ plane are of the form
\eq{\label{line}
a\left(x-\tfrac23\right)+by=0~;~~~a,b\in\mathbb{R}
~,}
i.e. straight lines passing by the point $(x,y)=\left(\tfrac23, 0\right)$. These straight lines correspond precisely to the 
solutions of \eqref{186}, and require $m=0$. More specifically, the slope of the line \eqref{line} is related to the constants in \eqref{186} via
\eq{
\frac{a}{b}= -\frac{3\sqrt{3}}{20}\frac{c_\phi}{c_A}
~.}

\subsubsection*{The  invariant plane $\mathcal{P}$}

On $\mathcal{P}$, the system  \eqref{eomds2} reduces to
\eq{\spl{
x&=\tfrac23+\tfrac{1}{5\sqrt{3}}y\\
y'&=\tfrac{2}{75}~\!y~\!(-125 + 20 \sqrt{3} y + 228 y^2)-5\sqrt{3}z^2\\
z'&=\tfrac{2}{75}~\! z~\!(25 + 200 \sqrt{3} y + 228 y^2) 
~.}}
All trajectories of the reduced system  spiral into the stable focus $p_1$.

\subsection{Case study IV: $\varphi, \chi$}
\label{case4}

Let us finally consider the case where $\varphi, \chi \neq 0$. The system of equations is given by
\eq{\spl{
x'&=x^2-3 x z^2+y^2+\frac{11}{2}z^2-1 \\
y'&= \sqrt{3} \left(x^2+y^2+z^2-1\right)+\frac{1}{2} \left(\sqrt{3}-6 y\right) z^2 \\
z'&= -\frac{1}{2} z \left(9 x+\sqrt{3} y+6 z^2-6\right)
~,}\label{spc}}
together with the constraint,
\begin{equation}
    -3 z^2 c_{\chi }^2 e^{-2 A-6 B+\frac{3 \phi }{2}}=c_{\varphi }^2 \left(x^2+y^2+z^2-1\right)~.
\end{equation}
The invariant plane is defined by
\eq{
 3 x- \sqrt{3}  y=4
~,}
and the acceleration condition is simply given by
\eq{
z>\sqrt{\frac{2}{3}} 
~,}
so that a trajectory undergoes acceleration whenever it passes above the $z=\sqrt{\frac{2}{3}}$ plane.
The equation of state parameter reads
\begin{equation}
  w=1-2z^2 \,.
\end{equation}
This system admits no fixed points, apart from those lying on the equator $\mathcal{C}$. Nevertheless, it is worth being discussed, in view of its connection with analytic solutions. 

As argued previously, the particular case $\varphi =0$ and $\chi = 0$ corresponds to one of the fixed points on $\mathcal{C}$, and coincide with the minimal solution \eqref{range55simple}. When only $\varphi$ is turned on, the solutions are restricted to the boundary $\mathcal{S}$ of the phase space (and their projection on the $z=0$ plane are straight lines); if both fluxes $\varphi$ and $\chi$ are turned on, trajectories live generically in the bulk and connect two fixed points of $\mathcal{C}$. In the former case, the trajectory that maximizes the number of e-foldings $N$ is the one passing by the North pole: it obviously maximizes its length inside the acceleration region, but also turns out to maximize its ``time'' spent inside the region (which is not necessarily the same). Numerically, this maximal number of e-foldings can be determined to be $N_{\rm max} = 0.30408$. 

\begin{figure}[H]
\begin{center}
\includegraphics[width=0.9\textwidth]{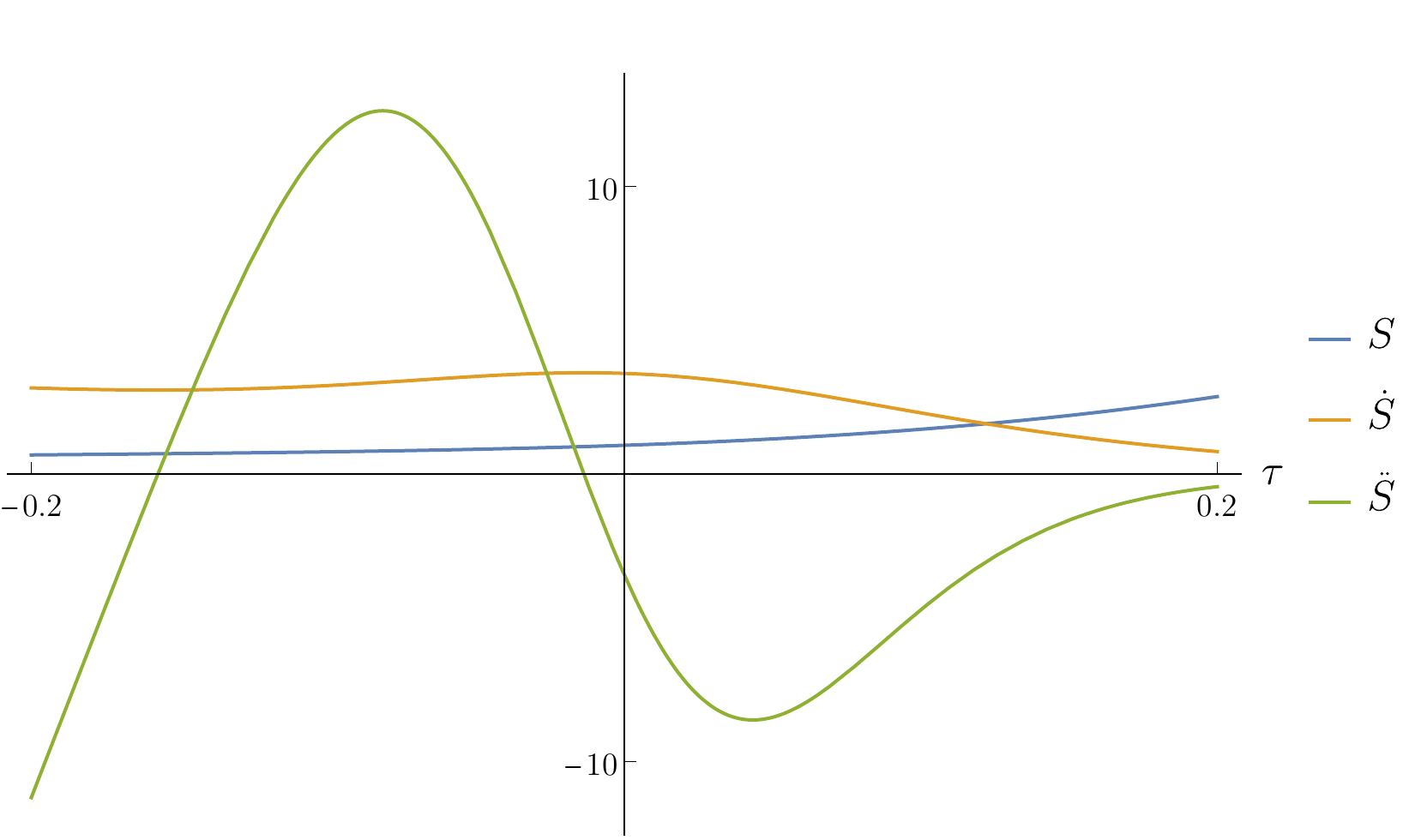}
\caption{Plot of the scale factor $S$ and its derivatives $\Sd, \Sdd$ as functions of $\tau$, with parameter $r=-\frac{9}{8}$. This special  value yields the extremal case with $N =N_{\rm max} = 0.30408$.}\label{fig:scaleder}
\end{center}
\end{figure}

\begin{figure}[H]
\begin{center}
\includegraphics[width=0.8\textwidth]{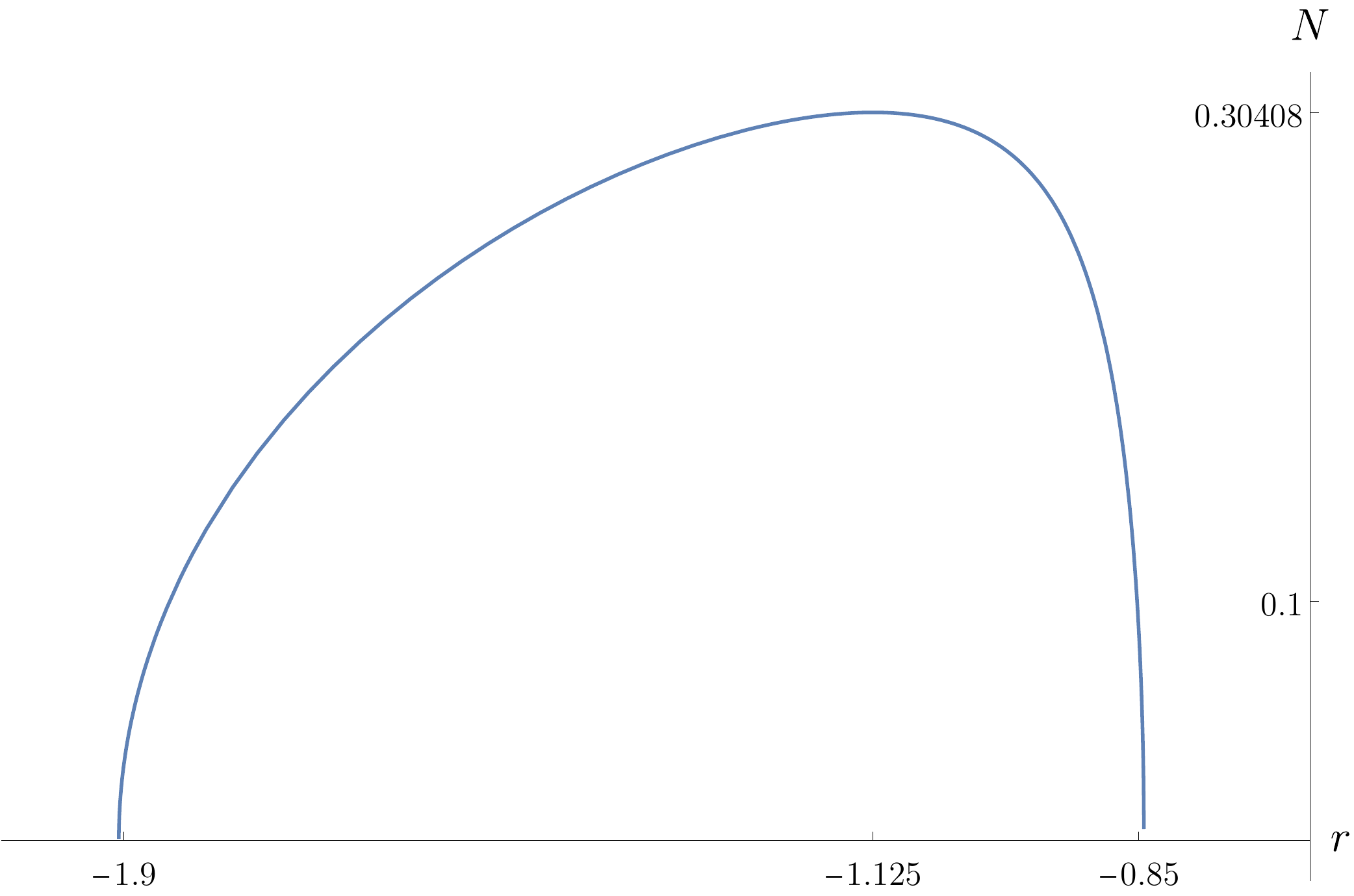}
\caption{Number of e-foldings $N$ as a function of the parameter $r$, for the one-flux solutions $\varphi \neq 0$. There is a restricted range permitting transient acceleration, and the maximal value is reached for $r_{\rm max} = -\frac{9}{8}$, giving $N_{\rm max} = 0.30408$.}\label{fig:Nr}
\end{center}
\end{figure}

Now, let us turn to the corresponding analytic solutions \eqref{CY4} found for $\varphi \neq 0, \chi = 0$. There exists a family of solutions parameterized by a real number\footnote{The solutions actually depend on two parameters, $r$ and $c_B$, but it turns out that the number of e-foldings only depends on the former, so we can restrict to $c_B=1$.} $r \leq -\frac{11}{16}$, with the following scale factor,
\begin{equation}
S(\tau)=e^{(4r+1) \tau} \cosh^{\frac{1}{4}} \left(2\sqrt{-33-48 \, r} \, \tau \right) ~.
\end{equation}
Having the explicit scale factor $S$ at hand allows us to compute the derivatives $\Sd, \Sdd$ and determine the values of $\tau$ for which accelerated expansion starts and stop (or in other words determine the values of $\tau$ for which one can satisfy both $\Sd > 0$ and $\Sdd > 0$), see Figure \ref{fig:scaleder}.
We can then readily compute the number of e-foldings $N(r)$ and extremize it with respect to $r$, see Figure \ref{fig:Nr}.
The maximal value is reached for $r_{\rm max} = -\frac{9}{8}$, which precisely gives the same $N_{\rm max} = 0.30408$ as above, and is of course in line with the bound $N_{\rm max} \leq 0.59980$ found in Section \ref{sec:oneflux} for one-flux compactifications.

\section{Conclusions}\label{sec:concl}

It has been known for some time that transient accelerated expansion is not difficult to achieve in flux compactifications arising from string-theory effective 10d supergravities. 
What our work is suggesting is that cosmologies featuring eternal or semi-eternal acceleration,  or 
a parametric control on the number of e-foldings 
are also generic! 
The necessary ingredients in all instances thereof seem to be a negative spatial 4d curvature (open universe), and a fixed point on the boundary 
of the acceleration region in the interior of the phase space. 

To our knowledge, this is also the first time where 
examples of spiraling cosmologies with an infinite number of cycles alternating between accelerated and decelerated expansion have been shown 
to arise from compactification of string-theory effective 10d supergravities. 
This so-called ``rollercoaster cosmology''  has been argued to be a potentially viable alternative for inflation \cite{DAmico:2020euu}.

In all examples exhibiting a parametric control on the number of e-foldings,  the acceleration vanishes asymptotically at future infinity, where spacetime approaches a Milne universe with angle defect.~Moreover, these are all captured by the 4d consistent truncation \eqref{consttr} with potential given in \eqref{16}, and require turning on  one of the parameters $\lambda$, $m$, $c_\varphi$, $c_0$  or $c_f$. Smooth accellerating cosmologies, without Big Bang singularities,  
are also possible, and they correspond to unique fine-tuned (therefore unstable) trajectories in phase space. Instead of a singularity at $T=0$, the spacetime approaches 
de Sitter space (in hyperbolic slicing).~This, however,  is not an asymptotic de Sitter, as $T=0$ is reached at finite proper time.~These solutions can  be geodesically completed  to $T<0$  in the past, as explained previously. 

As we have shown, 
the  techniques of the present paper allow us to straightforwardly translate the trajectories in phase space to the explicit form of the scale factor $S(T)$ of the corresponding 
FLRW  solution, as a function of cosmological time.~From this, the other cosmological observables can all be readily computed.

Our approach has been to work with the 10d equations of motion, not with a 4d effective potential. 
Nevertheless, in certain cases a graviton plus two-scalar (dilaton and warp factor) consistent truncation to a 4d theory $S_{4\d}$
is possible, such that  all cosmological solutions  
of $S_{4\d}$ lift to ten-dimensional solutions of IIA supergravity.  
This does not mean that solutions of $S_{4\d}$ uplift to 10d solutions with only dilaton and warp factor: 
all information about the flux 
(which is generically non-vanishing from the ten-dimensional point of view) enters the 4d potential of $S_{4\d}$ via certain  constants. 

The action  $S_{4\d}$ is in fact a consistent sub-truncation of the universal CY consistent truncation of  \cite{Terrisse:2019usq, Tsimpis:2020ysl}. 
Indeed, it was shown in those references that {\it the 4d effective theory of the universal sector of CY type II  compactifications is also a consistent truncation of 10d type II supergravity}.~We thus expect the consistent truncation $S_{4\d}$
of~\eqref{consttr} to be part of the 4d effective action, and thus subject to e.g.~the analysis and constraints presented recently in \cite{Calderon-Infante:2022nxb}.

A general stability analysis of the cosmological solutions presented here would require considering (small) perturbations in the space of all 10d fields. We shall leave this important point for future work.

The dynamical system techniques as applied in the present paper, limit us to solutions with a maximum of 
two species of flux. It would be desirable to overcome this limitation and explore 
richer solutions with all possible fluxes turned on, in each universal compactification class.

A lot of work has been done recently on classical de Sitter solutions using smeared orientifolds \cite{Andriot:2018ept, Andriot:2019wrs, Andriot:2020wpp, Andriot:2020vlg, Andriot:2022way, Andriot:2022yyj, Andriot:2022bnb}. 
Setting aside the still unresolved conceptual issues associated with the latter, including 
smeared orientifolds in our analysis would certainly enrich the structure of the phase space of 
the dynamical systems presented here, potentially leading to the appearance of    fixed points 
corresponding to de Sitter solutions. It would be interesting to explore this possibility further.

\section*{Acknowledgments}
We would like to thank Alfred Bovon for initial collaboration on some of the analytic solutions presented here \cite{alfred}, and David Andriot for useful discussions. 
D.T. would like to thank Jorge Russo and  Paul Townsend for useful email exchanges circa 2020.

\appendix

\section{Type IIA supergravity}\label{sec:iiasugra}

The ten-dimensional IIA action with Romans mass $m$ reads
\eq{\spl{\label{action3}S= \frac{1}{2\kappa_{10}^2}\int\d^{10}x\sqrt{{g}}\Big(
&-{R}+\frac12 (\partial\phi)^2+\frac{1}{2\cdot 2!}e^{3\phi/2}F^2\\
&+\frac{1}{2\cdot 3!}e^{-\phi}H^2+\frac{1}{2\cdot 4!}e^{\phi/2}G^2
+\frac{1}{2}m^2e^{5\phi/2}\Big) 
+S_\mathrm{CS}
~,
}}
where $S_\mathrm{CS}$ is the Chern-Simons term. 
The resulting equations of motion (EOM)  read as follows:
\begin{itemize}
\item Einstein EOM's,
\eq{\spl{\label{beomf2}
{R}_{MN}&=\frac{1}{2}\partial_M\phi\partial_N\phi+\frac{1}{16}m^2e^{5\phi/2}{g}_{MN}
+\frac{1}{4}e^{3\phi/2}\Big(  2F^2_{MN} -\frac{1}{8} {g}_{MN}  F^2 \Big)\\
&+\frac{1}{12}e^{-\phi}\Big(  3H^2_{MN} -\frac{1}{4} {g}_{MN}  H^2 \Big)
+\frac{1}{48}e^{\phi/2}\Big(   4G^2_{MN} -\frac{3}{8} {g}_{MN}  G^2 \Big)~,}}
where it is understood that $\Phi^2_{MN}:=\Phi_{MM_2\dots M_p}\Phi_N{}^{M_2\dots M_p}$, for any $p$-form $\Phi$. 

\item Dilaton EOM,
\eq{\spl{\label{beomf1}
0&=-{\nabla}^2\phi+\frac{3}{8}e^{3\phi/2}F^2-\frac{1}{12}e^{-\phi}H^2+\frac{1}{96}e^{\phi/2}G^2 +\frac{5}{4}m^2e^{5\phi/2}
~.
}}
\item Forms EOM's,
\eq{\spl{\label{beomf3}
0&=\d {\star}\big( e^{3\phi/2}F  )
+e^{\phi/2} H\wedge {\star}  G  \\
0 &= \d{\star} \big( e^{-\phi}H\big)
 +e^{\phi}F\wedge {\star}  G 
  -\frac{1}{2}G\wedge G
+ e^{3\phi/2}m {\star} F  
\\
0&=\d
{\star} 
\big(
e^{\phi/2}G\big)
-H\wedge G
~.
}}
Additionally the forms obey the  Bianchi identities,
\eq{\label{bi}
\d F= mH~;~~~\d H=0~;~~~\d G=H\wedge F
~.}
\end{itemize}

\section{Analytic solutions}
\label{app:sol}

\subsection{Compactification on Calabi-Yau manifolds}\label{sec:ctcy}

We will now search for analytic solutions of ten-dimensional IIA supergravity. We follow the ansatz of the consistent truncation of \cite{Terrisse:2019usq},  
but we work directly with the 10d equations of motion. 
We will furthermore restrict to the case without fermion condensates. 

The ansatz for the ten-dimensional two-form $F$, three-form $H$ and four-form $G$ reads
\eq{
F=\d\alpha~;~~~ H=\d\chi \wedge J+\d\beta~;~~~
G=\varphi\text{vol}_4+  J\wedge (\d\gamma - \alpha\wedge \d\chi)-\frac{1}{2}\d\xi\wedge\text{Im}\Omega
-\frac{1}{2}\d\xi'\wedge\text{Re}\Omega
~,}
where   $\varphi$, $\chi$, $\xi$, $\xi'$ are 4d scalars, $\alpha$, $\gamma$ are 4d one-forms, and $\beta$ is a 4d two-form.  We shall also introduce the 
4d three-form $h:=\d\beta$. 
The ansatz for the ten-dimensional metric {\it in 10d Einstein frame} reads
\eq{\d s^2_{10} =e^{2A(x)}\left(e^{2B(x)} g_{\mu\nu}\d x^{\mu}\d x^{\nu}+g_{mn}\d y^m\d y^n 
\right)~,
}
where the scalars $A$, $B$ only depend on the four-dimensional coordinates $x^\mu$, while $y^m$ are the CY coordinates. 
The equations of motion  are as follows,
\eq{\spl{\label{et1}
0&=e^{-8A-2B}\nabla^{\mu}\left(
e^{8A+2B}\partial_{\mu}A
\right)
-\frac{1}{32} e^{3\phi/2-2A-2B} \d\alpha^2
+\frac18e^{-\phi-4A}(\partial\chi)^2
-\frac{1}{48}e^{-\phi-4A-4B}h^2\\
&-\frac{1}{32}e^{\phi/2-6A-2B}(\d\gamma - \alpha\wedge \d\chi)^2
+\frac{1}{16}e^{\phi/2-6A}\Big[(\partial\xi)^2
+(\partial\xi')^2\Big]+\frac{3}{16} 
e^{\phi/2-6A-6B}\varphi^2
~,}}
coming from 
the internal $(m,n)$-components of the ten-dimensional Einstein 
equations. 
The external $(\mu,\nu)$-components give rise to
\eq{\spl{\label{et2}
R^{(4)}_{\mu\nu}&=
g_{\mu\nu}\left(\nabla^{2}A+\nabla^{2} B+
8(\partial A)^2+2(\partial B)^2+10\partial A\cdot \partial B\right)
\\
&-8\partial_{\mu}A\partial_{\nu}A-2\partial_{\mu}B\partial_{\nu}B
-16\partial_{(\mu}A\partial_{\nu)}B+8\nabla_{\mu}\partial_{\nu}A+2\nabla_{\mu}\partial_{\nu}B\\
&+\frac32 e^{-\phi-4A} \partial_{\mu}\chi\partial_{\nu}\chi
+\frac12 e^{3\phi/2 -2A-2B} \d\alpha^2_{\mu\nu}
+\frac14 e^{\phi-4A-4B} h^2_{\mu\nu}
+\frac12 \partial_{\mu}\phi\partial_{\nu}\phi\\
&+\frac{1}{2} e^{\phi/2-6A}(\partial_{\mu}\xi\partial_{\nu}\xi+\partial_{\mu}\xi'\partial_{\nu}\xi')
+\frac{3}{2} e^{\phi/2-6A-2B}(\d\gamma - \alpha\wedge \d\chi)^2_{\mu\nu}
 \\
&+\frac{1}{16} g_{\mu\nu}\Big(  
- \frac{1}{2} e^{3\phi/2-2A-2B} \d\alpha^2
-\frac{1}{3}e^{\phi-4A-4B}h^2
-3e^{\phi/2-6A}\Big[(\partial\xi)^2
+(\partial\xi')^2
\Big]\\
&-6e^{-\phi-4A}(\partial\chi)^2
-5e^{\phi/2-6A-6B}\varphi^2 
 -\frac{9}{2}e^{\phi/2-6A-2B}(\d\gamma - \alpha\wedge \d\chi)^2
\Big)
~,}}
while the mixed $(\mu,m)$-components are automatically satisfied. 
The dilaton equation reduces to
\eq{\spl{\label{et3}
0&=e^{-10A-4B}\nabla^{\mu}\left(
e^{8A+2B}\partial_{\mu}\phi
\right)
-\frac{1}{4}e^{\phi/2-8A-2B}\Big[(\partial\xi)^2
+(\partial\xi')^2
\Big]
-\frac{3}{8}e^{3\phi/2-4A-4B} \d\alpha^2
\\
&+\frac32e^{-\phi-6A-2B}(\partial\chi)^2
+\frac{1}{12}e^{-\phi-6A-6B}h^2
+\frac{1}{4}
e^{\phi/2-8A-8B}\varphi^2 -\frac{3}{8}e^{\phi/2-8A-4B}(\d\gamma - \alpha\wedge \d\chi)^2
~.}}
The $F$-form equation of motion 
reduces to the condition
\eq{\label{7}
\d(e^{3\phi/2+6A} \star_4 \d\alpha) = \varphi e^{\phi/2+2A-4B}\d\beta - 3e^{\phi/2+2A}\d\chi\wedge\star_4(\d\gamma - \alpha\wedge \d\chi)
~.}
The $H$-form equation gives rise to two equations,
\eq{\spl{\label{hfeom}
\d\left(
e^{-\phi+4A+2B}\star_4\d\chi
\right)  &=   (\d\gamma - \alpha\wedge \d\chi)\wedge(\d\gamma - \alpha\wedge \d\chi)-e^{\phi/2+2A} \d\alpha\wedge\star_4(\d\gamma - \alpha\wedge \d\chi)
~,}}
and, 
\eq{\label{seqh}
\d\left(
e^{-\phi+4A-2B}\star_4 \d\beta\right)
=  - \d\xi\wedge\d\xi' + e^{\phi/2+2A-4B} \varphi \d\alpha
~.}
The $G$-form equation  reduces to
\eq{\spl{
\label{gfeom1}
\d\left(e^{\phi/2+2A+2B}\star_4\d\xi\right)  &=  h\wedge\d\xi'\\
\d\left(e^{\phi/2+2A+2B}\star_4\d\xi'\right) &= -h\wedge\d\xi\\
\d\left(e^{\phi/2+2A} \star_4(\d\gamma - \alpha\wedge \d\chi)\right) &=  2\d\chi\wedge\d\gamma 
~,}}
together with
\eq{\label{gfeom2}
0=\d\left(
 \varphi e^{\phi/2+2A-4B} 
\right)
~,}
which can be integrated to
\eq{ 
\label{12}
\varphi= e^{-\phi/2-2A+4B} c_{\varphi}~,}
for some constant $c_{\varphi}$.

\subsection*{Cosmological ansatz}\label{sec:cosmcy}

We will now make a cosmological ansatz for all  fields, i.e.~one that is (in general) only  invariant under the isometries of the 3d spatial part of the metric. 
We  assume that all scalars are functions of the time coordinate $t$ alone. The one-forms are  assumed to be of the form
\eq{
\alpha=\alpha(t)\d t~;~~~\gamma=\gamma(t)\d t
~,}
for some scalars $\alpha$, $\gamma$ which may be time-dependent in general. 
With some abuse of notation, we have denoted by the same letter the one-forms and the corresponding scalars. 
In the following all equations of motion will be expressed exclusively in terms of the scalars, so hopefully no confusion will occur. 
The three-form $h$ is assumed to  be of the form
\eq{h=c_h\sqrt{\gamma}~\d x^1\wedge\d x^2\wedge\d x^3~,}
where $c_h$ is a constant. 
The unwarped 4d metric~\eqref{2}  is assumed to be of the form of eq.~\eqref{metricansatz}, while the 
 4d Einstein frame metric is thus given by \eqref{efm8}.


Substituting the ansatz above into the form equations of motion \eqref{7}-\eqref{12}, we obtain the following system: eq. \eqref{7} reduces to
\eq{\label{7b}
c_\varphi c_h =0~.
}
Eq. \eqref{hfeom} reduces to
\eq{ \label{hfeomb}
 \chi
  =   c_\chi    
  \int^t \d t' e^{\phi(t')-4A(t')-2B(t')}
   +d_\chi
~,}
where $c_\chi$, $d_\chi$  are constants. 
Eq.~\eqref{seqh} is automatically satisfied. 
Eq.~\eqref{gfeom1} reduces to
\eq{\spl{
\label{gfeom1b}
\d_t\left(e^{\phi/2+2A+2B} \d_t\xi\right)  &=  c_h\d_t\xi'\\
\d_t\left(e^{\phi/2+2A+2B}\d_t\xi'\right) &= -c_h\d_t\xi 
~,}}
where $\d_t$ denotes the derivative with respect to $t$. 
Note  that if eq.~\eqref{gfeom1b} is satisfied for some $\xi, \xi'\neq0$, this  implies
\eq{\label{xi42}
 (\d_t\xi)^2+   (\d_t\xi')^2 =2c_{\xi\xi'}^2 e^{-\phi-4A-4B}
~,}
for some arbitrary real constant $c_{\xi\xi'}$. 

The  system \eqref{gfeom1b} can be integrated to give
\eq{\spl{
\label{gfeom1intb}
e^{\phi/2+2A+2B} \d_t\xi   &=  c_h\xi' +c_{\xi}\\
 e^{\phi/2+2A+2B}\d_t\xi' &= -c_h\xi +c_{\xi'}
~,}}
for some constants $c_{\xi}$, $c_{\xi'}$. Let us define a new time variable $\nu$ by
\eq{\nu(t):=\int^t \d t' e^{-[\phi(t')/2+2A(t')+2B(t')]}
~.
}
If $c_h\neq 0$, the solution to \eqref{gfeom1intb} reads
\eq{
\label{gfeom1solb}
 \xi   =  \sin(c_h\nu+d_{\xi}) +\frac{c_{\xi'}}{c_h}~;~~~
\xi'   =  \cos(c_h\nu+d_{\xi}) -\frac{c_{\xi}}{c_h}
~,}
where $d_{\xi}$ is an arbitrary constant.  If $c_h= 0$, the solution to \eqref{gfeom1intb} reads instead
\eq{
 \xi   =  c_{\xi} \nu+d_{\xi}~;~~~
\xi'   =  c_{\xi'} \nu+d_{\xi'}
~,}
where $d_{\xi}$, $d_{\xi'}$ are are arbitrary constants.  
This then concludes the solution of all  equations of motion for the forms.

\subsubsection*{Adapted time coordinate}

In the following we will use the time variable $\tau$ defined in \eqref{taudef111}. 
The internal Einstein and dilaton equations,~\eqref{et1}, \eqref{et3} reduce to
\eq{\spl{\label{e1}
\d_\tau^2 A&=\tfrac{3}{16}c^2_{\varphi} e^{-\phi/2+6A+6B} -\tfrac{1}{8}c^2_{h} e^{-\phi+12A}
-\tfrac{1}{8}c^2_{\chi} e^{\phi+4A}-\tfrac{1}{8}c^2_{\xi\xi'} e^{-\phi/2+6A}\\
 \d_\tau^2 \phi&= \tfrac{1}{4}c^2_{\varphi} e^{-\phi/2+6A+6B}+
 \tfrac{1}{2}c^2_{h} e^{-\phi+12A}
 -\tfrac{3}{2}c^2_{\chi} e^{\phi+4A}+\tfrac{1}{2}c^2_{\xi\xi'} e^{-\phi/2+6A}
~.}}
The external Einstein equations \eqref{et2} 
reduce to
\eq{\spl{\label{e2}
-2ke^{16A+4B} -\tfrac{1}{2}c^2_{\varphi} e^{-\phi/2+6A+6B}+\tfrac{1}{2}c^2_{h} e^{-\phi+12A}+ \tfrac{1}{2}c^2_{\chi} e^{\phi+4A}+ \tfrac{1}{2}c^2_{\xi\xi'} e^{-\phi/2+6A}
&= \d_\tau^2B
\\
-12ke^{16A+4B} +c^2_{\varphi} e^{-\phi/2+6A+6B}+c^2_{h} e^{-\phi+12A}+3c^2_{\chi} e^{\phi+4A}+2c^2_{\xi\xi'} e^{-\phi/2+6A}&=
144(\d_\tau A)^2+12(\d_\tau B)^2\\
&+96\d_\tau A\d_\tau B- (\d_\tau\phi)^2 
~.}}
The second line above is a {constraint}, consistently propagated by the remaining equations of motion. Indeed the 
$\tau$-derivative of this equation is automatically satisfied by virtue of \eqref{e1} and the first equation in \eqref{e2}.



\subsubsection*{IIB solution}

Although the main focus of the present paper is on IIA supergravity, let us note that the minimal solution can be easily generalized to include 
a non-trivial axion $C$. The   ten-dimensional  Lagrangian, 
\eq{
\mathcal{L}=R-\tfrac12(\partial\phi)^2-\tfrac12e^{2\phi}(\partial C)^2
~,}
together with the ansatz \eqref{2} for the metric, now lead to the following 
equations of motion: for the axio-dilaton we have
\eq{
\d_\tau C=c_a e^{-2\phi}
~~~
\d^2_\tau\phi=c_a^2 e^{-2\phi}
~,}
for some real constant $c_a$. The Einstein equations reduce to
\eq{
\d_\tau A=c_A
~~~
\d_\tau B=c_B
~,}
for some real constants $c_A$, $c_B$, together with the constraint
\eq{\label{constre60}
12c_A^2+c_B^2+8c_A c_B-\tfrac{1}{12}(\d_\tau\phi)^2-\tfrac{1}{12}c_a^2e^{-2\phi}=0
~.}
The solution for the axio-dilaton reads
\eq{
 C=\frac{c_\phi}{c_a} ~\!\text{tanh}(c_\phi\tau+d_\phi)+d_a~;
~~~
 \phi= 
 \tfrac{1}{2}
\ln \frac{c_a^2}{c_\phi^2}    +\text{ln}\big[\text{cosh}(c_\phi \tau+d_\phi)  \big] 
~,}
for some constants $d_a$, $d_\phi$. 
Taking the above into account,  \eqref{constre60} reduces to the constraint \eqref{54constra}. It readily follows that, apart from the axio-dilaton,  
the IIB solution is identical to the minimal solution.

\subsection*{\boldmath$k\neq0$\unboldmath}\label{sec:4.2}

As before, we will take
\eq{
\varphi=h=0~;~~~\chi=d_\chi~;~~~\xi=d_\xi~;~~~\xi'=d_{\xi'}
~,}
which still solves the form equations \eqref{7}-\eqref{12} in the case of $k\neq0$.
Let us consider the remaining equations of motion. 
The internal Einstein and 
the dilaton equation give, cf.~\eqref{e1},
\eq{\label{40a}
A=c_A\tau+d_A~;~~~\phi=c_\phi\tau+d_\phi
~,
}
for some constants $d_A$, $d_\phi$, 
exactly as for the minimal solution. 
From the first equation in~\eqref{e2}, we obtain the solution for $B$, 
\eq{\label{60bht}
B=-4(c_A\tau+d_A) +\frac14 f(\tau)
~,}
where,
\eq{\label{42a}
f(\tau)=\left\{ 
\begin{array}{cll}  \ln  \big[  \tfrac{ c_B^2}{4k} ~\!\text{sech}^2({c_B}\tau+d_B)  \big]& ~,~~&k>0\\
\\
\ln  \big[  \tfrac{ c_B^2}{4|k|}    \text{csch}^2({c_B}\tau+d_B)  \big]  &~,~~ & k<0
\end{array}\right.
~,}
for some constants $c_B$, $d_B$. Moreover,  the second equation in~\eqref{e2} (the  constraint)  reduces to an algebraic condition,
\eq{\spl{\label{43a} c_\phi^2+48c_A^2&=3c_B^2 ~,}}
for   $k$ of either sign. In particular this implies, $r\leq|\tfrac14|$, cf.~\eqref{rdef}. In the following we consider the two cases in \eqref{42a} in more detail.

\subsubsection*{Type I solution: closed universe}\label{sec:kp}
 
Let us set  $k>0$. The 4d Einstein-frame  metric reads
\eq{\spl{\label{full61}
\d s^2_{4E} 
 &= 
 \text{sech}^3({c_B}\tau) 
\Big( - \d \tau^2+\frac{4k}{c_B^2}\text{cosh}^2({c_B}\tau)\d\Omega_k^2\Big)
~,
}}
where we have  set $d_A$, $d_B=0$ for simplicity.  
Since the sign of $c_B$ can be absorbed in the definition of $\tau$, we may suppose $c_B\geq 0$ without loss of generality; the inequality is saturated in case $c_A$, $c_B$, $c_\phi$ 
all vanish. 
In terms of coordinates $T_\pm\propto e^{\mp\frac{3}{2} c_B\tau }$ the metric asymptotically   
takes the form of \eqref{intro1}, 
where $a_\pm= \frac{1}{3}$. 
For $c_B\geq 0$, $\tau\rightarrow\pm\infty$ corresponds to $T_\pm\rightarrow0$, where a singularity is reached at finite proper time.

\subsubsection*{Type II solution: open universe}\label{sec:interpolating2k}\label{sec:kneg}

Let us now set  $k<0$. 
The 4d Einstein-frame   metric reads
\eq{\spl{
\d s^2_{4E} 
 &= 
 \text{csch}^3({c_B}\tau) 
\Big( - \d \tau^2+\frac{4|k|}{c_B^2}~\!\text{sinh}^2({c_B}\tau)\d\Omega_k^2\Big)
~,
}\label{CY3}}
where we have   set $d_A$, $d_B=0$ for simplicity.  
Since the sign of $c_B$ can be absorbed in the definition of $\tau$, we may suppose $c_B\geq 0$ without loss of generality.

The $\tau\rightarrow\pm\infty$ asymptotics of the warp factors are exactly the same as for $k>0$. For $\tau\rightarrow 0$, on the other hand, 
the function $f$ tends to $f\rightarrow -\ln(4|k|\tau^2)$, while the constraint imposes $c_A$, $c_\phi=0$. 
Hence  in the $\tau\rightarrow 0$ limit the solution reads
\eq{\label{67milne}
A=d_A~;~~~B=-4d_A-\tfrac14\ln(4|k|\tau^2)~;~~~
\phi=d_\phi
~,}
Note that comoving geodesics reach  $\tau=0$ at infinite proper time. 
The metric becomes asymptotically that of a regular Milne universe,\footnote{\label{foot4} Recall that the spatial 3d part of the metric 
is locally isometric to a hyperbolic   space of 
scalar  curvature $6k$, cf.~\eqref{spatialmetric}. An explicit parametrization is given by,
\eq{
\d\Omega_k^2=\frac{1}{|k|}
\big(
\d\chi^2+\text{sinh}^2\chi~\!\d\sigma^2
\big)
~,\nn}
with $\d\sigma^2$ the line element of the two-sphere.
}
\eq{\label{milnek}
\d s^2 =  -\d T^2+|k|~\!T^2\d\Omega_k^2
~,}
where we have defined 
$T=2 (  c_B^3\tau)^{-\frac12}$.  The warp factor of the internal space and the dilaton are constant.

\subsection*{ \boldmath$\varphi\neq0$\unboldmath} \label{sec:cosmmink1}

Let us now take
\eq{
\label{CY4}
\varphi=e^{-\phi/2-2A+4B} c_{\varphi}~;~~~h=0~;~~~\chi=d_\chi~;~~~\xi=d_\xi~;~~~\xi'=d_{\xi'}
~,}
which still solves the form equations \eqref{7}-\eqref{12} in the case of $\varphi\neq0$.
Let us consider the remaining equations of motion. 
Eqs.~\eqref{e1}  reduce to
\eq{\label{et3pbb}
\d_\tau^2 A=\tfrac{3}{16}c^2_{\varphi} e^{-\phi/2+6A+6B}~;~~~\d_\tau^2 \phi=\tfrac{1}{4}c^2_{\varphi} e^{-\phi/2+6A+6B}
~,}
while \eqref{e2}  
reduce to the following two  equations,
\eq{\spl{\label{red5pbb}
-\tfrac{1}{2}c^2_{\varphi} e^{-\phi/2+6A+6B}
&= \d_\tau^2B
\\
c^2_{\varphi} e^{-\phi/2+6A+6B}&=
144(\d_\tau A)^2+12(\d_\tau B)^2+96\d_\tau A\d_\tau B- (\d_\tau\phi)^2 
~.}}
To obtain an analytic solution we proceed as follows. We define $f:=-\phi/2+6A+6B$, which by virtue of the equations of motion \eqref{et3pbb},\eqref{red5pbb} obeys,
$\d_\tau^2f=-2c^2_\varphi e^f$. The solution to this  equation reads
\eq{\label{37}
f=\ln\tfrac{c_\phi^2}{c_\varphi^2}-2\ln  \big[\text{cosh}({c_\phi}\tau+d_\phi)  \big]
~,}
for some constants $c_\phi$, $d_\phi$. Plugging back into \eqref{et3pbb}, and the first line of \eqref{red5pbb}, we obtain the solution
\eq{\spl{\label{sdjkdflk}
A&=c_A\tau+d_A-\tfrac{3}{32}f\\
B&=c_B\tau+d_B+\tfrac{1}{4}f\\
\phi&=12(c_A+c_B)\tau+12(d_A+d_B)-\tfrac{1}{8}f~,
}}
for some constants $c_A$, $c_B$,  $d_A$, $d_B$, and $f$ as  given in \eqref{37}. Moreover, the second line of \eqref{red5pbb} imposes the constraint
\eq{\label{cccc}
c_\phi^2=-12c_B(16c_A+11c_B)
~,}
which implies $r \leq -\tfrac{11}{16}$. 
The 4d Einstein-frame metric reads
\eq{\spl{
\d s^2_{4E}
 &= 
e^{ (24c_A+6c_B)\tau  }\text{cosh}^{\frac32}(c_\phi\tau)  \left(-\d \tau^2
+e^{ -(16c_A+4c_B)\tau  }\text{sech}(c_\phi\tau)  \d\vec{x}^2\right)
~,
}}
where we have set $d_A$, $d_B$, $d_\phi=0$ for simplicity.

Let us assume $c_\phi\geq0$, without loss of generality, since the sign of $c_\phi$ can be absorbed in the definition of $\tau$.  In terms of coordinates 
$T_\pm\propto e^{(12c_A+3c_B\pm\frac{3}{4}c_\phi)\tau}$, the metric asymptotically 
takes the form of \eqref{intro1}, where $a_\pm = \frac{1}{3}$.

\begin{figure}[H]
\begin{center}
\includegraphics[width=0.8\textwidth]{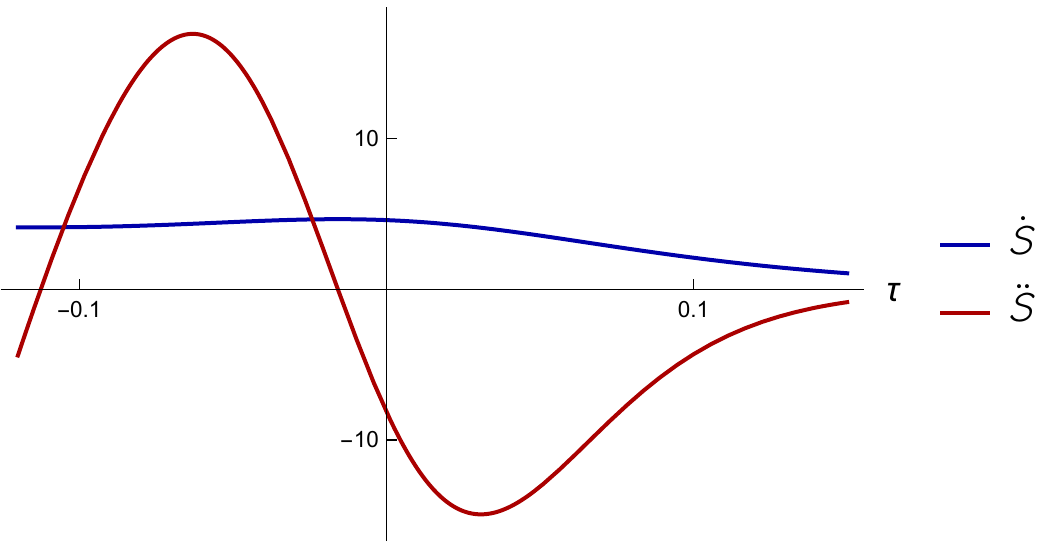}
\caption{Plot of $\dot{S}(T), \ddot{S}(T)$ as a function of  $\tau$, for $c_B=-1$, $r=-1.4$. }\label{fig:acceleration1}
\end{center}
\end{figure}

If $c_B>0$, $\tau=-\infty$ corresponds to $T_-=\infty$ while $\tau=\infty$ corresponds to $T_+=0$, where a singularity is reached at finite proper time.   The situation 
is reversed for $c_B<0$: $\tau=-\infty$ corresponds to $T_-=0$, where a singularity is reached at finite proper time, while $\tau=\infty$ corresponds to $T_+=\infty$. 
For a certain range of parameters, this model exhibits transient accelerated expansion, where $\dot{S}(T), \ddot{S}(T)>0$, cf. Figure \ref{fig:acceleration1}.

\subsubsection*{Critical solution: $\varphi\neq0$, $k<0$}\label{sec:crit}

Let 
us now consider $k\neq 0$. The equations of motion \eqref{e1}, \eqref{e2} admit the following solution,
\eq{\label{242bb}
A=-\tfrac{3}{56}\ln |\tau|+d_A+3d_\phi~;~~~B=-\tfrac{2}{7}\ln|\tau|-4d_A+16d_\phi~;~~~\phi=-\tfrac{1}{14}\ln|\tau|-36d_A+4d_\phi
~,
}
for an arbitrary constant $d_A$, and $k$, $d_\phi$ subject to the conditions
\eq{\label{243bb}
 d_\phi=-\tfrac{1}{112}\ln(\tfrac{7}{2}c_\varphi^2)~;~~~
k=-\tfrac34 c_\varphi^2 <0
~.}
The 4d metric is that 
of a {\it singular} Milne universe {\it with angular defect}, cf. Footnote~\ref{foot4},
\eq{
\d s^2=-\d T^2+\tfrac{7}{6} |k|~\! T^{2}\d\Omega_k^2
~,}
where we set  $T\propto|\tau|^{-\frac{1}{2}}$ and $\d\Omega_k^2$ is the line element of a locally hyperbolic three-space.  
The warp factor of the internal space and the dilaton scale as $e^A\propto T^{3/28}$, $e^\phi\propto T^{1/7}$ respectively.

\subsubsection*{Type II solution: $\varphi\neq0$, $k<0$}

We will now construct a solution interpolating between  future or past infinity $\tau\rightarrow\pm \infty$, and the above critical solution  in the $\tau\rightarrow0$ limit. 
Let us use an ansatz of the form
\eq{
A=c_A\tau+d_A+f(\tau)~;~~~
B=c_B\tau+d_B+e_Bf(\tau)~;~~~
\phi=c_\phi\tau+d_\phi+e_\phi f(\tau) 
~,}
for some (non-linear) function $f$ and constants $c$, $d$, $e$. Moreover we will require that the equations of motion \eqref{e1} and the first line of \eqref{e2} reduce to 
a single differential equation for $f$. 
These requirements impose the following condition on the spatial curvature,
\eq{
k=-\tfrac34 c_\varphi^2~,
}
(open universe) and, after some redefinitions of the constants, lead to the solution,
\eq{\spl{\label{79phi}
A&=c_A\tau+d_A +3g(\tau)\\
B&=-4(c_A\tau+d_A) +16g(\tau)\\
\phi&=-36(c_A\tau+d_A) +4g(\tau) 
~,}}
where
\eq{\label{106}
g=\tfrac{1}{112}
\ln \tfrac{2c_\phi^2}{7c_\varphi^2}   - \tfrac{1}{112}\text{ln}\big[\text{sinh}^2(c_\phi \tau+d_\phi)  \big] 
~.}
The second line of the constraint then reduces to
\eq{
c_\phi=\pm 28\sqrt{\tfrac{3}{5} }~\! c_A
~.}
The 4d Einstein metric reads
\eq{\spl{
\d s^2_{4E}
 &= 
 \text{csch}^{ 3}(c_\phi\tau)  \Big(-\d \tau^2
+\frac{7 c_\varphi^2}{2 c_\phi^2} ~\!\text{sinh}^2(c_\phi\tau)  \d\Omega_k^2\Big)
~,
}}
where we have set $d_A$,  $d_\phi=0$ for simplicity. 
The metric asymptotically 
takes the form \eqref{intro1},
 where $a_\pm = \frac{1 }{3  }$. 
On the other hand, for $\tau\rightarrow 0$  we have
\eq{\label{108}
   g\rightarrow  -\tfrac{1}{112} \ln (\tfrac{7}{ 2} c_\varphi^2  )  -\tfrac{1}{56} \ln |\tau|
~.
}
In this limit  the solution \eqref{79phi} thus takes the form given in eqs.~\eqref{242bb}, \eqref{243bb}. 
In other words, for $\tau\rightarrow 0$ the solution asymptotes the critical solution. 
Comoving geodesics reach  $\tau=0$ at infinite proper time.

\subsection*{ \boldmath$\chi\neq0$\unboldmath}\label{sec:chineq0}

Let us now take
\eq{
\d_t\chi= e^{\phi-4A-2B} c_\chi~;~~~\xi=d_\xi~;~~~\xi'=d_{\xi'}  ~;~~~
\varphi=0~;~~~h=0
~,}
which still solves the form equations \eqref{7}-\eqref{12} in the case of $\chi\neq0$.
Let us consider the remaining equations of motion. 
Equations \eqref{e1} and the first line of \eqref{e2} are solved by
\eq{\spl{
A&=c_A\tau+d_A+\tfrac{1}{16}f \\
B&=c_B\tau+d_B-\tfrac{1}{4}f\\
\phi&=-4(c_A\tau+d_A)+\tfrac{3}{4}f~,
}\label{CYchi}}
with
\eq{
f:=\ln\tfrac{c_\phi^2}{c_\chi^2}-2\ln  \big[\text{cosh}({c_\phi}\tau+d_\phi)  \big]
~,}
while the  second line of \eqref{e2} reduces to
\eq{\label{51}
3c_\phi^2=128c_A^2+96c_Ac_B+12c_B^2
~.}
The constraint 
imposes in particular,  $r\leq -\tfrac{\sqrt{3}}{8}(1+\sqrt{3})$ or   $r\geq \tfrac{\sqrt{3}}{8}(1-\sqrt{3})$.

The 4d  Einstein metric reads
\eq{
\d s^2_{4E} 
 = 
-e^{ (24c_A+6c_B)\tau  }  \d \tau^2
+e^{ (8c_A+2c_B)\tau  }  \d\vec{x}^2 
~,
}
where we have set $d_A$, $d_B$, and $d_{\phi}=0$ for simplicity.  
In terms of the coordinate  
$T \propto  e^{ (12c_A + 3c_B)\tau}$,   $T\in[0, \infty)$, the metric  
takes the form \eqref{intro1}, 
 with $a_\pm =  \frac13$.

\subsubsection*{Type I solution :  $k> 0$ }

Let us now consider $k\neq 0$. 
Equations \eqref{e1} can be solved to give 
\eq{\spl{
A&=c_A\tau+d_A +\tfrac{1}{16}g(\tau)\\
\phi&=-4(c_A\tau+d_A) + \tfrac{3}{4}g(\tau) 
~,}\label{CYchi2}}
for some constants $c_A$, $d_A$, where
\eq{\label{128}
g=\ln\tfrac{c_\phi^2}{c_\chi^2}-2\ln  \big[\text{cosh}({c_\phi}\tau+d_\phi)  \big]
~.}
Taking the above into account, the first of \eqref{e2} can be  solved for $B$, 
\eq{{}
B=-4A+\tfrac14 f ~,}
where
\eq{\label{130}
f=\left\{ 
\begin{array}{cll}  \ln  \big[\frac{c_B^2}{4k} ~\!\text{sech}^2({c_B}\tau+d_B)  \big]& ~,~~&k>0\\
\\
\ln  \big[\frac{c_B^2}{4|k|}\text{csch}^2({c_B}\tau+d_B)  \big]  &~,~~ & k<0
\end{array}\right.
~,}
for some constants $c_B$, $d_B$. Plugging the solution into the second line of \eqref{e2} we obtain the constraint
\eq{\label{cccccc}
c_B^2=\tfrac{64}{3}c_A^2+c_\phi^2
~,}
for either sign of $k$. 

Let us first consider the case of a closed universe ($k>0$). 
The 4d  metric reads, with $d_A$, $d_B$, $d_{\phi} = 0 $ for simplicity,
\eq{\spl{
\d s^2_{4} 
 &=  \text{sech}^{3} ({c_B}\tau)  
 \Big(
 - \d \tau^2  +\frac{4k}{c_B^2} \text{cosh}^{2}({c_B}\tau) ~\!\d\Omega_k^2
 \Big)
~.
}}

\subsubsection*{Type II and critical solutions: $k<0$}\label{sec:cfegsec}

Let us now set  $k<0$. The 4d Einstein metric  reads
\eq{\spl{
\d s^2_{4E} 
 &=  \text{csch}^{3} ({c_B}\tau)  
 \Big(
 - \d \tau^2 +\frac{4|k|}{c_B^2}  \text{sinh}^{2}({c_B}\tau) ~\!\d\Omega_k^2
 \Big)
~.
}\label{CYchi3}}

\subsection*{ \boldmath$\xi, \xi' \neq0$\unboldmath}

Let us assume that eq.~\eqref{xi42} is satisfied for some  arbitrary real constant $c_{\xi\xi'}$, and 
let us take
\eq{
\chi= d_\chi~;~~~
\varphi=0~;~~~h=0
~.}
This ansatz thus solves the form equations \eqref{7}-\eqref{12} in the case of $\xi, \xi'\neq0$. 
Let us consider the remaining equations of motion. Equations \eqref{e1} and the first of \eqref{e2} are solved by
\eq{\spl{
A&=c_A\tau+d_A+\tfrac{1}{8}f \\
B&=c_B\tau+d_B-\tfrac{1}{2}f\\
\phi&=12(c_A\tau+d_A)-\tfrac{1}{2}f~,
}\label{CYxi1}}
with
\eq{
f:=\ln\tfrac{2c_\phi^2}{c^2_{\xi\xi'}}-2\ln  \big[\text{cosh}({c_\phi}\tau+d_{\xi\xi'})  \big]
~,}
while the  second line of \eqref{e2} reduces to
\eq{\label{51x}
c_{\phi}^2=24c_Ac_B+3c_B^2
~,}
which implies in particular,  $r\geq -\tfrac{1}{8}$. 
The 4d  Einstein metric  reads
\eq{
\d s^2_{4E} 
 = 
-e^{ (24c_A+6c_B)\tau  }  \d \tau^2
+e^{ (8c_A+2c_B)\tau  }  \d\vec{x}^2 
~.
}

\subsubsection*{Type I solution:  $k>0$}

Let us now take $k\neq0$. 
Equations \eqref{e1} can be solved to give  
\eq{\spl{
A&=c_A\tau+d_A +\tfrac{1}{8}g(\tau)\\
\phi&=12(c_A\tau+d_A) - \tfrac{1}{2}g(\tau) 
~,}\label{CYxi2}}
for some constants $c_A$, $d_A$,  
where
\eq{\label{145}
g:=\ln  \big[\tfrac{2c^2_\phi}{c^2_{\xi\xi'}}~\!\text{sech}^2({c_{\phi}}\tau+d_{\phi})  \big]~,
}
Moreover  the first of \eqref{e2} can be  solved for $B$, 
\eq{{}
B=-4A+\tfrac14 f ~,}
where
\eq{\label{147}
f=\left\{ 
\begin{array}{cll}  \ln  \big[\frac{c_B^2}{4k} ~\!\text{sech}^2({c_B}\tau+d_B)  \big]& ~,~~&k>0\\
\\
\ln  \big[\frac{c_B^2}{4|k|}\text{csch}^2({c_B}\tau+d_B)  \big]  &~,~~ & k<0
\end{array}\right.
~,}
for some constants $c_B$, $d_B$. Plugging the solution into the second line of \eqref{e2} we obtain the constraint 
\eq{\label{ccccccxk}
3c_B^2=192c_A^2+4c_{\phi}^2
~,}
for either sign of $k$, which imposes   $|r|\leq \tfrac{{1}}{8}$. 
Let us first consider the case of closed universe (k>0). 
%
%
The 4d   Einstein metric reads,
\eq{\spl{
\d s^2_{4E}  &=  \text{sech}^{3} (c_B\tau)\Big(
-\d \tau^2
+ \frac{4k}{c_B^2}~\!  \text{cosh}^2(c_B\tau)  \d\Omega_k^2
\Big)
~,
}}
where we have set $d_A$, $d_B$, $d_{\phi}=0$ for simplicity.

\subsubsection*{Type II and critical solutions:  $k<0$}

Let us now set  $k<0$. The 4d part of the Einstein metric  reads
\eq{\spl{
\d s^2_{4E}  &=  \text{csch}^{3} (c_B\tau)\Big(
-\d \tau^2
+  \frac{4|k|}{c_B^2}~\!    \text{sinh}^2(c_B\tau)  \d\Omega_k^2
\Big)
~.
}\label{CYxi3}}

\subsection*{ \boldmath$\chi,\xi, \xi' \neq0$\unboldmath}

Let us now take,
\eq{
\chi= e^{\phi-4A-2B} c_\chi~;~~~
\varphi=0~;~~~h=0
~.}
This ansatz thus solves the form equations \eqref{7}-\eqref{12} in the case of $\chi, \xi, \xi'\neq0$. 
Let us consider the remaining equations of motion. 
Equations \eqref{e1} and the first of \eqref{e2} are solved by
\eq{\spl{
A&=\tfrac{1}{8}f+\tfrac{1}{16}g \\
B&=c_B\tau+d_B-\tfrac{1}{2}f-\tfrac{1}{4}g\\
\phi&=-\tfrac{1}{2}f+\tfrac{3}{4}g~,
}}
 with
\eq{f:=\ln  \big[\tfrac{2c^2_A}{c^2_{\xi\xi'}}~\!\text{sech}^2(c_{A}\tau+d_{A})  \big]~;~~~
g:=\ln  \big[\tfrac{c^2_\phi}{c^2_{\chi}}~\!\text{sech}^2({c_\phi}\tau+d_{\phi})  \big]~,
}
while the  second line of \eqref{e2} reduces to
\eq{\label{51x5}
12c_B^2=4c_A^2+3c_{\phi}^2 
~,}
which imposes   $|r|\leq \sqrt{3}$.  
The 4d  Einstein metric  reads
\eq{\spl{
\d s^2_{4E} 
 &= -
e^{ 6c_B\tau  } 
\d \tau^2
+e^{ 2c_B\tau  }  \d\vec{x}^2 
~,
}}
where we have set $d_A$, $d_B$, and $d_{\phi}=0$ for simplicity.

\subsubsection*{Type I solution:  $k>0$}\label{sec:3.6.1}

Let us now assume that  $k>0$. Equations \eqref{e1}   are solved by
\eq{\spl{
A&=\tfrac{1}{8}f+\tfrac{1}{16}g ~;~~~
\phi=-\tfrac{1}{2}f+\tfrac{3}{4}g~,
}\label{CYxck1}}
 with,
\eq{\label{162}
f:=\ln  \big[\tfrac{2c^2_A}{c^2_{\xi\xi'}}~\!\text{sech}^2(c_{A}\tau+d_{A})  \big]~;~~~
g:=\ln  \big[\tfrac{c^2_\phi}{c^2_{\chi}}~\!\text{sech}^2({c_\phi}\tau+d_{\phi})  \big]~,
}
Moreover the first equation of \eqref{e1} can be combined with the first of  \eqref{e2} to solve for $B$, 
\eq{\label{80xk}
B=-4A+\tfrac14 h ~,}
where
\eq{
h=\left\{ \label{164}
\begin{array}{cll}  \ln  \big[\frac{c_B^2}{4k} ~\!\text{sech}^2({c_B}\tau+d_B)  \big]& ~,~~&k>0\\
\\
\ln  \big[\frac{c_B^2}{4|k|}\text{csch}^2({c_B}\tau+d_B)  \big]  &~,~~ & k<0
\end{array}\right.
~,}
for some constants $c_B$, $d_B$. 
Plugging the solution into the second line of  \eqref{e2} we obtain the constraint
\eq{\label{51x5k}
3c_B^2=4c_A^2+3c_{\phi}^2 
~,}
which imposes   $|r|\leq \tfrac{\sqrt{3}}{2}$. 
Let us first consider the case of a closed universe $(k>0)$. 
The 4d Einstein metric   reads
\eq{\spl{
\d s^2_{4E} 
 &= 
  \text{sech}^{3} (c_{B}\tau)  \Big(
-\d \tau^2
+\frac{4k}{c_B^2}~\! \text{cosh}^{2} (c_{B}\tau) \d\Omega_k^2 
\Big)
~,
}}
where we have set $d_A$, $d_B$, and $d_{\phi}=0$ for simplicity.

\subsubsection*{Type II and critical solutions: $k<0$}

Let us now set  $k<0$. The 4d Einstein metric  reads
\eq{\spl{
\d s^2_{4E}  &=  \text{csch}^{3} (c_{B}\tau) \Big(
-\d \tau^2
+\frac{4|k|}{c_B^2}~\!  \text{sinh}^{2} (c_{B}\tau) \d\Omega_k^2 
\Big)
~.
}\label{CYxck2}}

\subsection*{\boldmath$h, \chi,\xi, \xi' \neq0$\unboldmath}

Let us  take
\eq{
\chi= e^{\phi-4A-2B} c_\chi~;~~~
\varphi=0
~.}
This ansatz thus solves the form equations \eqref{7}-\eqref{12} in the case of $h, \chi, \xi, \xi'\neq0$. 
Let us consider the remaining equations of motion. Equations \eqref{e1}   are solved by
\eq{\spl{
A=\tfrac{1}{8}f+\tfrac{1}{16}g ~;~~~
\phi=-\tfrac{1}{2}f+\tfrac{3}{4}g~,
}}
 with
\eq{\label{183}
f:=\ln  \Big[    \frac{ 2c_A^2~\!e^{c_{A}\tau+d_{A}} }{
(c^2_{\xi\xi'}+e^{c_{A}\tau+d_{A}})^2+c_A^2c_h^2
}   \Big]
~;~~~
g:=\ln  \big[\tfrac{c^2_\phi}{c^2_{\chi}}~\!\text{sech}^2({c_\phi}\tau+d_{\phi})  \big]~,
}
Moreover the first equation of \eqref{e1} can be combined with the first of  \eqref{e2} to solve for $B$, 
\eq{\label{80xkfh}
B=-4A+\tfrac14 h ~,}
where,
\eq{\label{185}
h=\left\{ 
\begin{array}{cll}  \ln  \big[\frac{c_B^2}{4k} ~\!\text{sech}^2({c_B}\tau+d_B)  \big]& ~,~~&k>0\\
\\
\ln  \big[\frac{c_B^2}{4|k|}\text{csch}^2({c_B}\tau+d_B)  \big]  &~,~~ & k<0
\end{array}\right.
~,}
for some constants $c_B$, $d_B$. 
Plugging the solution into the second line of  \eqref{e2} we obtain the constraint
\eq{\label{51x5kfh}
3c_B^2=c_A^2+3c_{\phi}^2 
~,}
for either sign of $k$. 
Let us first consider the case of a  closed universe $(k>0)$. 
The 4d   Einstein metric   reads
\eq{\spl{
\d s^2_{4E} 
 &=  
   \text{sech}^{3} (c_{B}\tau)   \Big(
-\d \tau^2
+ \frac{4k}{c_B^2}~\! \text{cosh}^{2} (c_{B}\tau) \d\Omega_k^2 
\Big)
~,}}
where we have set $d_A$, $d_B$,  $d_{\phi}=0$ for simplicity.

\subsubsection*{Type II and critical solutions: $k<0$}

Let us now set  $k<0$. The 4d Einstein metric  reads
\eq{\spl{
\d s^2_{4E} 
 &=  
   \text{csch}^{3} (c_{B}\tau)   \Big(
-\d \tau^2
+ \frac{4|k|}{c_B^2}~\! \text{sinh}^{2} (c_{B}\tau) \d\Omega_k^2 
\Big)
~.}\label{CYhxc}}

\subsection*{Compactification with background flux} 

The cosmological ansatz can be easily modified to accommodate a non-vanishing background flux for the three- and four-forms, as in 
\cite{Tsimpis:2020ysl},\footnote{We have redefined: $\xi\rightarrow c_0+4\omega\xi$ with respect to \cite{Tsimpis:2020ysl}.}
\eq{
  H=\d\chi \wedge J+\d\beta+\tfrac12 b_0\text{Re}\Omega~;~~~
G=\varphi\text{vol}_4+  J\wedge (\d\gamma - \alpha\wedge \d\chi)+\tfrac12 c_0 J\wedge J-\tfrac{1}{2}\d\xi\wedge\text{Im}\Omega
-\tfrac{1}{2}D\xi'\wedge\text{Re}\Omega
~,}
with background charges $b_0$, $c_0\in\mathbb{R}$,  
where the covariant derivative is defined as: $D\xi'=\d\xi'+b_0\alpha$. 
The internal $(m,n)$-components of the Einstein 
equations now read
\eq{\spl{\label{et12dr}
0&=e^{-8A-2B}\nabla^{\mu}\left(
e^{8A+2B}\partial_{\mu}A
\right)  
+\frac18e^{-\phi-4A}(\partial\chi)^2
-\frac{1}{48}e^{-\phi-4A-4B}h^2\\
& 
+\frac{1}{16}e^{\phi/2-6A}
\Big[ (\partial \xi)^2+(D\xi')^2\Big]
+\frac{3}{16} 
e^{\phi/2-6A-6B}\varphi^2
+18 e^{-\phi-4A+2B}b_0^2
+\frac{7}{16}e^{\phi/2-6A+2B} c_0^2
~.}}

The external $(\mu,\nu)$-components read
\eq{\spl{\label{et22dr}
R^{(4)}_{\mu\nu}&=
g_{\mu\nu}\left(\nabla^{2}A+\nabla^{2} B+
8(\partial A)^2+2(\partial B)^2+10\partial A\cdot \partial B\right)
\\
&-8\partial_{\mu}A\partial_{\nu}A-2\partial_{\mu}B\partial_{\nu}B
-16\partial_{(\mu}A\partial_{\nu)}B+8\nabla_{\mu}\partial_{\nu}A+2\nabla_{\mu}\partial_{\nu}B
\\
&+\frac32 e^{-\phi-4A} \partial_{\mu}\chi\partial_{\nu}\chi
+\frac14 e^{\phi-4A-4B} h^2_{\mu\nu}
+\frac12 \partial_{\mu}\phi\partial_{\nu}\phi
+\frac{1}{2} e^{\phi/2-6A} ( \partial_{\mu}\xi\partial_{\nu}\xi+D_{\mu}\xi' D_{\nu}\xi')
 \\
&+\frac{1}{16} g_{\mu\nu}\Big(  
-\frac{1}{3}e^{\phi-4A-4B}h^2
-3e^{\phi/2-6A} \Big[ (\partial \xi)^2+(D\xi')^2\Big]-6e^{-\phi-4A}(\partial\chi)^2\\
&-5e^{\phi/2-6A-6B}\varphi^2  
-288e^{-\phi-4A+2B}b_0^2
-9 c_0^2e^{\phi/2-6A+2B}  
\Big)
~,}}
while the mixed $(\mu,m)$-components are automatically satisfied. 
The dilaton equation reads
\eq{\spl{\label{et32dr}
0&=e^{-10A-4B}\nabla^{\mu}\left(
e^{8A+2B}\partial_{\mu}\phi
\right)
-\frac{1}{4}e^{\phi/2-8A-2B}\Big[ (\partial \xi)^2+(D\xi')^2\Big]
+\frac32e^{-\phi-6A-2B}(\partial\chi)^2\\
&+\frac{1}{12}e^{-\phi-6A-6B}h^2
+\frac{1}{4}
e^{\phi/2-8A-8B}\varphi^2
+72 e^{-\phi-6A}b_0^2
-\frac{3}{4}c_0^2 e^{\phi/2-8A} 
~.}}
The $F$-form equation of motion 
reduces to the condition
\eq{\spl{\label{112dr}
0 &= \varphi  \d\beta  
-12b_0 e^{6B}\star_4  D\xi'
~.}}
The $H$-form equation reduces to
\eq{\spl{\label{hfeom2dr}
\d\left(
e^{-\phi+4A+2B}\star_4\d\chi
\right)  &=  c_0 \varphi 
 ~\!\text{vol}_4  
~.}}
The $G$-form equation of motion reduces to
\eq{\spl{
\label{gfeom12dr}
\d\left(e^{\phi/2+2A+2B}\star_4\d\xi\right)  &=  h\wedge D\xi'
+12 b_0\varphi ~\!\text{vol}_4
\\
\d\left(e^{\phi/2+2A+2B}\star_4 D\xi'\right) &= - h\wedge\d\xi\\
 0&=   c_0 ~\! \d\beta
~,}}
together with the constraint
\eq{\label{gfeom22dr}
0=\d\left(
 \varphi e^{\phi/2+2A-4B} + 3c_0\chi+ 12 b_0\xi
\right)
~.}
The latter can be  integrated to solve for $\varphi$ in terms of the other fields,
\eq{\label{fg2dr}
\varphi=\left(C -  3c_0\chi-12 b_0\xi\ \right)e^{-2A+4B-\phi/2}
~,}
where $C$ is an arbitrary constant.

\subsubsection*{Type I solution: \boldmath$b_0, c_0\neq0$\unboldmath}

If we want $b_0$, $c_0\neq0$, then we must set $h$, $ D\xi'=0$, as follows from \eqref{112dr}, \eqref{gfeom12dr}. It is then consistent 
to take $\chi$, $\xi=\text{constant}$, and $\varphi=0$. This ansatz solves all form equations of motion. 
Let us consider the remaining equations of motion. 
Eq.~\eqref{et12dr} and 
the dilaton equation~\eqref{et32dr} reduce to
\eq{\label{et3bbbf}
\d_\tau^2 A=18b_0^2  e^{-\phi+12A+6B}
+\tfrac{7}{16}c_0^2 e^{\phi/2+10A+6B}
~;~~~
\d_\tau^2 \phi=72b_0^2  e^{-\phi+12A+6B}
-\tfrac{3}{4}c_0^2 e^{\phi/2+10A+6B}
~.}
The external Einstein equations \eqref{et22dr} now
reduce to the following two  equations,  
\eq{\spl{\label{red5bbbf}
-36b_0^2  e^{-\phi+12A+6B}
- c_0^2 e^{\phi/2+10A+6B}
&= \d_\tau^2B
\\
+144b_0^2  e^{-\phi+12A+6B}
+3 c_0^2 e^{\phi/2+10A+6B}
&=
144(\d_\tau A)^2+12(\d_\tau B)^2+96\d_\tau A\d_\tau B- (\d_\tau\phi)^2 
~.}}
The second line above is a {constraint}, consistently propagated by the remaining equations of motion. Indeed the 
$\tau$-derivative of this equation is automatically satisfied by virtue of \eqref{et3bbbf}.

\begin{figure}[H]
\begin{center}
\includegraphics[width=0.8\textwidth]{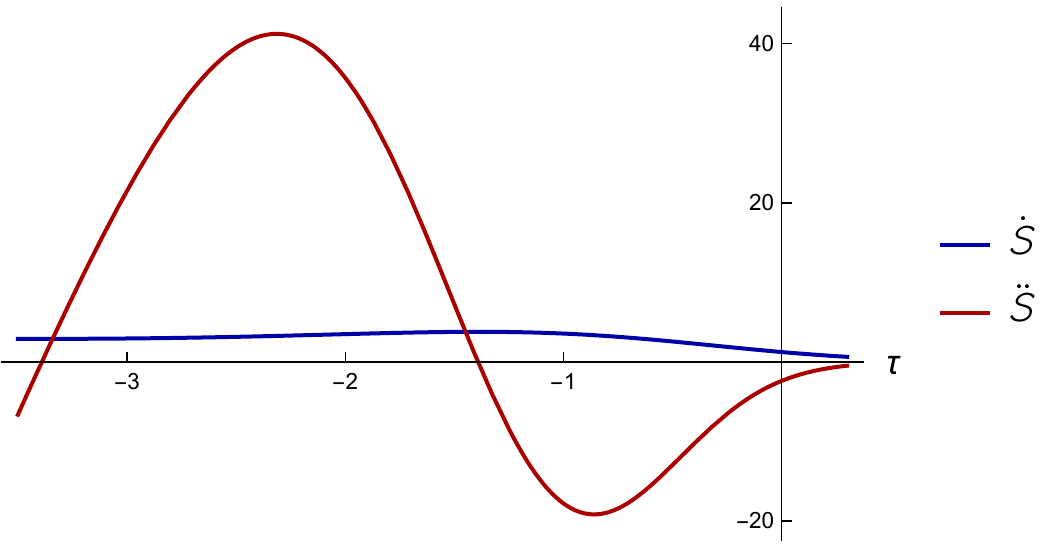}
\caption{Plot of $\dot{S}(T), \ddot{S}(T)$ as a function of  $\tau$, for $c_B=-1$, $r=0.67$. }\label{fig:acceleration2}
\end{center}
\end{figure}

Equations \eqref{et3bbbf} and the first of \eqref{red5bbbf} are solved by
\eq{\spl{
A&=-\tfrac{9}{16}(c_B\tau+d_B)-\tfrac{1}{4}f-\tfrac{7}{32}g \\
B&=c_B\tau+d_B+\tfrac{1}{2}f+\tfrac{1}{2}g\\
\phi&=-\tfrac{3}{4}(c_B\tau+d_B)- f+\tfrac{3}{8}g~,
}\label{CYbg}}
 with,
\eq{f:=\ln  \big[\tfrac{c^2_A}{36b_0^2}~\!\text{sech}^2(c_{A}\tau+d_{A})  \big]~;~~~
g:=\ln  \big[\tfrac{c^2_\phi}{c^2_{0}}~\!\text{sech}^2({c_\phi}\tau+d_{\phi})  \big]~,
}
while the  second line of \eqref{red5bbbf} reduces to
\eq{\label{51x5yy}
3c_B^2=4c_A^2+3c_{\phi}^2 
~,}
which imposes   $|r|\leq \tfrac{\sqrt{3}}{2}$. 
The 4d Einstein metric reads
\eq{\spl{
\d s^2_{4E} 
 &= -e^{ -\frac{15}{2} c_B\tau  } \text{cosh}^6(c_A\tau) \text{cosh}^{\frac{9}{2}} (c_{\phi}\tau)~\!\d \tau^2
+ e^{ -\frac{5}{2} c_B\tau  } \text{cosh}^2(c_A\tau) \text{cosh}^{\frac{3}{2}}(c_{\phi}\tau) \d\vec{x}^2
~,
}}
where we have set $d_A$, $d_B$ and $d_{\phi}=0$ for simplicity. 
In terms of the coordinates $T_\pm=e^{( -\frac{15}{2} c_B\pm3c_A\pm\frac94c_\phi)\tau}$, 
the metric asymptotically 
takes the form \eqref{intro1},
 where $a_\pm =\tfrac13$. 
For a certain range of parameters, this model exhibits transient accelerated expansion, where $\dot{S}(T), \ddot{S}(T)>0$, cf. Figure \ref{fig:acceleration2}.


\subsection{Compactification on Einstein manifolds}\label{sec:em}

We will now consider (massive) IIA backgrounds for which the internal 6d manifold is Einstein, 
\eq{\label{150}
R_{mn}=\lambda g_{mn}~,}
where $R_{mn}$ is the Ricci tensor associated to $g_{mn}$, and $\lambda\in\mathbb{R}$.  
The 10d metric is as in \eqref{2}. We assume the following form ansatz,
\eq{\spl{
F=0~;~~~ H=0~;~~~
G&=\varphi\text{vol}_4 ~,}}
where $\varphi$ is a 4d scalar.

The resulting equations of motion are as follows. 
The internal $(m,n)$-components of the Einstein 
equations  read
\eq{\spl{\label{et1em}
e^{2B}\lambda=e^{-8A-2B}\nabla^{\mu}\left(
e^{8A+2B}\partial_{\mu}A
\right)+\tfrac{1}{16}m^2e^{5\phi/2+2A+2B}
+\tfrac{3}{16} 
e^{\phi/2-6A-6B}\varphi^2
~,}}
where $m$ is the Romans' mass. 
The external $(\mu,\nu)$-components read
\eq{\spl{\label{et2em}
R^{(4)}_{\mu\nu}&=
g_{\mu\nu}\left(\nabla^{2}A+\nabla^{2} B+
8(\partial A)^2+2(\partial B)^2+10\partial A\cdot \partial B\right)
\\
&-8\partial_{\mu}A\partial_{\nu}A-2\partial_{\mu}B\partial_{\nu}B
-16\partial_{(\mu}A\partial_{\nu)}B+8\nabla_{\mu}\partial_{\nu}A+2\nabla_{\mu}\partial_{\nu}B
+\frac12 \partial_{\mu}\phi\partial_{\nu}\phi
\\
&+\tfrac{1}{16} g_{\mu\nu}\big(  
-5e^{\phi/2-6A-6B}\varphi^2 
+ m^2e^{5\phi/2+2A+2B}
\big)
~,}}
while the mixed $(\mu,m)$-components are automatically satisfied. 
The dilaton equation reads
\eq{\spl{\label{et3em}
0&=e^{-10A-4B}\nabla^{\mu}\left(
e^{8A+2B}\partial_{\mu}\phi
\right)
-\tfrac{5}{4}m^2e^{5\phi/2}
+\tfrac{1}{4}
e^{\phi/2-8A-8B}\varphi^2
~.}}
The $F$, $H$-form equations are automatically satisfied, 
while the $G$-form equation of motion reduces to
\eq{\label{fg}
\varphi= c_\varphi e^{-2A+4B-\phi/2}
~,}
where $c_\varphi$ is an arbitrary constant.

As before, we will assume that the unwarped  4d metric is of the form \eqref{metricansatz} 
and moreover that  $A$, $B$, $\phi$ only depend on time. 
Eq.~\eqref{et1em} and the dilaton equation~\eqref{et3em}  reduce to
\eq{\spl{\label{et3bbem}
\d_\tau^2 A&=
-\lambda e^{16A+6B}
+\tfrac{1}{16}m^2 e^{5\phi/2+18A+6B}
+\tfrac{3}{16}c^2_{\varphi} e^{-\phi/2+6A+6B}\\
\d_\tau^2 \phi&=
-\tfrac{5}{4}  m^2e^{5\phi/2+18A+6B}
+\tfrac{1}{4}c^2_{\varphi} e^{-\phi/2+6A+6B}
~.}}
The external Einstein equations \eqref{et2em} 
reduce to the following two  equations,  
\eq{\spl{\label{red5bbem}
\lambda e^{16A+6B}
-2ke^{16A+4B}
-\tfrac{1}{2}c^2_{\varphi} e^{-\phi/2+6A+6B}&= \d_\tau^2B
\\
-12\lambda e^{16A+6B}
-12ke^{16A+4B}
+m^2 e^{5\phi/2+18A+6B}
+c^2_{\varphi} e^{-\phi/2+6A+6B}&=
144(\d_\tau A)^2+12(\d_\tau B)^2\\
&+96\d_\tau A\d_\tau B-(\d_\tau\phi)^2  
~.}}
The second line above is the {constraint}, consistently propagated by the remaining equations of motion.


\subsection*{\boldmath$\lambda\neq 0$\unboldmath }\label{sec:5.2}

For $m$, $k$, $\varphi=0$,  
the system of equations in \eqref{et3bbem} and the first of \eqref{red5bbem} can be solved to give
\eq{\spl{\label{186}
A&= c_A\tau +d_A+\tfrac{1}{10} g(\tau) \\
B&=-\tfrac83(c_A\tau +d_A)-\tfrac{1}{10} g(\tau)  \\
\phi&=c_\phi\tau +d_\phi
~,}}
for some constants $c_A$, $d_A$, $c_\phi$, $d_\phi$, where
\eq{\label{223}
g=\left\{ 
\begin{array}{cll}      \ln  \big[\tfrac{c_B^2}{5\lambda}~\!\text{sech}^2(c_B~\!\tau + d_B)  \big]  & ~,~~&\lambda>0\\
\\
 \ln  \big[\frac{c_B^2}{5|\lambda|}\text{csch}^2({c_B}\tau+d_B)  \big]  &~,~~ & \lambda<0
\end{array}\right.
~,}
for some constants $c_B$, $d_B$. Plugging the solution into the second line of \eqref{red5bbem} we obtain the constraint
\eq{\label{115kl}
\tfrac{12}{5}c_B^2=c_\phi^2+\tfrac{80}{3}  c_A^2
~,}
for either sign of  $\lambda$. The constraint imposes   $|r|\leq \tfrac{{3}}{10}$.

\subsubsection*{Type I solution: $\lambda>0$}

Let us first consider an internal space of positive curvature $(\lambda>0)$. 
The 4d Einstein metric \eqref{2} reads
\eq{\spl{
\d s^2_{4E} 
 &= -e^{ 8c_A\tau  } \text{sech}^{\frac{18}{5}} (c_B\tau)\d \tau^2
+ e^{ \frac{8}{3} c_A\tau  }  \text{sech}^{\frac{6}{5}} (c_B\tau) \d\vec{x}^2
~,
}\label{El1}}
where we have set $d_A$,  and $d_{\phi}=0$ for simplicity. 
To examine the asymptotic behavior of the metric it suffices to consider $c_B\geq0$ (the other cases are obtained by inverting the sign of $\tau$). 
The metric asymptotically 
takes the form of \eqref{intro1}, 
where $a_\pm = \tfrac13$.


\subsubsection*{Type II and critical solutions: $\lambda<0$}\label{sec:critrec}

\begin{figure}[H]
\begin{center}
\includegraphics[width=0.8\textwidth]{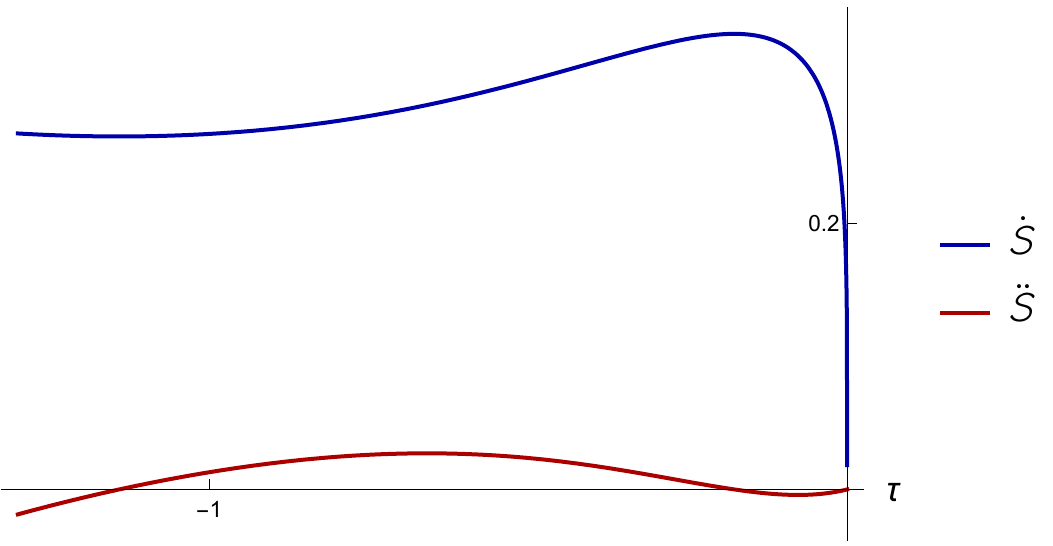}
\caption{Plot of $\dot{S}(T), \ddot{S}(T)$ as a function of  $\tau$, for $c_B=-1$, $r=-0.26$. 
In the $\tau\rightarrow 0$ limit, we have  $\dot{S}(T), \ddot{S}(T)\rightarrow 0$.}\label{fig:acceleration3}
\end{center}
\end{figure}

Let us now set  $\lambda<0$. 
The 4d Einstein metric  reads
\eq{\spl{
\d s^2_{4E} 
 &= -e^{ 8c_A\tau  } \text{csch}^{\frac{18}{5}} (c_B\tau)\d \tau^2
+ e^{ \frac{8}{3} c_A\tau  } \text{csch}^{\frac{6}{5}} (c_B\tau) \d\vec{x}^{~\!2}
~,
}\label{El2}}
The $\tau\rightarrow\infty$ asymptotics of the warp factors are the same as for $\lambda>0$. For $\tau\rightarrow 0$, on the other hand, 
the function $g$ in \eqref{223} tends to $g\rightarrow -\ln(5|\lambda|\tau^2)$. Moreover, the constraint imposes $c_A$, $c_\phi=0$. 
Hence, in the $\tau\rightarrow 0$ limit the solution reads
\eq{
\label{El3}
A=d_A -\tfrac{1}{10}\ln(5|\lambda|\tau^2)
 ~;~~~B=-\tfrac83d_A+\tfrac{1}{10}\ln(5|\lambda|\tau^2)~;~~~
\phi=d_\phi
~.}
Asymptotically, at infinite proper time, the metric reaches the form
\eq{
\d s^2_{4E}\rightarrow  -\d T^2+T^{\frac32}\d\vec{x}^{~\!2}
~,}
where we have defined $T\propto \tau^{-\frac45}$. 
The warp factor of the internal space   
scales as $e^A\propto T^{1/4}$, while 
 the dilaton is constant.
 This  is also an exact solution of the theory in its own right. 
For a certain range of parameters, this model exhibits transient accelerated expansion, where $\dot{S}(T), \ddot{S}(T)>0$, cf. Figure \ref{fig:acceleration3}.


\subsection*{Type I solution: \boldmath$m\neq 0$\unboldmath}\label{sec:cosmmink2}

Setting $k$, $\lambda$, $\varphi=0$, 
the system of equations in \eqref{et3bbem} and the first of \eqref{red5bbem} can be solved to give
\eq{\spl{\label{193}
A&=c_A\tau + d_A-\tfrac{1}{32}g(\tau)\\
B&=c_B\tau + d_B  \\
\phi&=-\tfrac{12}{5}(3c_A + c_B)\tau - \tfrac{12}{5}(3d_A + d_B) +\tfrac58 g(\tau) 
~,}}
for some constants $c_A$, $d_A$, $c_B$, and $d_B$ where
\eq{\label{245}
g=  \ln  \big[\tfrac{c_\phi^2}{m^2}~\!\text{sech}^2(c_\phi\tau + d_\phi)  \big]
~,}
for some constant $d_\phi$. 
Plugging the solution into the second line of  \eqref{red5bbem} we obtain the constraint
\eq{\label{115iii}
 \tfrac{25}{12}c_\phi^2=192c_A^2+128c_Ac_B+13c_B^2
~,}
which  imposes   $r\leq -\tfrac{{13}}{24}$ or $r\geq -\tfrac{{1}}{8}$ .

\begin{figure}[H]
\begin{center}
\includegraphics[width=0.8\textwidth]{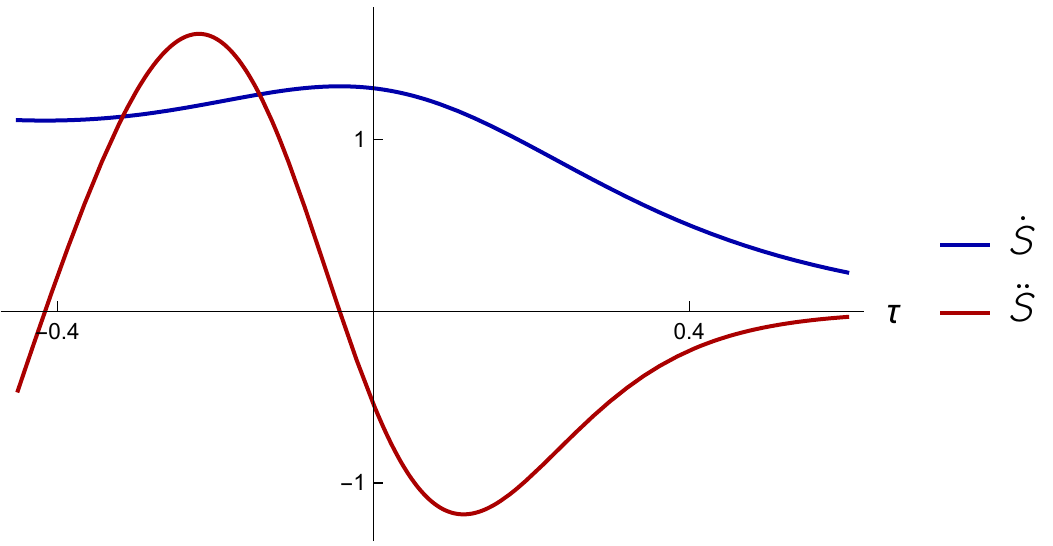}
\caption{Plot of $\dot{S}(T), \ddot{S}(T)$ as a function of  $\tau$, for $c_B=1$, $r=0.075$. }\label{fig:acceleration4}
\end{center}
\end{figure}

The 4d Einstein metric \eqref{2} reads
\eq{\spl{
\d s^2_{4E} 
 &=- e^{ (24c_A + 6c_B)\tau  } \text{cosh}^{\frac{3}{2}}(c_{\phi}\tau) \d \tau^2
+ e^{ (8c_A + 2c_B)\tau  } \text{cosh}^{\frac{1}{2}} (c_{\phi}\tau)\d\vec{x}^{~\!2}
~,
}}
where we have set $d_A$, $d_B$ and $d_{\phi}=0$ for simplicity. 
To examine the asymptotic behavior of the metric it suffices to consider $c_\phi\geq0$. 
The metric asymptotically 
takes the form  \eqref{intro1}, 
where $a_\pm = \frac13$. 
For a certain range of parameters, this model exhibits transient accelerated expansion, where $\dot{S}(T), \ddot{S}(T)>0$, cf. Figure \ref{fig:acceleration4}.

\subsection*{\boldmath$\lambda, k\neq0$\unboldmath}\label{sec:lkneq0}

Assuming $B$ is constant (we may set $B=0$ for simplicity) and setting
\eq{\label{198lal}
k=\tfrac12\lambda~;~~~
m, \varphi=0
~,}
the system of equations in \eqref{et3bbem} and the first of \eqref{red5bbem} can be solved to give $B=0$, $\phi=c_\phi\tau +d_\phi$, 
for some constants  $c_\phi$, $d_\phi$, and
\eq{\label{asolu}
A=   \left\{ 
\begin{array}{cll}   \tfrac{1}{16}  \ln  \big[\tfrac{c_A^2}{8\lambda}~\!\text{sech}^2(c_A~\!\tau + d_A)  \big]  & ~,~~&\lambda>0\\
\\
\tfrac{1}{16} \ln  \big[\frac{c_A^2}{8|\lambda|}\text{csch}^2({c_A}\tau+d_A)  \big]  &~,~~ & \lambda<0
\end{array}\right.
~,}
for some constants $c_A$, $d_A$. Plugging the solution into the second line of \eqref{red5bbem} we obtain the constraint  
\eq{\label{115kls}
c_\phi^2 = \tfrac{9}{4}  c_A^2
~,}
for either sign of $\lambda$.

\subsubsection*{Type I  solution: $\lambda>0$}

Let us first consider the case $\lambda>0$. 
The 4d Einstein  metric  reads
\eq{\spl{
\d s^2_{4E} 
 &= \text{sech}^{3}(c_A\tau)   \Big(-\d \tau^2
+ \frac{16k}{c_A^2}  ~\!\text{cosh}^2(c_A\tau) \d\Omega_k^2\Big)
~,
}\label{Elk1}}
where we have set $d_A$,  and $d_{\phi}=0$ for simplicity, and we took \eqref{198lal} into account. 
For $\tau\rightarrow\pm\infty$  the metric asymptotically 
takes the form of a power-law expansion \eqref{intro1} for $a_\pm=\tfrac13$.


\subsubsection*{Type II and critical solutions: $\lambda<0$}\label{sec:5.4.2}

Let us now set  $\lambda<0$. 
The 4d Einstein metric  reads
\eq{\spl{
\d s^2_{4E} 
 &= \text{csch}^{3}(c_A\tau)   \Big(-\d \tau^2
+ \frac{16|k|}{c_A^2}~\! \text{sinh}^2(c_A\tau)  \d\Omega_k^2\Big)
~,
}\label{Elk2}}
where we took \eqref{198lal} into account. 
The $\tau\rightarrow\infty$ asymptotics of the warp factors are the same as for $\lambda>0$. For $\tau\rightarrow 0$, on the other hand, 
the function $A$ in \eqref{asolu} tends to $A\rightarrow -\tfrac{1}{16}\ln(8|\lambda|\tau^2)$. Moreover, the constraint imposes   $c_\phi=0$. 
Hence, in the $\tau\rightarrow 0$ limit, which is reached at infinite proper time,  the solution reads
\eq{\label{Elk3}
A=-\tfrac{1}{16}\ln(8|\lambda|\tau^2)~;~~~B=0~;~~~
\phi=d_\phi
~.}
The metric asymptotes a singular Milne universe with angular defect,
\eq{
\d s^2\rightarrow  \d T^2+4|k|~\!T^2\d\Omega_k^2
~,}
where we have defined $T\propto \tau^{-\frac12}$. 
The warp factor of the internal space scales as $e^A\propto T^{1/4}$, while the dilaton is constant. 
This is also an exact solution of the theory in its own right.


\subsection*{Critical solution: \boldmath$\lambda, m\neq0$\unboldmath}\label{sec:34}

Setting $k$, $c_\varphi=0$, 
the system of equations in \eqref{et3bbem}, \eqref{red5bbem} admits the solution
\eq{\label{242b}
A=-\tfrac{25}{128}\ln|\tau|+d_A~;~~~B=\tfrac{3}{16}\ln|\tau|+d_B~;~~~\phi=\tfrac{5}{32}\ln|\tau|+d_\phi
~,
}
for arbitrary constants $d_A$, $d_B$, $d_\phi$ subject to the conditions
\eq{\label{243b}
9d_A+3d_B+\tfrac54 d_\phi=-\tfrac12\ln({8}m^2)~;~~~
\lambda=-\tfrac32 m^2e^{2d_A+\tfrac52 d_\phi}<0
~.}
The 4d Einstein metric  takes the form
\eq{
\d s^2_{4E}=-\d T^2+T^{\frac{38}{25}}\d\vec{x}^{~\!2}
~,}
where $T\propto\tau^{-\frac{25}{32}}$.
The warp factor of the internal space and the dilaton scale as $e^A\propto T^{1/4}$, $e^\phi\propto T^{-1/5}$ respectively.

\subsection*{Type I solution: \boldmath$\varphi, m\neq0$\unboldmath}

\begin{figure}[H]
\begin{center}
\includegraphics[width=0.8\textwidth]{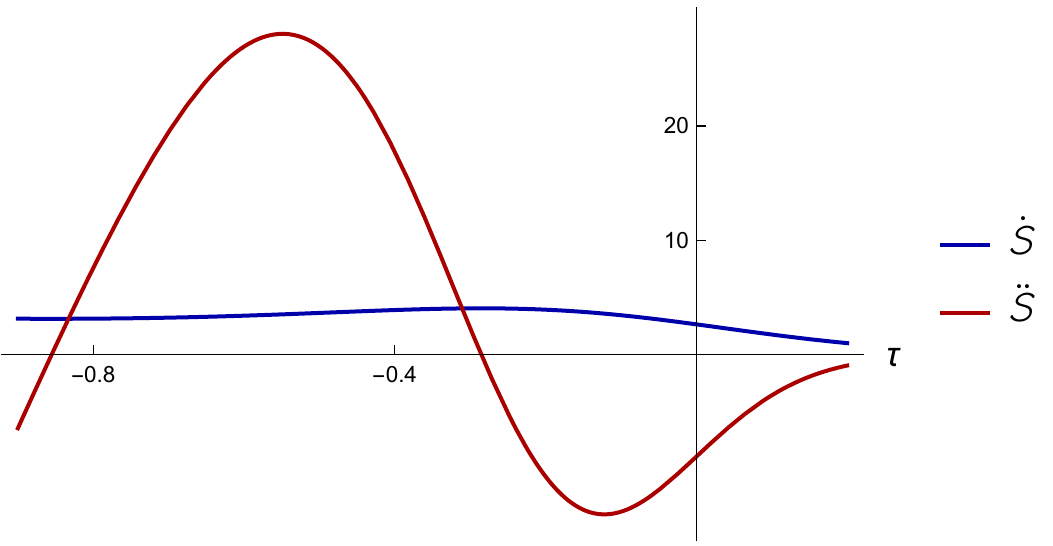}
\caption{Plot of $\dot{S}(T), \ddot{S}(T)$ as a function of  $\tau$, for $c_A=1$. }\label{fig:acceleration5}
\end{center}
\end{figure}

Setting $c_\varphi^2=m^2$, 
the system of equations in \eqref{et3bbem} and the first of \eqref{red5bbem} can be solved to give
\eq{\spl{
A&=c_A\tau + d_A-\tfrac{1}{4}g(\tau)\\
B&=-\tfrac43(c_A\tau + d_A) +\tfrac{1}{2}g(\tau) \\
\phi&=-4A
~,}\label{Epm1}}
for some constants $c_A$, $d_A$, $c_B$, and $d_B$ where
\eq{
g=  \ln  \big[\tfrac{2c_\phi^2}{m^2}~\!\text{sech}^2(c_\phi\tau + d_\phi)  \big]
~,}
for some constant $d_\phi$. 
Plugging the solution into the second line of  \eqref{red5bbem} we obtain the constraint
\eq{\label{115}
  c_\phi=\pm   \tfrac{4}{\sqrt{3}}  c_A
~.}
The 4d Einstein metric  reads
\eq{\spl{
\d s^2_{4E} 
 &=- e^{ 16c_A \tau  } \text{cosh}^{6}(c_{\phi}\tau) \d \tau^2
+ e^{ \frac{16}{3} c_A \tau  }  \text{cosh}^{2}(c_{\phi}\tau) \d\vec{x}^{~\!2}
~,
}}

where we have set $d_A$,  and $d_{\phi}=0$ for simplicity. 
The metric asymptotically 
takes the form \eqref{intro1}, with $a_\pm = \tfrac13$. 
For a certain range of parameters, this model exhibits transient accelerated expansion, where $\dot{S}(T), \ddot{S}(T)>0$, cf. Figure \ref{fig:acceleration5}.

\subsection*{Type II and critical solution: \boldmath$k, m\neq0$\unboldmath}\label{sec:critsolkm}

Setting $k=-\tfrac34 m^2$, 
the system of equations in \eqref{et3bbem} and the first of \eqref{red5bbem} can be solved to give
\eq{\spl{
A&=\tfrac{5}{12}(c_A\tau + d_A)-\tfrac{1}{112} \ln  \big[\tfrac{7m^2}{2c_\phi^2}~\!\text{sinh}^2(c_\phi\tau + d_\phi)  \big]  \\
B&=-\tfrac{35}{3}(c_A\tau + d_A)+24A\\
\phi&=\tfrac{28}{3}(c_A\tau + d_A)-20A~,}\label{Ekm1}}
for some constants $c_A$, $d_A$, $c_\phi$ and $d_\phi$.  Plugging the solution into the second line of  \eqref{red5bbem} we obtain the constraint
\eq{\label{1152}
  c_\phi=\pm   \tfrac{7}{\sqrt{15}}  c_A
~.}
The scale factor  reads
\eq{
S^2=\sqrt{\frac{2}{7}}\left| \frac{c_\phi}{m} ~\!\text{csch}(c_\phi\tau + d_\phi)\right|
~.}
In the $\tau\rightarrow 0$ limit, which is reached at infinite proper time,  the solution reads
\eq{\spl{
A&=\tfrac{5}{12}  d_A-\tfrac{1}{112} \ln  \big(\tfrac{7m^2}{2} \tau^2 \big) \\
B&=-\tfrac{35}{3}d_A+24A\\
\phi&=\tfrac{28}{3}d_A-20A~.}\label{Ekm2}}
The metric asymptotes a singular Milne universe with angular defect,
\eq{
\d s^2=  \d T^2+\tfrac{7}{6}|k|~\!T^2\d\Omega_k^2
~,}
where we have defined $T\propto \tau^{-\frac12}$. This is also an exact solution of the theory in its own right.

\subsection*{Critical solution: \boldmath$k, \lambda, m\neq0$\unboldmath}

Setting $c_{\varphi}=0$, equations \eqref{et3bbem}, \eqref{red5bbem} admit the solution
\eq{\label{Eklm}
A=-\tfrac18\ln |\tau|+A_0~;~~~B=B_0~;~~~\phi=\tfrac{1}{10}\ln|\tau|+\phi_0~,
}
for $A_0$, $B_0$, $\phi_0$ arbitrary real constants 
and
\eq{
k=-\tfrac{3}{50}e^{-16 A_0 - 4 B_0}~;~~~m^2=\tfrac{2}{25}e^{-18 A_0 - 6 B_0 - 5 \phi_0/2}~;~~~\lambda=-\tfrac{3}{25}e^{-16 A_0 - 6 B_0}~.
}
We thus obtain a singular Milne universe,
\eq{
\d s^2_{4E}=\d T^2+\tfrac{25}{6} |k| T^2\d\Omega_k^2
~.}

\subsection*{Critical solution: \boldmath$k, \varphi, m\neq0$\unboldmath}

Setting $\lambda=0$, equations \eqref{et3bbem}, \eqref{red5bbem} admit the solution 
\eq{\label{Ekpm}
A=-\tfrac{1}{16}\ln |\tau|+A_0~;~~~B=-\tfrac{1}{4}\ln|\tau|+B_0~;~~~\phi=\tfrac{1}{4}\ln|\tau|+\phi_0~,
}
for $A_0$, $B_0$, $\phi_0$ arbitrary real constants 
and
\eq{
k=-\tfrac{3}{16}e^{-16 A_0 - 4 B_0}~;~~~m^2=\tfrac{1}{4}e^{-18 A_0 - 6 B_0 - 5 \phi_0/2}~;~~~c_\varphi^2=\tfrac{1}{4}e^{-6 A_0 - 6 B_0 + \phi_0/2}~.
}
We thus obtain a singular Milne universe,
\eq{
\d s^2_{4E}=\d T^2+\tfrac{4}{3} |k| T^2\d\Omega_k^2
~.}

\subsection*{AdS solution: \boldmath$k, \lambda, m, \varphi\neq0$\unboldmath}

A different analytic solution is obtained by setting $A$, $\phi=0$, and $\lambda=m^2$, $c_\varphi=\pm\sqrt{5}m$. 
Equations \eqref{et3bbem}, \eqref{red5bbem} then reduce to
\eq{\spl{\label{kl3red5bbem}
-\tfrac{3}{2}m^2 e^{6B}
-2ke^{4B}
&= \d_\tau^2B
\\
-\tfrac{1}{2}m^2 e^{6B}
-ke^{4B}
&=
(\d_\tau B)^2 
~.}}
Let us define a new time variable $\sigma$ via
\eq{\label{216note}
\tfrac{\d\sigma}{\d\tau}=e^{2B}
~,}
in terms of which the equations read
\eq{\spl{\label{skl3red5bbem}
-\tfrac{1}{2}m^2 e^{2B}
&= \d_\sigma^2B
\\
-\tfrac{1}{2}m^2 e^{2B}
-k
&=
(\d_\sigma B)^2 
~.}}
These can be integrated to give
\eq{\label{218note}
 B= \tfrac{1}{2}  \ln  \big[\tfrac{2c_B^2}{m^2}~\!\text{\text{sech}}^2(c_B\sigma + d_B)  \big]~~~;
 ~k=  -c_B^2
~,}
for some real constants $c_B$, $d_B$. 
The 4d Einstein metric  reads
\eq{\spl{
\d s^2_{4E}&= \tfrac{2c_B^2}{m^2}~\!\text{\text{sech}}^2(c_B\sigma)  (-\d \sigma^2
+  \d\Omega_k^2) 
~,
}}
where we set $d_B=0$ for simplicity. 
This is an AdS$_4$ metric in conformal coordinates. To see this, we may define a new coordinate $T$, such that 
$\text{cos}~\!T=\text{sech}(c_B\sigma)$, in terms of which the metric takes the well-known form,
\eq{\spl{
\d s^2_{4E}&=    \tfrac{2 }{m^2}(-\d T^2
+ |k| ~\!\text{cos}^2T \d\Omega_k^2) 
~. 
}}
For $T\in[-\tfrac{\pi}{2},\tfrac{\pi}{2}]$, this parameterizes  AdS$_4$ in global coordinates and hyperbolic slicing.

\subsection*{\boldmath$\lambda, \varphi \neq0$\unboldmath}

Setting $k$, $m=0$,   
the system of equations in \eqref{et3bbem} and the first of \eqref{red5bbem} can be solved to give
\eq{\spl{
A&=-\tfrac{1}{48}(c_A\tau+d_A) +\tfrac{3}{16}f(\tau)+\tfrac{1}{10}g(\tau)\\
B&=\tfrac{1}{18}(c_A\tau+d_A) -\tfrac{1}{2}f(\tau)-\tfrac{1}{10}g(\tau)  \\
\phi&=\tfrac{5}{12}(c_A\tau+d_A) +\tfrac14 f(\tau) 
~,}}
for some constants $c_A$, $d_A$, where
\eq{\label{290}
f=-\tfrac12\ln  \big[\tfrac{c^2_B}{c^2_\varphi}~\!\text{sech}^2(c_B\tau + d_B)  \big]~;~~~
g=\left\{ 
\begin{array}{cll}     \ln  \big[\tfrac{c_\phi^2}{5\lambda}~\!\text{sech}^2(c_\phi\tau + d_\phi)  \big] & ~,~~&\lambda>0\\
\\
  \ln  \big[\tfrac{c_\phi^2}{5|\lambda|}~\!\text{csch}^2(c_\phi\tau + d_\phi)  \big]  &~,~~ & \lambda<0
\end{array}\right.
~,}
for some constants $d_B$, $d_\phi$.  
Plugging the solution into the second line of  \eqref{red5bbem} we obtain the constraint
\eq{\label{118}
\tfrac{12}{5}c_\phi^2=\tfrac{5}{27}c_A^2+c_B^2
~.}
for either sign of $\lambda$. The constraint allows any $r\in\mathbb{R}$.

\subsubsection*{Type I solution: $\lambda>0$}

Let us first consider the case $\lambda>0$. 
The 4d Einstein metric \eqref{2} reads
\eq{\spl{
\d s^2_{4E} 
 &= -e^{ -\frac{1}{6}c_A\tau  } \text{cosh}^{\frac{3}{2}} (c_{B}\tau) \text{sech}^{\frac{18}{5}}(c_{\phi}\tau) \d \tau^2
+ e^{ -\frac{1}{18}c_A\tau  } \text{cosh}^{\frac{1}{2}} (c_{B}\tau) \text{sech}^{\frac{6}{5}}(c_{\phi}\tau) \d\vec{x}^{~\!2}
~,
}\label{Elp1}}
where we have set $d_A$, $d_\phi$ and $d_{B}=0$ for simplicity. 
To examine the asymptotic behavior of the metric it suffices to consider $c_\varphi\geq0$. 
The metric asymptotically 
takes the form \eqref{intro1}, with $a_\pm =  \tfrac13$.


\subsubsection*{Type II solution: $\lambda<0$}\label{sec:exception}

\begin{figure}[H]
\begin{center}
\includegraphics[width=0.8\textwidth]{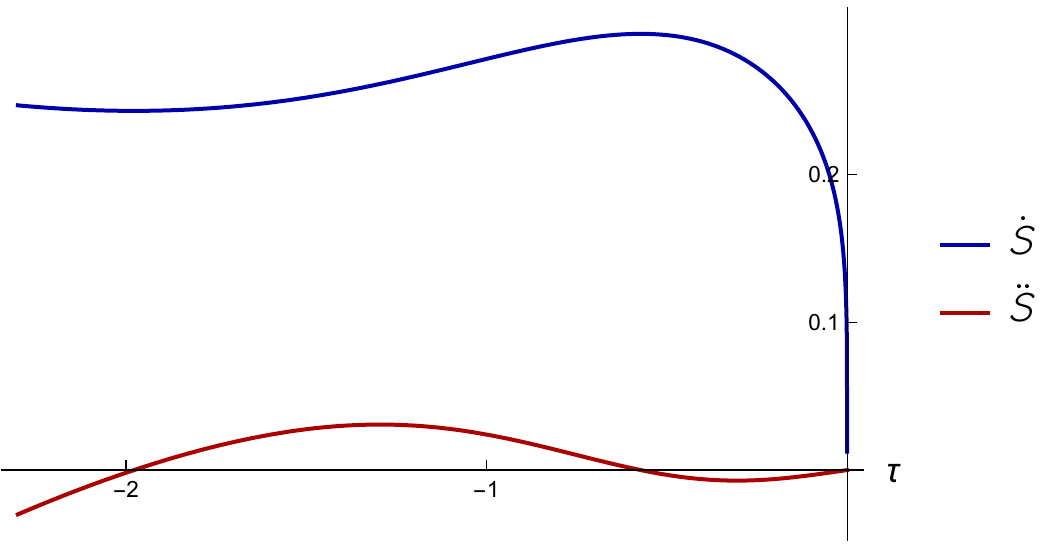}
\caption{Plot of $\dot{S}(T), \ddot{S}(T)$ as a function of  $\tau$, for $c_A=1$. In the $\tau\rightarrow 0$ limit, we have  $\dot{S}(T), \ddot{S}(T)\rightarrow 0$.}\label{fig:acceleration6}
\end{center}
\end{figure}

Let us now set  $\lambda<0$. 
The 4d Einstein metric  reads
\eq{\spl{
\d s^2_{4E} 
 &= -e^{ -\frac{1}{6}c_A\tau  } \text{cosh}^{\frac{3}{2}} (c_{B}\tau) \text{csch}^{\frac{18}{5}}(c_{\phi}\tau) \d \tau^2
+ e^{ -\frac{1}{18}c_A\tau  } \text{cosh}^{\frac{1}{2}} (c_{B}\tau)  \text{csch}^{\frac{6}{5}}(c_{\phi}\tau)\d\vec{x}^{~\!2}
~.
}\label{Elp2}}
The $\tau\rightarrow\pm\infty$ asymptotics of the warp factors are the same as for $\lambda>0$. For $\tau\rightarrow 0$, on the other hand, 
the function $f$ in \eqref{290} tends to a constant, while   $g\rightarrow -\ln(5|\lambda|\tau^2)$. Moreover, the constraint imposes $c_A$, $c_B=0$. 
Hence, in the $\tau\rightarrow 0$ limit, which is reached at infinite proper time, the solution reads
\eq{
A=d_A -\tfrac{1}{10}\ln(5|\lambda|\tau^2)
 ~;~~~B=-\tfrac83d_A+\tfrac{1}{10}\ln(5|\lambda|\tau^2)~;~~~
\phi=d_\phi
~.}
The metric takes the form
\eq{
\d s^2\rightarrow  -\d T^2+T^{\frac32}\d\vec{x}^2
~,}
where we have defined $T\propto \tau^{-\frac45}$. This is {\it not} an exact solution of the theory, unless $c_\varphi=0$. In this case we recover precisely 
the critical solution with $\lambda<0$. Hence the model interpolates between two metrics of the form 
 \eqref{intro1}, one with $a =  \tfrac13$ for $|\tau|\rightarrow \infty$, and one with $a =  \tfrac34$ for $\tau\rightarrow 0$.  
 For a certain range of parameters, we can have transient accelerated expansion, where $\dot{S}(T), \ddot{S}(T)>0$, cf. Figure \ref{fig:acceleration6}.

\subsection{Compactification on Einstein-K\"{a}hler manifolds}\label{sec:ekm}

We will now consider  (massive) IIA backgrounds for which the internal 6d manifold is Einstein-K\"{a}hler, 
\eq{R_{mn}=\lambda g_{mn}~,}
where $R_{mn}$ is the Ricci tensor associated to $g_{mn}$, and $\lambda\in\mathbb{R}$.  
In addition there is a real, closed K\"{a}hler two-form $J$, $\d J=0$.

Let us first consider the massless limit, and we take the form ansatz,
\eq{
  F=0 ~;~~~
  H=\d\chi \wedge J+\d\beta ~;~~~
G=\varphi\text{vol}_4+  J\wedge \d\gamma  +\tfrac12 c_0 J\wedge J 
~.}
The internal $(m,n)$-components of the Einstein 
equations now read
\eq{\spl{
e^{2B}\lambda&=e^{-8A-2B}\nabla^{\mu}\left(
e^{8A+2B}\partial_{\mu}A
\right)  
+\tfrac18e^{-\phi-4A}(\partial\chi)^2
-\tfrac{1}{48}e^{-\phi-4A-4B}h^2\\
& 
+\tfrac{3}{16} 
e^{\phi/2-6A-6B}\varphi^2
+\tfrac{7}{16}e^{\phi/2-6A+2B} c_0^2 
~.}}

The external $(\mu,\nu)$-components read
\eq{\spl{
R^{(4)}_{\mu\nu}&=
g_{\mu\nu}\left(\nabla^{2}A+\nabla^{2} B+
8(\partial A)^2+2(\partial B)^2+10\partial A\cdot \partial B\right)
\\
&-8\partial_{\mu}A\partial_{\nu}A-2\partial_{\mu}B\partial_{\nu}B
-16\partial_{(\mu}A\partial_{\nu)}B+8\nabla_{\mu}\partial_{\nu}A+2\nabla_{\mu}\partial_{\nu}B
\\
&+\tfrac32 e^{-\phi-4A} \partial_{\mu}\chi\partial_{\nu}\chi
+\tfrac14 e^{\phi-4A-4B} h^2_{\mu\nu}
+\tfrac12 \partial_{\mu}\phi\partial_{\nu}\phi
 \\
&+\tfrac{1}{16} g_{\mu\nu}\Big(  
-\tfrac{1}{3}e^{\phi-4A-4B}h^2
 -6e^{-\phi-4A}(\partial\chi)^2-5e^{\phi/2-6A-6B}\varphi^2  
-9 c_0^2e^{\phi/2-6A+2B}   
\Big)
~,}}
while the mixed $(\mu,m)$-components are automatically satisfied. 
The dilaton equation reads
\eq{\spl{
0&=e^{-10A-4B}\nabla^{\mu}\left(
e^{8A+2B}\partial_{\mu}\phi
\right)
+\tfrac32e^{-\phi-6A-2B}(\partial\chi)^2\\
&+\tfrac{1}{12}e^{-\phi-6A-6B}h^2
+\tfrac{1}{4}
e^{\phi/2-8A-8B}\varphi^2
-\tfrac{3}{4}c_0^2 e^{\phi/2-8A} 
~.}}
The $F$-form equation of motion 
reduces to the condition
\eq{\spl{
0 &= \varphi ~\!  h
~.}}
The $H$-form equation reduces to
\eq{\spl{
\d\left(
e^{-\phi+4A+2B}\star_4\d\chi
\right)  &=  c_0~\!  \varphi 
 ~\!\text{vol}_4  
~.}}
The $G$-form equation of motion reduces to
\eq{\spl{
 0&=   c_0 ~\! h
~,}}
together with the constraint
\eq{
0=\d\left(
 \varphi e^{\phi/2+2A-4B} + 3c_0\chi 
\right)
~.}
The latter can be  integrated to solve for $\varphi$ in terms of the other fields,
\eq{
\varphi=\left(C -  3c_0\chi\ \right)e^{-2A+4B-\phi/2}
~,}
where $C$ is an arbitrary constant.

\subsection*{\boldmath$c_0=0$\unboldmath}

In this case, we set 
\eq{
\varphi= c_\varphi e^{-2A+4B-\phi/2}~;~~~
\d_t\chi=c_\chi  e^{\phi-4A-2B}
~,}
and in addition we  impose
\eq{
c_{\varphi}  c_{h}=0
~.}
All form equations are then satisfied. 
The internal Einstein and dilaton equations reduce to
\eq{\spl{\label{intek}
\d_\tau^2 A&=-\lambda e^{16A+6B}+
\tfrac{3}{16}c^2_{\varphi} e^{-\phi/2+6A+6B} -\tfrac{1}{8}c^2_{h} e^{-\phi+12A}
-\tfrac{1}{8}c^2_{\chi} e^{\phi+4A} \\
 \d_\tau^2 \phi&= \tfrac{1}{4}c^2_{\varphi} e^{-\phi/2+6A+6B}+
 \tfrac{1}{2}c^2_{h} e^{-\phi+12A}
 -\tfrac{3}{2}c^2_{\chi} e^{\phi+4A} 
~,}}
The external Einstein equations 
reduce to
\eq{\spl{ \label{extek}
\lambda e^{16A+6B}-2ke^{16A+4B} -\tfrac{1}{2}c^2_{\varphi} e^{-\phi/2+6A+6B}+\tfrac{1}{2}c^2_{h} e^{-\phi+12A}+ \tfrac{1}{2}c^2_{\chi} e^{\phi+4A} 
&= \d_\tau^2B
\\
-12\lambda e^{16A+6B}-12ke^{16A+4B} +c^2_{\varphi} e^{-\phi/2+6A+6B}+c^2_{h} e^{-\phi+12A}+3c^2_{\chi} e^{\phi+4A} &=
\\
&\hskip-6cm 144(\d_\tau A)^2+12(\d_\tau B)^2+96\d_\tau A\d_\tau B- (\d_\tau\phi)^2 
~.}}

\subsubsection*{Critical solution: $h$, $\lambda\neq0$}

Setting $k$, $c_\chi$, $c_\varphi=0$, equations \eqref{intek}, \eqref{extek} admit the solution
\eq{\label{EK1}
A=-\tfrac{11}{64}\ln |\tau|+A_0~;~~~B=\tfrac{1}{8}\ln|\tau|+B_0~;~~~\phi=-\tfrac{1}{16}\ln|\tau|+\phi_0~,
}
for $A_0$, $B_0$, $\phi_0$ arbitrary real constants 
and
\eq{
c_h^2=\tfrac{1}{8}e^{-12 A_0 + \phi_0}~;~~~\lambda=-\tfrac{3}{16}e^{-16 A_0 - 6 B_0}~.
}
We thus obtain a power-law, flat universe expansion,
\eq{
\d s^2_{4E}=\d T^2+T^{\frac{18}{11}} \d\vec{x}^2
~,}
where we have set $T\propto |\tau|^{-\frac{11}{16}}$.  
The warp factor of the internal space and the dilaton scale as $e^A\propto T^{1/4}$, $e^\phi\propto T^{1/11}$ respectively.

\subsection*{\boldmath$c_0\neq 0$\unboldmath}

In this case, we reinstate the Romans mass ($m\neq 0$) and we set
\eq{
\varphi= 0~;~~~
 \chi  =0~;~~~h=0
~.}
All form equations are then satisfied. 
The internal Einstein and dilaton equations reduce to
\eq{\spl{\label{intek2}
\d_\tau^2 A&=-\lambda e^{16A+6B}  +\tfrac{7}{16}c_0^2 e^{\phi/2+10A+6B}+\tfrac{1}{16}m^2 e^{5\phi/2+18A+6B}\\
 \d_\tau^2 \phi&= -\tfrac{3}{4}c_0^2 e^{\phi/2+10A+6B} -\tfrac{5}{4}m^2 e^{5\phi/2+18A+6B}
~.}}
The external Einstein equations 
reduce to
\eq{\spl{ \label{extek2}
\lambda e^{16A+6B}-2ke^{16A+4B}  - c_0^2 e^{\phi/2+10A+6B}
&= \d_\tau^2B
\\
-12\lambda e^{16A+6B}-12ke^{16A+4B} +3 c_0^2 e^{\phi/2+10A+6B} +m^2 e^{5\phi/2+18A+6B} &=
144(\d_\tau A)^2+12(\d_\tau B)^2\\
&+96\d_\tau A\d_\tau B- (\d_\tau\phi)^2 
~.}}

\subsubsection*{Critical solution: $k$, $c_0\neq0$}

Setting $m$,  $\lambda=0$, equations \eqref{intek2}, \eqref{extek2} admit the solution
\eq{\label{EK2}
A=-\tfrac{7}{104}\ln \tau+A_0~;~~~B=-\tfrac{3}{13}\ln \tau+B_0~;~~~\phi=\tfrac{3}{26}\ln\tau+\phi_0~,
}
for $A_0$, $B_0$, $\phi_0$ arbitrary real constants 
and
\eq{
c_0^2=\tfrac{2}{13}e^{-10 A_0-6B_0 - \phi_0/2}~;~~~k=-\tfrac{5}{26}e^{-16 A_0 - 4 B_0}~.
}
We thus obtain a singular Milne universe,
\eq{
\d s^2_{4E}=\d T^2+\tfrac{13}{10} |k|T^2\d\Omega_k^2
~.}

\subsection*{\boldmath$c_f\neq 0$\unboldmath}

A different ansatz with non-vanishing two-form flux is also possible, 
\eq{\label{f2an}
  F=c_f J ~;~~~
  H=0 ~;~~~
G=\varphi\text{vol}_4 
~,}
where $c_f$ is a constant. This automatically satisfies the Bianchi identities and the form equations of motion provided
\eq{
\varphi= c_\varphi e^{-2A+4B-\phi/2}~,
}
for some constant $c_\varphi$. 
The remaining equations of motion 	are as in \eqref{eintd}, \eqref{eext12}, 
with potential given by
\eq{\label{wap}
U=
\tfrac12 c^2_{\varphi} e^{-\phi/2+6A+6B}+\tfrac32c_f^2e^{3\phi/2+14A+6B}
  -6ke^{16A+4B}-6\lambda e^{16A+6B}
~.}

\subsubsection*{Critical solution: $k$, $c_f\neq0$}

The equations of motion with potential \eqref{wap} admit the solution
\eq{\label{EK27}
A=-\tfrac{5}{104}\ln \tau+A_0~;~~~B=-\tfrac{4}{13}\ln \tau+B_0~;~~~\phi=\tfrac{9}{26}\ln\tau+\phi_0~,
}
for $A_0$, $B_0$, $\phi_0$ arbitrary real constants 
and
\eq{
c_f^2=\tfrac{2}{13}e^{-14 A_0-6B_0 - 3\phi_0/2}~;~~~k=-\tfrac{5}{26}e^{-16 A_0 - 4 B_0}~.
}
We thus obtain a Milne universe with angle defect,
\eq{
\d s^2_{4E}=\d T^2+\tfrac{13}{10} |k|T^2\d\Omega_k^2
~.}
%


 \section{Two-flux dynamical systems}
 \label{app:sys}
 
 In this appendix, we provide some more details on the dynamical systems arising from 
 compactifications with two species of fluxes turned on. For each of them, we give the system of equations together with the constraint, the equation defining the invariant plane $\mathcal{P}$ (when it exists), the equation defining the acceleration region, the equation of state parameter $w$, as well as the list of critical points together with their stability analysis whenever it is relevant. 

\subsection{Compactification on Calabi-Yau manifolds}

\subsection*{\boldmath$\varphi, \xi \neq 0$\unboldmath}

System of equations, 
\eq{\spl{
x'&=  \frac{3}{2} \left(x^2-2 x z^2+y^2+4 z^2-1\right) \\
y'&=  \frac{1}{2} \left(-\sqrt{3} x^2-\sqrt{3} y^2-6 y z^2+\sqrt{3}\right) \\
z'&=  -\frac{1}{2} z \left(9 x+\sqrt{3} y+6 z^2-6\right)
~.}\label{spx}}
Constraint,
\begin{equation}
    -2 z^2 c_{\xi \xi '}^2 e^{-6 B}= c_{\varphi }^2 \left(x^2+y^2+z^2-1\right) \,.
\end{equation}
Invariant plane,
\eq{
 x+ \sqrt{3}  y=2
~.}
Acceleration condition,
\eq{
z>\sqrt{\frac{2}{3}} 
~.}
Equation of state parameter,
\begin{equation}
    w=1-2z^2 \,.
\end{equation}
Critical points: $p_\mathcal{C}$.

\subsection*{\boldmath$\varphi, k \neq 0$\unboldmath}

System of equations, 
\eq{\spl{
x'&=  2 x^3+x \left(2 y^2-z^2-2\right)+\frac{9}{2}z^2 \\
y'&=  y \left(2 x^2-z^2-2\right)+2 y^3+\frac{\sqrt{3}}{2}z^2 \\
z'&= -\frac{1}{2} z \left(-4 x^2+9 x-4 y^2+\sqrt{3} y+2 z^2-2\right)
~.}\label{spk}}
Constraint, 
\begin{equation}
    12 k z^2 e^{10 A-2 B+\frac{\phi }{2}}= c_{\varphi }^2 \left(x^2+y^2+z^2-1\right)~. 
\end{equation}
Invariant plane,
\eq{
 x- 3\sqrt{3}  y=0
~.}
Acceleration condition,
\eq{
z^2>2(x^2+y^2)
~,}
Equation of state parameter,
\eq{
w=\frac{x^2+y^2-z^2}{x^2+y^2+z^2}~. 
}
Critical points, 
\eq{
p_\mathcal{C}~,~~~ p_0~,~~~ p_1 = \left(\frac{3}{14},\frac{1}{14 \sqrt{3}},\sqrt{\frac{2}{21}}\right)~.}
The point $p_1$ lies on the boundary of the acceleration region. 
The linearized system at $p_1$ has one real and two complex eigenvalues: $-2, -1\pm i \sqrt{\frac{17}{7}}$. It follows that 
$p_1$ is a stable focus (resp. node) for trajectories within (resp. orthogonal to) $\mathcal{P}$. This model is qualitatively very similar to the one analyzed in detail in Section \ref{case2}.

\subsection*{\boldmath$\chi, \xi \neq 0$\unboldmath}

System of equations, 
\eq{\spl{
x'&=  \frac{1}{2} \left(3 x^2+3 y^2+z^2-3\right) \\
y'&=  -\frac{1}{2} \sqrt{3} \left(x^2+y^2+3 z^2-1\right) \\
z'&= z \left(x+\sqrt{3} y\right)
~.}\label{scx}}
Constraint, 
\begin{equation}
   -\frac{2}{3} z^2 e^{2 A-\frac{3 \phi }{2}} c_{\xi \xi '}^2= c_{\chi }^2 \left(x^2+y^2+z^2-1\right)~. 
\end{equation}
The acceleration condition is 
impossible to satisfy, and $w=1$. \\ \\
Critical points:  $p_\mathcal{C}$.

\subsection*{\boldmath$\chi, h \neq 0$\unboldmath}

System of equations, 
\eq{\spl{
x'&=  3 x^2+3 y^2+2 z^2-3 \\
y'&=  -\sqrt{3} \left(x^2+y^2+2 z^2-1\right) \\
z'&= z \left(x+\sqrt{3} y\right)
~.}\label{sch}}
Constraint, 
\begin{equation}
-\frac{1}{3} z^2  c_h^2 e^{8 A-2 \phi }= c_{\chi }^2 \left(x^2+y^2+z^2-1\right)~. 
\end{equation}
The acceleration condition is impossible to satisfy, and $w=1$. \\ \\
Critical points:  $p_\mathcal{C}$.

\subsection*{\boldmath$\chi, k \neq 0$\unboldmath}

System of equations, 
\eq{\spl{
x'&=  2 x^3+2 x \left(y^2+z^2-1\right)-z^2 \\
y'&=  2 y \left(x^2+y^2+z^2-1\right)-\sqrt{3} z^2 \\
z'&= z \left(2 x^2+x+2 y^2+\sqrt{3} y+2 z^2-2\right)
~.}\label{sck}}
Constraint, 
\begin{equation}
4 k z^2 e^{12 A+4 B-\phi }= c_{\chi }^2 \left(x^2+y^2+z^2-1\right)~. 
\end{equation}
Invariant plane,
\eq{
 3x- \sqrt{3}  y=0
~.}
Acceleration condition impossible to satisfy, and  $w=1$. \\ \\
Critical points:  $p_\mathcal{C}$, $p_0$.

\subsection*{\boldmath$\xi, k \neq 0$\unboldmath}

System of equations, 
\eq{\spl{
x'&=  2 x \left(x^2+y^2+z^2-1\right)-\frac{3}{2} z^2 \\
y'&=  2 y \left(x^2+y^2+z^2-1\right)+\frac{\sqrt{3}}{2} z^2 \\
z'&= \frac{1}{4} z \left(8 \left(x^2+y^2+z^2-1\right)+6 x-2 \sqrt{3} y\right)
~.}\label{sxk}}
Constraint, 
\begin{equation}
6 k z^2 e^{10 A+4 B+\frac{\phi }{2}}=c_{\xi \xi '}^2 \left(x^2+y^2+z^2-1\right)~. 
\end{equation}
Invariant plane,
\eq{
x+ \sqrt{3}  y=0
~.}
Acceleration condition impossible to satisfy, and $w=1$. \\ \\ 
Critical points:  $p_\mathcal{C}$, $p_0$.

\subsection*{\boldmath$\xi, h \neq 0$\unboldmath}

System of equations, 
\eq{\spl{
x'&=  \frac{3}{2} \left(2 x^2+2 y^2+z^2-2\right) \\
y'&=  -\frac{1}{2} \sqrt{3} \left(2 x^2+2 y^2+z^2-2\right) \\
z'&= \frac{1}{2} z \left(3 x-\sqrt{3} y\right)
~.}\label{sxh}}
Constraint, 
\begin{equation}
-\frac{1}{2} z^2 c_h^2 e^{6 A-\frac{\phi }{2}} =c_{\xi \xi '}^2 \left(x^2+y^2+z^2-1\right)~. 
\end{equation}
Acceleration condition impossible to satisfy, and  $w=1$. \\ \\
Critical points:  $p_\mathcal{C}$.

\subsection*{\boldmath$h, k \neq 0$\unboldmath}

System of equations, 
\eq{\spl{
x'&=  2 x \left(x^2+y^2+z^2-1\right)-3 z^2 \\
y'&=  2 y \left(x^2+y^2+z^2-1\right)+\sqrt{3} z^2 \\
z'&= z \left(2 x^2+3 x+2 y^2-\sqrt{3} y+2 z^2-2\right)
~.}\label{shk}}
Constraint, 
\begin{equation}
12 k z^2 e^{4 A+4 B+\phi }= c_h^2 \left(x^2+y^2+z^2-1\right)~. 
\end{equation}
Invariant plane,
\eq{
x+ \sqrt{3}  y=0
~.}
Acceleration condition impossible to satisfy, and  $w=1$. \\ \\
Critical points:  $p_\mathcal{C}$, $p_0$.

\subsection*{\boldmath$c_0, b_0 \neq 0$\unboldmath}

System of equations, 
\eq{\spl{
x'&=  \frac{1}{2} \left(6 (x-1) \left(x^2+y^2-1\right)+z^2\right) \\
y'&=  -\left(\sqrt{3}-3 y\right) \left(x^2+y^2\right)-3 y+\frac{1}{2} \sqrt{3} \left(2-3 z^2\right) \\
z'&= \frac{1}{2} z \left(x (6 x-7)+y \left(6 y+\sqrt{3}\right)\right)
~.}\label{sczbz}}
Constraint, 
\begin{equation}
-72 b_0^2 z^2 e^{2 A-\frac{3 \phi }{2}}=\frac{3}{2} c_0^2 \left(x^2+y^2+z^2-1\right)~. 
\end{equation}
Invariant plane,
\eq{
9x+ \sqrt{3} y = 10
~.}
Acceleration condition,
\eq{
3\left(x^2+y^2\right)<1 
~.}
Equation of state parameter,
\begin{equation}
w=-1 + 2 \left(x^2 + y^2\right) ~.
\end{equation}
Critical points,
\eq{
p_\mathcal{C}~,~~~ p_1=\left(1,\tfrac{1}{\sqrt{3}},0\right)~,}
all outside the acceleration region.The point $p_1$ lies in the region exterior to $\mathcal{S}$.

\subsection{Compactification on Einstein manifolds}

\subsection*{\boldmath$\varphi$, $\lambda\neq 0$\unboldmath}

System of equations, 
\eq{\spl{
x'&=(3 x-2) (x^2 + y^2-1) + \tfrac52 z^2\\
y'&=3 y ( x^2 + y^2-1) +\tfrac{\sqrt{3}}{2} z^2\\
z'&=\tfrac12 z\left[ -9 x + 6 x^2 + y \left( 6 y-\sqrt{3}\right)\right]
~.}\label{spl}}
Constraint, 
\begin{equation}
6 \lambda  z^2 e^{10 A+\frac{\phi }{2}}=\frac{1}{2} c_{\phi }^2 \left(x^2+y^2+z^2-1\right)~. 
\end{equation}
Invariant plane,
\eq{
 3 x - 5 \sqrt{3} y-2=0
~.}
Acceleration condition,
\eq{
\tfrac13 > x^2 + y^2 
~.}
Equation of state parameter,
\begin{equation}
w=-1 + 2 \left(x^2 + y^2\right) ~.
\end{equation}
Critical points, 

\eq{
p_\mathcal{C} ~,  ~~~p_1=\left(\tfrac23,0,0\right)~,
} 
all outside of the acceleration region. In particular $p_1$ lies on the invariant plane.


\subsection*{\boldmath$\varphi$, $m\neq 0$\unboldmath}

System of equations, 
\eq{\spl{
x'&=\tfrac32 (-1 + 2 x) \left(-1 + x^2 + y^2\right) + 3 z^2\\
y'&=\tfrac12 \left[-6 y + \left(5 \sqrt{3}+ 6 y\right) \left(x^2 + y^2\right) + \sqrt{3} \left(-5 + 6 z^2\right)\right] \\
z'&=\tfrac12 z\left[ -9 x + 6 x^2 + y \left( 6 y-\sqrt{3}\right)\right]
~.}\label{spm}}
Constraint,
\begin{equation}
    -\frac{1}{2} m^2 z^2 e^{12A+ 3\phi}= \frac{1}{2} c_{\phi }^2 \left(x^2+y^2+z^2-1\right)~.
\end{equation}
Invariant plane,
\eq{
 3 x -  \sqrt{3} y-4=0
~.}
Acceleration condition,
\eq{
\tfrac13 > x^2 + y^2 
~.}
Equation of state parameter,
\begin{equation}
w=-1 + 2 \left(x^2 + y^2\right) ~.
\end{equation}
Critical points,
\eq{
p_\mathcal{C}~,~~~ p_1=\left(\tfrac12,-\tfrac{5}{2\sqrt{3}},0\right)~,}
all outside the acceleration region. In particular $p_1$ lies on the invariant plane and in the region exterior to $\mathcal{S}$.

\subsection*{\boldmath$\chi$, $\lambda\neq 0$\unboldmath}

System of equations, 
\eq{\spl{
x'&=(3 x-2) ( x^2 + y^2-1 ) + 3 (x-1 ) z^2 \\
y'&=- \sqrt{3}  z^2 + 3 y (x^2 + y^2 + z^2-1) \\
z'&=z \left[ x + \sqrt{3} y+3( x^2 +  y^2+z^2-1)  \right]
~.}\label{scl}}
Constraint,
\begin{equation}
6 \lambda  z^2 e^{12 A+6 B-\phi }=\frac{3}{2} c_{\chi }^2 \left(x^2+y^2+z^2-1\right)~.
\end{equation}
Invariant plane,
\eq{
 3 x -  \sqrt{3} y-2=0
~.}
Acceleration condition,
\eq{
\tfrac13 > x^2 + y^2+ z^2
~,}
Equation of state parameter,
\begin{equation}
w=-1 + 2 (x^2 + y^2+z^2) ~.
\end{equation}
Critical points,
\eq{
p_\mathcal{C}~,~~~p_1=\left(\tfrac23,0,0\right) 
~.
}
The point $p_1$  lies  on the invariant plane,  outside the acceleration region.


\subsection*{\boldmath$h$, $\lambda\neq 0$\unboldmath}

System of equations, 
\eq{\spl{
x'&= (-2 + 3 x) (-1 + x^2 + y^2) + (-5 + 3 x) z^2 \\
y'&= \sqrt{3}  z^2 + 3 y (-1 + x^2 + y^2 + z^2) \\
z'&= z \left[ 3x - \sqrt{3} y+3( x^2 +  y^2+z^2-1)  \right]
~.}\label{shl}}
Constraint,
\begin{equation}
6 \lambda  z^2 e^{4 A+6 B+\phi }=\frac{1}{2} c_h^2 \left(x^2+y^2+z^2-1\right)~.
\end{equation}
Invariant plane,
\eq{
 x +  \sqrt{3} y-\tfrac23=0
~.}
Acceleration condition,
\eq{
\tfrac13 >x^2 + y^2 + z^2
~.}
Equation of state parameter,
\begin{equation}
w=-1 + 2 (x^2 + y^2+z^2) ~.
\end{equation}
Critical points,
\eq{
p_\mathcal{C}~,~~~ 
p_1=\left(\tfrac23,0,0\right)~,~~~ 
p_2=\left(\tfrac{11}{18},\tfrac{1}{18\sqrt{3}},\tfrac29\sqrt{\tfrac{2}{3} }\right)
~.
}
The points $p_1$ and $p_2$ both lie on the invariant plane, in the region inside   the invariant sphere $\mathcal{S}$,  outside the acceleration region.


\subsection*{\boldmath$c_0$, $m\neq 0$\unboldmath}

System of equations, 
\eq{\spl{
x'&= \tfrac32 (-1 + 2 x) (-1 + x^2 + y^2) + 2 z^2 \\
y'&= \tfrac12 \left[-6 y + (5  \sqrt{3} + 6 y) (x^2 + y^2) +  \sqrt{3} (-5 + 4 z^2)\right] \\
z'&= \tfrac12 z \left[ x (-7 + 6 x) + y (  \sqrt{3} + 6 y) \right]
~.}\label{sczm}}
Constraint,
\begin{equation}
-\frac{1}{2} m^2 z^2 e^{8 A+2 \phi }=\frac{3}{2} c_0^2 \left(x^2+y^2+z^2-1\right)~.
\end{equation}
Invariant plane,
\eq{
 3 x -  \sqrt{3}  y-4 =0
~.}
Acceleration condition,
\eq{
\tfrac13 >x^2 + y^2  
~.}
Equation of state parameter,
\begin{equation}
    w=-1 + 2 (x^2 + y^2 )~.
\end{equation}
Critical points, 
\eq{
p_\mathcal{C}~,~~~ p_1=\left(\tfrac12,-\tfrac{5}{2\sqrt{3}},0\right)~,}
all outside the acceleration region. In particular $p_1$ lies on the invariant plane and in the region exterior to $\mathcal{S}$.


\subsection*{\boldmath$c_0$, $k\neq 0$\unboldmath}

System of equations, 
\eq{\spl{
x'&=2 x (-1 + x^2 + y^2) + \tfrac12 (7 - 2 x) z^2 \\
y'&= -\tfrac12  (\sqrt{3}  + 6 y) z^2 + 2 y (-1 + x^2 + y^2 + z^2) \\
z'&= \tfrac12  z \left[ 2 + x (-7 + 4 x) + y (\sqrt{3} + 4 y) - 2 z^2 \right]
~.}\label{sczk}}
Constraint,
\begin{equation}
6 k z^2 e^{6 A-2 B-\frac{\phi }{2}}=\frac{3}{2} c_0^2 \left(x^2+y^2+z^2-1\right)~.
\end{equation}
Invariant plane,
\eq{
 3 x + 7 \sqrt{3}  y=0
~.}
Acceleration condition,
\eq{
z^2 >2(x^2 + y^2)  
~.}
Equation of state parameter,
\begin{equation}
    w=-\tfrac13 + \tfrac43 (x^2 + y^2 )-\tfrac23 z^2~.
\end{equation}
Critical points, 
\eq{
p_\mathcal{C}~,~~~p_0~,~~~ p_1=\left(\tfrac{7}{26},-\tfrac{\sqrt{3}}{26},\sqrt{\tfrac{2}{13}}\right)~.}
The point $p_1$   lies in the region inside  the invariant sphere $\mathcal{S}$. Both $p_0$ and $p_1$ lie on the boundary of the acceleration region and on the invariant plane. 
The linearized system at $p_1$ has one real and two complex eigenvalues: $-2, -1\pm 3i \sqrt{\frac{3}{13}}$. It follows that 
$p_1$ is a stable focus (resp. node) for trajectories within (resp. orthogonal to) $\mathcal{P}$. 
This model is qualitatively very similar to the one analyzed in detail in Section \ref{case2}.


\subsection*{\boldmath$c_0$, $\lambda\neq 0$\unboldmath}

System of equations, 
\eq{\spl{
x'&=(-2 + 3 x) (-1 + x^2 + y^2) + \tfrac32 z^2 \\
y'&=3 y (-1 + x^2 + y^2) - \tfrac{\sqrt{3}}{2} z^2     \\
z'&= \tfrac12 z \left[ x (-7 + 6 x) + y (  \sqrt{3} + 6 y) \right]
~.}\label{sczl}}
Constraint,
\begin{equation}
6 \lambda  z^2 e^{6 A-\frac{\phi }{2}}=\frac{3}{2} c_0^2 \left(x^2+y^2+z^2-1\right)~.
\end{equation}
Invariant plane,
\eq{
 x +\sqrt{3}  y-\tfrac23 =0
~.}
Acceleration condition,
\eq{
\tfrac13 >x^2 + y^2  
~,}
with $w=-1 + 2 (x^2 + y^2 )$. Critical points, 
\eq{
p_\mathcal{C}~;~~~ p_1=(\tfrac23,0,0)~,}
all of them outside the acceleration region. $p_1$ lies in the interior of $\mathcal{S}$ and on the invariant plane.


\subsection{Compactification on Einstein-K\"{a}hler manifolds}


\subsection*{\boldmath$\varphi$, $c_f$ $\neq 0$\unboldmath}

System of equations, 
\eq{\spl{
x'&= \tfrac12 (-5 + 6 x) (-1 + x^2 + y^2) + 2 z^2 \\
y'&= \tfrac12 \left[-6 y + 3( \sqrt{3} + 2 y) (x^2 + y^2) +  \sqrt{3} (-3 + 4 z^2)\right] \\
z'&= \tfrac12 z \left[ -9 x + 6 x^2+ y (  -\sqrt{3} + 6 y) \right]
~.}\label{sfp}}
Constraint,
\begin{equation}
-\frac{3}{2}c_f^2 z^2 e^{8 A+2 \phi } =\frac{1}{2}c_{\phi }^2 \left(x^2+y^2+z^2-1\right)~.
\end{equation}
Invariant plane,
\eq{
 3 x -  \sqrt{3}  y-4 =0
~.}
Acceleration condition,
\eq{
\tfrac13 >x^2 + y^2  
~.}
Equation of state parameter,
\begin{equation}
w=-1 + 2 \left(x^2 + y^2\right) ~.
\end{equation}
Critical points,
\eq{
p_\mathcal{C}~,~~~ p_1=\left(\tfrac56,-\tfrac{\sqrt{3}}{2},0\right)~,}
all outside the acceleration region. In particular $p_1$ lies on the invariant plane and in the region exterior to $\mathcal{S}$.


\subsection*{\boldmath$k$, $c_f\neq 0$\unboldmath}

System of equations, 
\eq{\spl{
x'&=2 x \left(x^2+y^2-1\right)+\frac{1}{2} (5-2 x) z^2  \\
y'&=2 y \left(x^2+y^2-1\right)-y z^2-\frac{3 \sqrt{3} }{2}z^2 \\
z'&=\frac{1}{2} z \left(x (4 x-5)+4 y^2+3 \sqrt{3} y-2 z^2+2\right)
~.}\label{skf}}
Constraint,
\begin{equation}
6 k z^2 e^{2 A-2 B-\frac{3 \phi }{2}}=\frac{3}{2} c_f^2 \left(x^2+y^2+z^2-1\right)~.
\end{equation}
Invariant plane,
\eq{
 9 x + 5 \sqrt{3} y=0
~.}
Acceleration condition,
\eq{
z^2 >2(x^2 + y^2)  
~.}
Equation of state parameter,
\begin{equation}
    w=\frac{x^2+y^2-z^2}{x^2+y^2+z^2}~,
\end{equation}
Critical points,

\eq{
p_\mathcal{C}~,~~~p_0~,~~~
p_1=\left(\tfrac{5}{26},-\tfrac{3\sqrt{3}}{26},\sqrt{\tfrac{10}{13} }\right)~.
}
The point $p_1$ lies in the region inside of the invariant sphere $\mathcal{S}$. It lies on the boundary of 
the acceleration region, as well as on the invariant plane. The linearized system at $p_1$ has one real and two complex eigenvalues: $-2, -1\pm 3i \sqrt{\frac{3}{13}}$. It follows that 
$p_1$ is a stable focus (resp. node) for trajectories within (resp. orthogonal to) $\mathcal{P}$. This model is qualitatively very similar to the one analyzed in detail in Section \ref{case2}.

\subsection*{\boldmath$\lambda$, $c_f\neq 0$\unboldmath}

System of equations, 
\eq{\spl{
x'&=\tfrac12 \left[ (-5 + 6 x) (-1 + x^2 + y^2)- z^2 \right]  \\
y'&=\tfrac32 \left[ ( \sqrt{3}+ 2 y) (x^2 + y^2-1) + \sqrt{3} z^2 \right] \\
z'&=z \left[3( x^2 +  y^2) -2x \right]
~.}\label{slf}}
Constraint,
\begin{equation}
6 \lambda  \left(x^2+y^2+z^2-1\right)=\frac{3}{2} z^2 e^{\frac{3 \phi }{2}-2 A} c_f^2~.
\end{equation}
Invariant plane,
\eq{
 9 x +  \sqrt{3} y-6=0
~.}
Acceleration condition,
\eq{
\tfrac13 > x^2 + y^2
~.}
Equation of state parameter,
\begin{equation}
w=-1 + 2 \left(x^2 + y^2\right) ~.
\end{equation}
Critical points,
\eq{
p_\mathcal{C}~,~~~p_1=\left(\tfrac23,0,\tfrac{\sqrt{5}}{3}\right)~,~~~
p_2=\left(\tfrac{9}{14},\tfrac{\sqrt{3}}{14},\tfrac{4\sqrt{2}}{7} \right)~,~~~
p_3=\left(\tfrac56,-\tfrac{\sqrt{3}}{2},0\right)
~.
}
The points $p_2$ and $p_3$ lie in the region exterior to $\mathcal{S}$, while $p_1$ lies on the invariant sphere.  The points $p_1$, $p_2$, $p_3$ all lie on the invariant plane and 
outside the region of acceleration.

\newpage

\bibliography{refs}
\bibliographystyle{JHEP}
\end{document}